\documentclass[aps,prb,reprint,superscriptaddress,longbibliography]{revtex4-1}

\usepackage{etoolbox}
\usepackage{braket}
\patchcmd{\section}
{\centering}
{\raggedright}
{}
{}
\patchcmd{\subsection}
{\centering}
{\raggedright}
{}
{}

\usepackage{graphicx}
\usepackage{float}
\usepackage{color}
\usepackage{bm}
\usepackage{ulem}
\usepackage{amsmath}
\definecolor{darkblue}{rgb}{0,0,0.5}
\definecolor{darkgreen}{rgb}{0,0.5,0}
\definecolor{darkred}{rgb}{.7,0,0}
\definecolor{purple}{rgb}{0.5,0,0.6}
\definecolor{orange}{rgb}{1,0.5,0}
\definecolor{grey}{rgb}{.6,.6,.6}
\definecolor{lightpink}{rgb}{1,0.7,0.75}
\definecolor{pink}{rgb}{1,0.4,0.58}
\definecolor{deeppink}{rgb}{1,0.08,0.58}

\newcommand{\st}[1]{{\color{black}{#1}}}
\newcommand{\cb}[1]{{\color{black}{#1}}}

\newcommand{\pr}[1]{{\color{black}{#1}}}
\newcommand{\cg}[1]{{\color{black}{#1}}}

\begin{document}

%\title{Control of propagating quantum electronic states}
\title{Coherent control of single electrons: a review of current progress}
%\title{\st{: a review of recent progress}}

\author{Christopher B\"auerle}
\affiliation{Univ. Grenoble Alpes, CNRS, Grenoble INP, Institut N\'eel, 38000 Grenoble, France}

\author{D. Christian Glattli}
\affiliation{Service de Physique de l' \'Etat Condens\'e (CNRS UMR 3680), IRAMIS, CEA-Saclay, F-91191 Gif-Sur-Yvette, France}

\author{Tristan Meunier}
\affiliation{Univ. Grenoble Alpes, CNRS, Grenoble INP, Institut N\'eel, 38000 Grenoble, France}

\author{Fabien Portier}
\affiliation{Service de Physique de l' \'Etat Condens\'e (CNRS UMR 3680), IRAMIS, CEA-Saclay, F-91191 Gif-Sur-Yvette, France}

\author{Patrice Roche}
\affiliation{Service de Physique de l' \'Etat Condens\'e (CNRS UMR 3680), IRAMIS, CEA-Saclay, F-91191 Gif-Sur-Yvette, France}

\author{Preden Roulleau}
\affiliation{Service de Physique de l' \'Etat Condens\'e (CNRS UMR 3680), IRAMIS, CEA-Saclay, F-91191 Gif-Sur-Yvette, France}

\author{Shintaro Takada}
\affiliation{Univ. Grenoble Alpes, CNRS, Grenoble INP, Institut N\'eel, 38000 Grenoble, France}
\affiliation{National Institute of Advanced Industrial Science and Technology (AIST), National Metrology Institute of
Japan (NMIJ), Tsukuba, Ibaraki 305?8563, Japan}

\author{Xavier Waintal}
\affiliation{Univ. Grenoble Alpes, INAC-PHELIQS, F-38000 Grenoble, France}

\date{\today}

% --------------------------------------------------------------------------------------------------
% ----------------  Guide line of ROPP --------------------------------------
% --------------------------------------------------------------------------------------------------
%
%
% ---------------------------------------------------------------------------------------------------


\maketitle

\tableofcontents

\vspace{4cm}

\textbf{Abstract}

In this \cb{report we review} the present state of the art of the control of propagating quantum states at the single-electron level and its potential application to quantum information processing. 
We give an overview of the different approaches which have been developed over the last ten years in order to gain full control over a propagating single electron in a solid state system.
After a brief introduction of the basic concepts, we present experiments on flying qubit circuits for ensemble of electrons measured in the low frequency (DC) limit.
We then present the basic ingredients necessary to realise such experiments at the single-electron level. This includes a review of the various single electron sources which are compatible with integrated single electron circuits. This is followed by a review of recent key experiments on electron quantum optics with single electrons.
Finally we will present recent developments about the new physics that emerges using ultrashort voltage pulses.
We conclude our review with an outlook and future challenges in the field.

%\tableofcontents

%%%%%%%%%%%%%%%%%%%%
\section{Introduction:}   
\label{Introduction}
%%%%%%%%%%%%%%%%%%%%
In current semiconductor technology, where the integrated circuits are composed of transistors, which are nowadays as small as a few tens of nanometers in scale,
the electronic circuits are still operated with of a huge number of electrons. 
The ultimate goal, in this respect, is the realisation of integrated circuits at the single-electron level.
Over the past decade, an important effort has been made in the field of low-dimensional electronic conductors towards \textit{single electron electronics} with the goal to gain full control over single electrons in solid state devices. 
Nowadays it is possible to confine and manipulate single electrons in a very controlled way in semiconductor nanostructures such as nanowires or quantum dots \cite{kouwenhoven_QDreview_1997,vanderwiel_rmp_2002,fujisawa_ropp_2006,defranceschi_nnano_2010,kouwenhoven_rmp_2015}.
However, in order for the single electron circuits of the future to lead to useful applications, one requires a mechanism to transport and interconnect a single electron from one functional part of the circuit to another and to manipulate it in a very controlled way. 
In addition, the ability to control single electrons on-demand enables electron quantum optics experiments where single electrons emitted in a ballistic electronic interferometer play the role of single photons emitted in an optical medium in quantum optics. 

Coherent manipulation of single electrons in solid state devices are also attractive for quantum information purposes because they have a high potential for scalability. Depending on the system used, the charge or the spin may code 
binary qubit information. A particular appealing idea is to use a single \textit{flying} electron itself as the conveyor of quantum information \cite{bertoni_prl_2000, ionicioiu_ijmp_2001, barnes_prb_2000, beenakker_prl_2004, bertrand_nnano_2016}. Such electronic flying qubits allow performing quantum operations on qubits while they are being coherently transferred. 
Information processing typically takes place in the nodes of the quantum network on locally controlled qubits, but quantum networking would require \textit{flying} qubits to exchange information from one location to another \cite{divincenzo_fortschritt_2000}. 
It is therefore of prime interest to develop ways of transferring information from one node to the other. The availability of flying qubits would enable the possibility to develop new non-local architectures for quantum computing with possibly cheaper hardware overhead than e.g. surface codes \cite{fowler_pra_2012}.

Photons in vacuum are a natural choice for flying qubits due to their long coherence time.
Solid state electronic devices have advantage in terms of size and hence for possible scalability, however, with a serious drawback of a much shorter coherence time.
Electronic spin states are often chosen as spatially localised qubits as they can be easily confined to a small volume \cite{vanderwiel_rmp_2002,hanson_rmp_2007,zwanenburg_rmp_2013,awschalom_science_2013}. Coherent transport of quantum information has been demonstrated in solid state systems by transporting single electrons in multiple quantum dot networks \cite{hanno_ncom_2017,vandersypen_npjQI_2017} or by coupling of superconducting qubits to microwave photons \cite{wallraff_nature_2004,schoelkopf_nature_2007,roch_prl_2014}. 
Recent advances in the field of electron quantum transport have shown that solid-state flying qubits based on single electrons are also very promising as these systems have possible applications in electron interferometry and entanglement.

In this review we concentrate on integrated electronic circuits operated at the single-electron level in semiconductor heterostructures and outline their potential towards a flying qubit architecture. We will focus in particular on quantum experiments appropriate for electron quantum optics where the emitted single electrons play the role of flying charge qubits. We give a comprehensive review of the present state of the art and put emphasis on the connection to quantum information processing and the physical phenomena underlying realistic devices.
%
%%%%%%%%%%%%%%%%%%%%
%%%%%%%%%%%%%%%%%%%%
\section{Basic theoretical concepts for flying qubit architectures with single electrons}   
\label{basic concepts}
%%%%%%%%%%%%%%%%%%%%
%%%%%%%%%%%%%%%%%%%%

The flying qubits that will be discussed in this review aim to encode the quantum information into two different paths (or rails) that can be taken by an electron during its propagation.
Such a quantum rail can be defined by a one-dimensional channel along which a single-electron wave packet is propagating ballistically. 
This can be experimentally realised in a gate-defined nanostructure on top of a two-dimensional electron gas formed at the interface of a semiconductor heterostructure and will be described in detail in section \ref{Flying qubit circuits}.
In order to realise this flying qubit architecture with single flying electrons, it is necessary to be able to control the state of an electron via two independent qubit rotations.
%%%%%%%%%%%%%%%%%%%%
\subsubsection{Single-qubit operations}
\label{single operation}
\begin{figure}
\includegraphics[width=8.5cm]{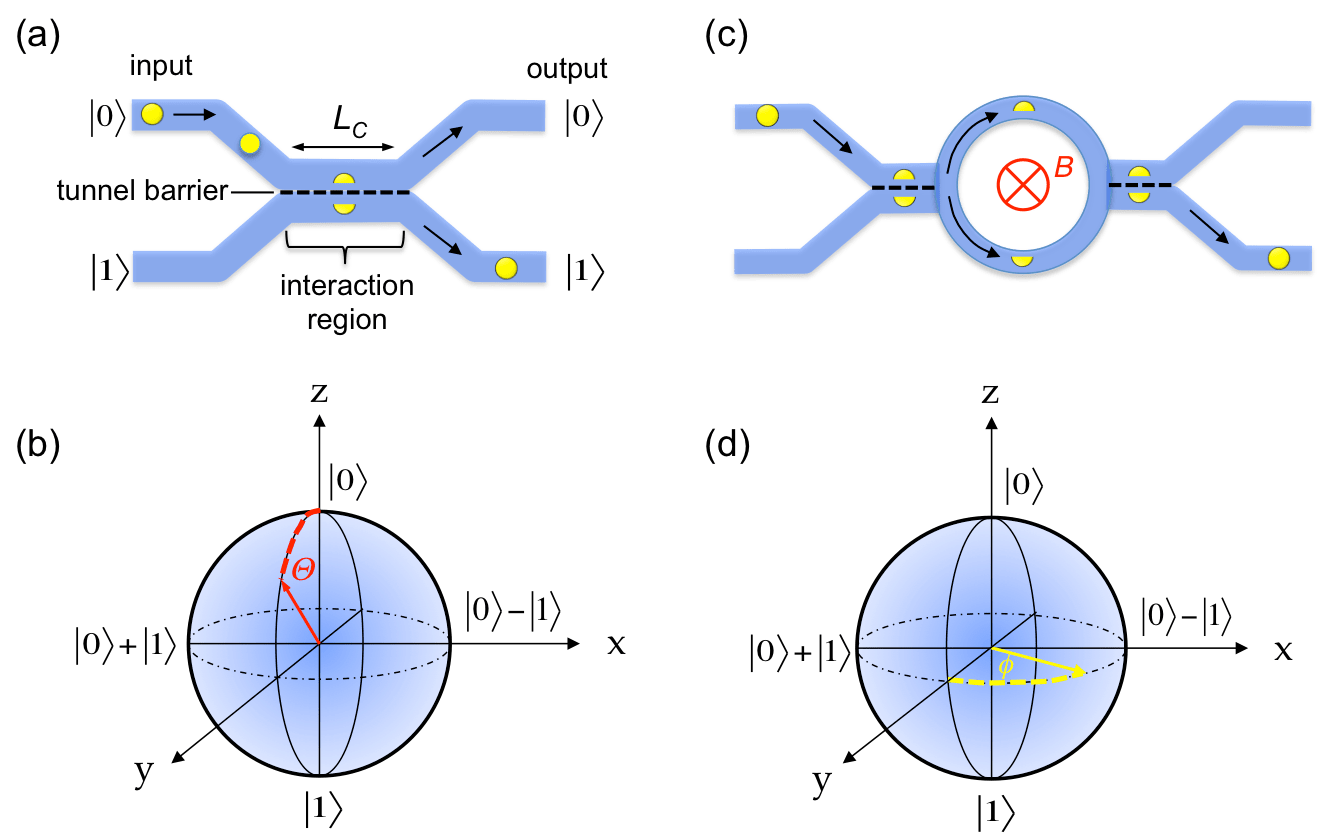}
\caption{{\bf{Implementation of qubit-rotations using single electron channels.}} (a) two single electron channels representing the qubit are brought together to an interaction region where they are tunnel-coupled over a length $L_C$. 
The tunnel-coupling energy of the two propagating quantum states induces a rotation of angle  $ \Theta$ around the x-axis as shown on the corresponding Bloch sphere (b). 
As a result, an electron (yellow dot) injected into channel $\ket{0}$ will oscillate in the interaction region between the upper and lower channel before being projected into the output channels. 
The two \textit{half} dots schematise a superposition of $1/\sqrt{2}$ ($\alpha \ket{0}$ + $ \beta \ket{1}$) within the interaction region.  The tunnel barrier indicated by the black dashed line allows varying the rotation angle $ \Theta$ by changing the tunnel-coupling between the two electron channels.
(c) Two path interferometer for single electrons. An electron is injected in the upper channel and passed through a beam splitter (tunnel-coupled wire) before entering the Aharonov-Bohm interferometer. 
The magnetic field induces a phase shift $\phi$ between the upper and lower path which allows realising a rotation around the z-axis on the Bloch sphere as shown in (d). }
\label{fig_TC+AB}
\end{figure}

%%%%%%%%%%%%%%%%%%%%
Let us define the two qubit states $\ket{0}$ and $\ket{1}$ on the Bloch sphere as shown in figure \ref{fig_TC+AB}.
A rotation around the x-axis with rotation angle $\theta$ is described by the following rotation matrix:

\begin{equation}
S_x(\theta)=
 \left(\begin{array}{cc} \cos \frac{\theta}{2} & i\sin \frac{\theta}{2} \\  i\sin \frac{\theta}{2} & \cos \frac{\theta}{2} \end{array} \right)
\end{equation}

%$$
%S_x(\theta)=
%\begin{pmatrix}
%\cos \frac{\theta}{2} & i \sin \frac{\theta}{2} \\
%i \sin \frac{\theta}{2} & \cos \frac{\theta}{2}
%\end{pmatrix}; 
%$$
while a rotation around the z-axis with rotation angle $\phi$ is given by : 

\begin{equation}
S_z(\phi) =
 \left(\begin{array}{cc} e^{-i \frac{\phi}{2}} & 0 \\  0 & e^{i \frac{\phi}{2}} \end{array} \right)
\end{equation}

Such rotation matrices\cb{ can also be expressed in terms of Pauli matrices: $S_x(\theta) = {\rm exp}(i\sigma_x \theta/2)$, $S_z(\phi) = {\rm exp}(-i\sigma_z \phi/2)$.
They} are usually referred to as single qubit rotations \cite{chuang_book}. In order to construct any arbitrary state on the Bloch sphere, it is enough to combine two of the three possible rotation matrices.
Combining the two rotation matrices, it is also possible to construct a universal transformation $U$ \cite{weinfurter_pra_1995,bertoni_prl_2000} 
%check Chuang book  
\begin{equation}
U (\alpha, \beta, \phi) = S_x(\alpha - \frac{\pi}{2}) S_z(\phi) S_x(\beta + \frac{\pi}{2})
\label{eq:single_rotation}
\end{equation}

\begin{figure}
\includegraphics[width=8.5cm]{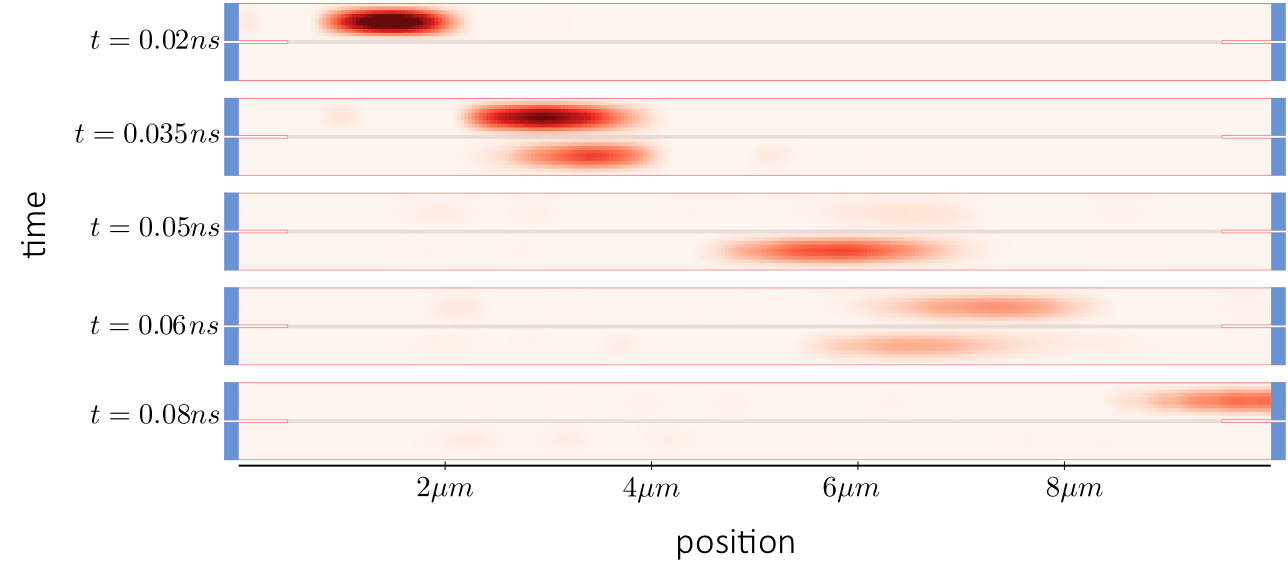}
\caption{{\bf{Tunnelling of an electron wave packet.}} Real time simulations of the propagation of a single-electron wave packet through a 10 $\mu$m long tunnel-coupled wire. Each individual quantum wire has a width of 200 nm. A charge pulse is injected into the upper quantum wire and tunnels across the tunnel barrier as it propagates along the quantum wire creating a coherent superposition of states. Snapshots are taken at different instances of time varying from 20 to 80\,ps. The colour code measures the additional electronic density with respect to equilibrium (dark red: high density; light red: low density). 
}
\label{simulation_TCW}
\end{figure}

% some general comments on qubits
Such a scheme can be implemented into a coherent nanoelectronic circuit by coupling two quantum rails \cite{ionicioiu_ijmp_2001} as schematised in figure \ref{fig_TC+AB}(a).
One can define two qubit states $\ket{0}$ and $\ket{1}$ by the presence of the electron in the lowest energy state of one of the two rails: 
\vspace{0.5cm}
\\
\hspace{1cm}  $\ket{0}$ = electron present in the upper rail \\
\hspace{1cm}  $\ket{1}$ = electron present in the lower rail
\vspace{0.5cm}
\\
When \st{the confinement potential of }the two quantum wires are\st{ identical in the region, where the two quantum wires are} coupled by a tunnel barrier (interaction region), hybridisation between these two states occurs and the new eigenbasis is given by the symmetric $\ket{S}$ and antisymmetric state $\ket{A}$ \cite{alamo_apl_1989,tsukuda_apl_1990,xu_jap_1993} :
\begin{equation}
\ket{S}=\frac{1}{\sqrt{2}} \Big(\ket{0}+\ket{1} \Big); \ket{A}=\frac{1}{\sqrt{2}} \Big( \ket{0}-\ket{1} \Big)
\end{equation}
By injecting an electron into the upper rail $\ket{0}$, the wave
function will evolve into a superposition of  $1/\sqrt{2}$ ($\ket{S}$ + $\ket{A}$). 
While travelling through the interaction region of length $L_C$, the wave function of the electron will then pick up a phase and will evolve into $1/\sqrt{2}$ ($e^{ik_S L_C} \ket{S}  + e^{ik_A L_C} \ket{A}$), with $k_S$ ($k_A$) being the wave vector of the symmetric (antisymmetric) state. 
Projecting back the wave function onto the two output channels allows obtaining the probability of finding an electron in the upper (lower) channel. Doing the same calculation by injecting an electron in state $\ket{1}$ (lower rail) one can then work out the complete transmission matrix which reads:
\begin{equation}
S_{tw}=
\exp(i\frac{k_S+k_A}{2}L) \left(\begin{array}{cc} \cos(\frac{k_S-k_A}{2}L)& i\sin(\frac{k_S-k_A}{2}L) \\  i\sin(\frac{k_S-k_A}{2}L) & \cos(\frac{k_S-k_A}{2}L) \end{array} \right)
\end{equation}
%-------------------------------------------------------------
%-------------------------------------------------------------
%------------------------------------ Fig. 1 ----------------
%-------------------------------------------------------------
%-------------------------------------------------------------
%-------------------------------------------------------------
%-------------------------------------------------------------
%------------------------------------ Fig. 1 ----------------
%-------------------------------------------------------------
%-------------------------------------------------------------
By comparing this to equation (1) one can immediately see that this matrix corresponds to a rotation matrix with rotation angle $\theta =  \Delta k \cdot L  = (k_S-k_A) \cdot L $. 
This means that the electron wave packet propagating through the tunnel-coupled wire will oscillate between states $\ket{0}$ and $\ket{1}$ (upper and lower wire) and hence represents a rotation around the x-axis of the Bloch sphere\cite{bautze_prb_2014}. 
Time resolved numerical simulations of the propagation of a single-electron wave packet through such tunnel-coupled wires for realistic experimental conditions can now be realised \cite{bertoni_prl_2000,waintal_phys-rep_2014, bautze_prb_2014}. An example of such real-time simulations is shown in figure \ref{simulation_TCW}.

As mentioned above, to attain any arbitrary final state on the Bloch sphere, another rotation has to be implemented. 
This can be done by connecting the wire to a ring geometry in order to pick up an additional phase due to the Aharonov-Bohm effect \cite{aharonov-bohm_prb_1959,tonomura_physics-today_2009} \st{as schematised in Fig.\,\ref{fig_TC+AB}(c)}.
Electron interference arises due to a phase difference between electrons passing through the upper or lower path. 
The phase difference is given by  
$\Delta \phi = \int {\bm{k}} \cdot {\rm d}{\bm l} - (e / \hbar) B \cdot dS $, 
where $\bm{k}$ is the wave vector, \st{${\bm l}$ the path along the ring}, $B$ the magnetic field and $S$ the surface area enclosed by the ring structure.
One way of changing $\Delta \phi $ is to simply change the Aharonov-Bohm flux. 
From an experimental point of view, this is however not very practical as the magnetic field cannot be changed on fast timescales. 
For an electron travelling at the Fermi velocity of approximately 1$\times 10^5$ m/s passing through a 10 $\mu$m long coherent quantum conductor will only take 100 ps. 
A more practical way is to modify the phase by changing the wave vector $\bm{k}$.
This can be done on a very fast time scale using an electrostatic gate\st{.
Such a modification of $\Delta \phi$ by an electrostatic gate} will be described in section \ref{MC0B}. 
Combining both single qubit rotations, the tunnel-coupled wire as well as the Aharonov-Bohm ring, one can then entirely control the phase of the electron and realise a flying qubit architecture \cite{yamamoto_nnano_2012}.
%-------------------------------------------------------------
%-------------------------------------------------------------
%------------------------------------ Fig. 2 ----------------
%-------------------------------------------------------------
%-------------------------------------------------------------
%-------------------------------------------------------------
%-------------------------------------------------------------
%------------------------------------ Fig. 2 ----------------
%-------------------------------------------------------------
%-------------------------------------------------------------
%%%%%%%%%%%%%%%%%%%%%%%%%%%%%%%%
\subsubsection{Two-qubit operations}
\label{two operations}
\begin{figure}
\includegraphics[width=8.5cm]{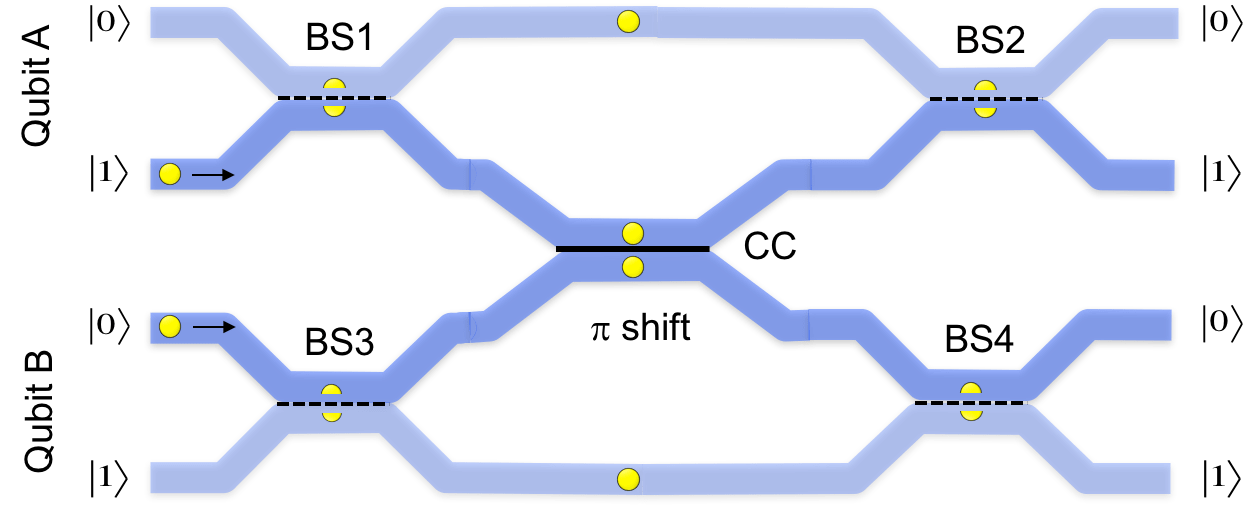}
\caption{ {\bf{Two-qubit gate using single electron channels.}} Possible experimental set-up for the implementation of a two-qubit gate for single electrons. 
The blue regions represent the one-dimensional channels defined by electrostatic gates (not shown) in the 2DEG. 
One of the two qubits can be used as the control qubit whereas the other is the target qubit. 
The black dashed lines are beam splitters (BS) and the interaction region (black solid line) acts as a Coulomb coupler (CC) between the two electrons launched simultaneously from qubit A and B.}
\label{fig_2-qubit-gate}
\end{figure}

%%%%%%%%%%%%%%%%%%%%%%%%%%%%%%%%
The next step is to combine the single qubit operations in order to perform a two-qubit operation \cite{ionicioiu_ijmp_2001}.
It is actually possible to use the interaction region to control the state of one qubit with a second qubit and to realise for instance a controlled phase gate. 
This quantum gate exploits the Coulomb interaction between two single electrons in two different pairs of coupled quantum wires. 
So far we have only considered one electron at a time in the interaction region. 
It is however possible to control the phase of an electron in one of the two rails by the presence of another electron in the other rail due to Coulomb interaction. 
%This scheme is usually termed a \textit{Coulomb coupler} (CC) and allows using a second \textit{control} qubit in order to control the phase of the \textit{target} qubit. 
This scheme is usually termed a \textit{Coulomb coupler} (CC) and allows controlling the phase of the \textit{target} qubit using a second \textit{control} qubit.
\st{A possible design for the realisation of a controlled phase gate for flying electrons using a CC is shown in figure \ref{fig_2-qubit-gate}.
Two qubits (A and B) are coupled in the center by a CC.
In addition four tunnel-coupled wires are added, which allow for tunnelling between the wires and hence controlling the rotation angle $\Theta$ of each individual qubit.
When set at $\Theta = \pi/2$, they act as beam splitters.
Other schemes have been proposed to implement two qubit gates such as ballistic Aharonov-Bohm qubits \cite{yu_ssc_2008,schomerus_njp_2007} or surface acoustic wave driven electrons \cite{barnes_prb_2000}. 
Here we concentrate on the most general case which can be applied to both systems.}

The induced phase $\chi$ on each electron\st{ in the CC} is proportional to the coupling strength \cb{$\Delta$} (capacitive coupling) between the two rails and the interaction time\st{ $t_{\rm c}$}.
In order to have a strong coupling between the two electrons in the interaction region, the two qubit rails have to be sufficiently close.
On the other hand, tunnelling between the two rails should be suppressed. 
A \st{CC} should therefore have a large potential barrier in order to prevent electron tunnelling from one rail to the other, and at the same time the two rails should be close enough to induce a significant phase shift. 
For the experimental systems we will describe in the following, this can be achieved by adjusting the barrier height induced by the electrostatic gate of the \st{CC} which separates the two rails as well as the length of the interaction region.
%-------------------------------------------------------------
%-------------------------------------------------------------
%------------------------------------ Fig. 3 ----------------
%-------------------------------------------------------------
%-------------------------------------------------------------
%-------------------------------------------------------------
%-------------------------------------------------------------
%------------------------------------ Fig. 3 ----------------
%-------------------------------------------------------------
%-------------------------------------------------------------
\st{The operation of the CC in figure \ref{fig_2-qubit-gate} can hence be written as:
\vspace{0.5cm}
\\
\hspace{5cm} $\ket{00} \rightarrow \ket{00}$  \\
$\ket{01} \rightarrow \ket{01}$  \\
$\ket{10} \rightarrow e^{-i\chi} \ket{10}$  \\
$\ket{11} \rightarrow \ket{11}$,  \\
\\
where $\chi = 2\Delta t_{\rm c}$.
This phase shift $\chi$ can be measured experimentally by performing an interference experiment and observe a change in the detection probability of an electron in one of the output ports.
To do that, the rotation angle for BS1 and BS2 is set to $\pi/2$, the one for BS3 and BS4 is set to $0$.
In this case the probability $P_{\rm 0} $ of detecting an electron in $A\ket{0} $ oscillates as one changes the barrier height of the CC, which modulates $\chi$.
The \textit{inverted} oscillation $P_{\rm 1} = 1 - P_{\rm 0}$ should be observed in the other output.
\cb{For the same circuit a controlled phase gate can be implemented by setting $\chi$ to $\pi$, the rotation angle $\Theta$ for BS1 and BS2 to $0$, the one for BS3 to $\pi$, and the one for BS4 to $3\pi$. }}

This device structure also allows to entangle two different qubits \cite{bertoni_epjd_2012}.
To do so one can for instance activate the beam splitters\st{ ($\Theta = \pi/2$ for all BS)} and 
adjust the Coulomb coupler in such a way that a $\pi$ phase shift is induced between the two propagating electrons. 
\st{By sending synchronously one electron into input  A1 and B0 the outcome of such a scheme is a maximally entangled Bell state \cite{ionicioiu_pra_2001} $1/\sqrt{2}(-i A\ket{0} B \ket{-} - A\ket{1}  B \ket{+} )$, where $\ket{+} = 1/\sqrt{2}(\ket{0}+i\ket{1})$ and $\ket{-} = 1/\sqrt{2}(\ket{0}-i\ket{1})$.}
Another interesting feature of this system is that it is easily scalable by simply adding several qubits in parallel to realise a multi qubit system.
\st{Combination of an arbitrary single qubit rotation ($U$ in Eq. \ref{eq:single_rotation}) and a controlled phase gate allows then to perform an arbitrary unitary operation for a n-qubit system \cite{chuang_book}. A similar approach to scale up the system is also used in linear quantum optics \cite{knill_nature_2001, obrien_nature_03,obrien_science_2008,obrien_science_2015}.}

The experimental realisation of such a system is by all means not trivial. 
Several important requirements have to be fulfilled such as high fidelity on-demand single electron injection as well as single-shot read-out of the electrons at the output ports. 
Another important requirement is that the electrons within the different quantum rails have to be synchronised at all times in order to properly perform two-qubit gating, as the two electrons have to reach simultaneously the Coulomb coupling window.
% check: CNOT vs CPhase
%beam splitters are used to realise a quantum superposition between two channels and one beam splitter is used to control the interaction between two flying electrons and induce the phase shift between the non-interacting and the interacting path.
All these issues will be addressed in detail in sections \ref{Flying qubit circuits}-\ref{Quantum optics like experiments} by giving an overview of the different approaches developed in the field to gain full control of electron transport at the single-electron level.
%
%%%%%%%%%%%%%%%%%%%%%%%%%%%%%%%%%%%%
%%%%%%%%%%%%%%%%%%%%%%%%%%%%%%%%%%%%
%\subsection{{\color{red} Electrons versus photons ( Christian + Preden)}}
%%%%%%%%%%%%%%%%%%%%%%%%%%%%%%%%%%%%
%%%%%%%%%%%%%%%%%%%%%%%%%%%%%%%%%%%%
%
% advantage over photons: important difference is the coupling strength. 
%The photon coupling is weaker than the electron coupling and therefore photons have a longer coherence time. However, due to this weak coupling, it is much more difficult to construct a 2 qubit gate which operates at the single photon level. 
% Either ones needs a huge third order susceptibilities  or concept of linear quantum computing.  
% Also it is much easier to detect and prepare single electron states than a single photon states. 
% another advantage is the large ...  
%
%%%%%%%%%%%%%%%%%%%%%%%%%%%%%%%%%%%%
%%%%%%%%%%%%%%%%%%%%%%%%%%%%%%%%%%%%
%\section{Flying qubit circuits in the stationary limit}
\section{Low-frequency transport in quantum coherent circuits}
\label{Flying qubit circuits}
%%%%%%%%%%%%%%%%%%%%%%%%%%%%%%%%%%%%
%%%%%%%%%%%%%%%%%%%%%%%%%%%%%%%%%%%%

\st{We have seen in section \ref{basic concepts} that beam splitters, phase shifters and interferometers are the basic elements needed to realise electronic flying qubits.
The most promising experimental systems to realise such electronic flying qubits are at present two-dimensional electron systems formed at the interface of a GaAs/AlGaAs heterostructure.
These systems are extremely well mastered and electron trajectories can be easily implemented by engineering the desired quantum rails using electrostatic gates deposited \cb{on} the surface of the GaAs heterostructure.
The phase coherence length $L_{\rm \phi}$ in such systems can attain several tens of micrometers at low temperatures ($< 100$ mK) \cite{roulleau_prl_2008,niimi_prl_2009,niimi_prb_2010}, which is sufficient to implement many gate operations on the fly.

T}he control and manipulation of the phase of an electron is of prime importance for the realisation of flying qubits at the single-electron level. 
In order to access the phase of an electron it is usually convenient to realise a two-path interference experiment with electrons\cite{electron-diffraction,tonomura-original,tonomura_physics-today_2009}, similar to the well known Young's double slit experiment for photons \cite{young_original,hecht_optics}.
A simple realisation is an Aharonov-Bohm ring of micrometer size, such that phase coherence of the electrons is ensured throughout the entire device \cite{webb_prl_1985, prober_prl_1985, webb_review_1986, yacoby_prl_1995,heiblum_nature_1998,ensslin_nlett_2008} 
when working at low temperatures.  
At the entrance of the ring structure the electron wave function is split into two paths and recombined at the output as depicted in figure \ref{fig_TC+AB}c. 
A phase difference is induced between the upper and lower arm of the ring by means of an externally applied magnetic field \footnote{more precisely, the electron couples to the vector potential \cite{aharonov-bohm_prb_1959} rather than the magnetic field, hence even if the electron does not experience a magnetic field, for instance when passing next to an infinite solenoidal coil, the electron picks up a phase $\phi = -e/\hbar \oint {\bm A} \cdot {\rm d}{\bm l}$ (see also ref. \cite{tonomura_physics-today_2009})}.
% maybe footnote to specify that it is the vector potential which counts
When scanning the magnetic field, the conductance oscillates as a function of magnetic field with a period which is proportional to the surface area enclosed by the two trajectories. As will be detailed below, the realisation of a \textit{true} two-path interferometer is quite challenging as electrons in solids behave quite differently than photons in vacuum. 
%Once the photons arrive at the screen where one can observe the interference pattern, they are absorbed and disappear. 
Contrary to photons, electrons in solids can backscatter and this complicates seriously the interpretation of the interference pattern \cite{buttiker-imry_prb_1985,buttiker-yeyati_prb_1995,yacoby_prb_1996}.

\begin{figure}[h]
\includegraphics[width=8.5cm]{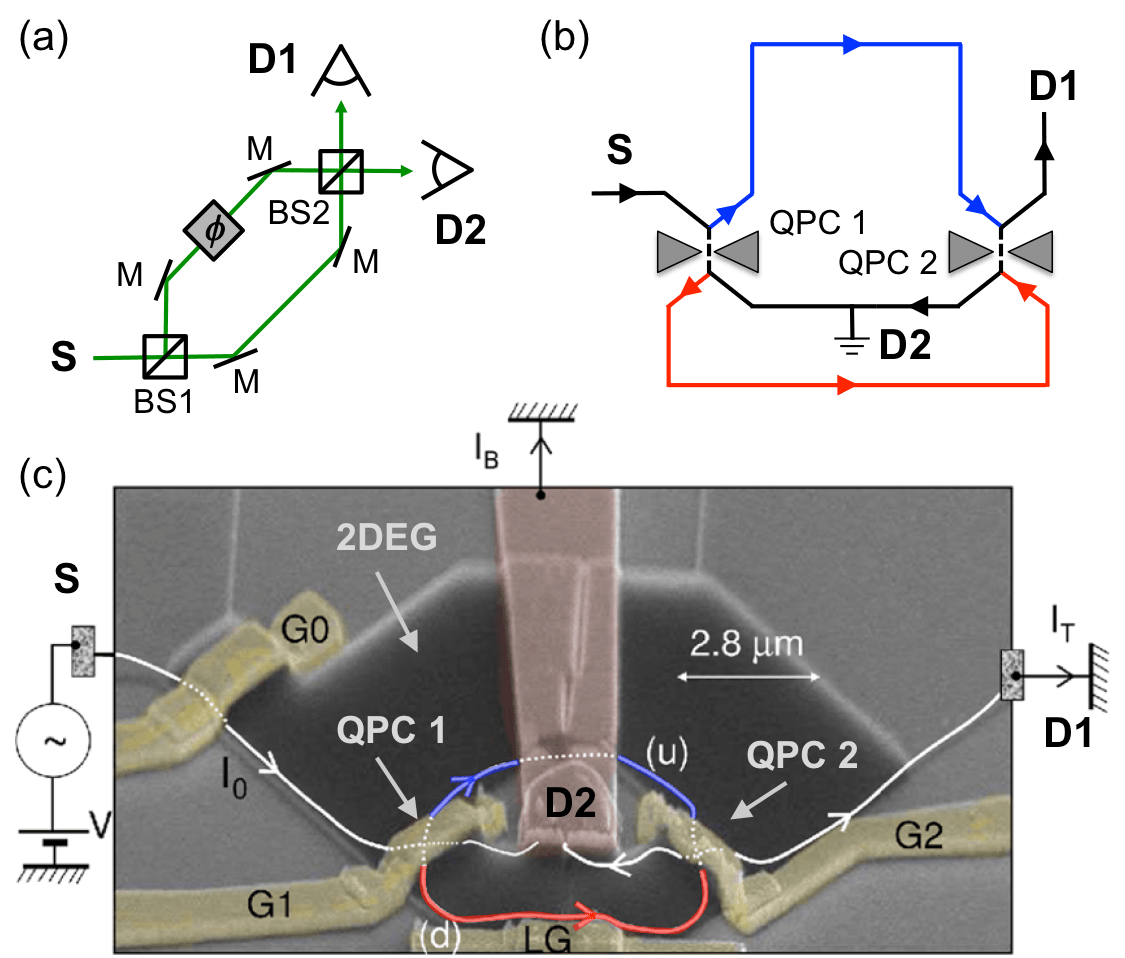}
\caption{\textbf{Mach--Zehnder interferometry.}
\cb{Mach--Zehnder interferometer (MZI) for photons (a) and for electrons (b). (a) Photons emitted from a photon source (S) are split by a beam splitter (BS) into two paths, guided with mirrors (M) towards a second beam splitter where they are recombined and then detected at the two detectors (D). 
A phase shifter allows to modulate the phase $\phi$ of the photon wave function. (b) Similar set-up as (a) but for electrons.
The beam splitters can be realised by means of quantum point contacts (QPCs) in the quantum Hall regime which leads to a 50 \% transmitted and 50 \% reflected beam. 
(c) Experimental realisation of an electronic MZI in the quantum Hall regime.  
The image shows a scanning electron micrograph (SEM) } with a schematic representation of the outer edge state. G0, G1, G2 are quantum point contacts which mimic beam splitters. The pairs of split gates defining a QPC are electrically connected via an Au metallic bridge deposited on an insulator (SU8). G0 allows to control the transmission of the impinging current, G1 and G2 are \cb{the two electrostatic gates which form the} beam splitters of the MZI. LG is a side gate which allows a variation of the length of the lower path (d) in red. The small ohmic contact in between the two arms (D2) collects the back scattered current I$_\textrm{B}$ to the ground through a long gold bridge. The two interfering electron trajectories are schematised in blue and red in (b) and (c).  (figure adapted from ref. \cite{roulleau_prl_2008}).}
\label{MZI.png}
\end{figure}

One possibility to overcome these unwanted effects is to work in the quantum Hall regime, where transport is realised along edge states. This system is particularly appealing as backscattering is suppressed due to the chirality of the system. It is for instance possible to realise the electronic analogue of the optical Mach --Zehnder interferometer (MZI). A beam splitter, made from a quantum point contact (QPC)  \cite{vanWees_prl_1988,wharam_jphysc_1988} splits an incoming electron beam into two independent beams which are guided towards a second beam splitter where they recombine and interfere \cite{heiblum_nature_2003,roulleau_prl_2008,strunk_prb_2008} 
as schematised in figure \ref{MZI.png}. In the past, this system has been used to study fundamental phenomena such as quantum coherence \cite{roche_prl_2012}, entanglement \cite{neder_prl_2007}, electron-electron interactions as well as two-particle interference \cite{neder_nature_2007}. 
% Fabry-Perot  \cite{camino_05,willet_pnas_09,halperin_11} 
%Over the last decade, a variety of interesting electronic devices have been proposed and realised to implement flying qubits 

Another possibility to realise a Mach--Zehnder (MZ) type interferometer is by combining an Aharonov-Bohm ring to two tunnel-coupled wires \cite{yamamoto_nnano_2012}. 
In this case (see figures \ref{fig_ABTunnelDevice} and \ref{fig_FQbitOperations} ) the tunnel-coupled wires \cite{debray_jpcm_2001, yamamoto_science_2006, gervais_science_2014} fulfill the function of the beam splitters as we have already seen in section \ref{basic concepts}. 
This system can be operated as a flying qubit MZ interferometer
by simply controlling the phase of the electrons via electrostatic gates \cite{bautze_prb_2014} at basically zero magnetic field. 
%and allows for instance to measure the phase shift of an electron across a quantum dot \cite{takada_prl_2014,takada_apl_2015}. 

\cb{Since Mach--Zehnder interferometry represents an essential ingredient for the realisation of electronic flying qubits, we review these two different Mach--Zehnder interferometers in more detail in the next two sections.} 
\cb{Let us emphasise that these experiments were done in the DC limit by applying a voltage at low frequencies  and billions of coherent electrons are passing through the nano-device. 
Experiments at the single-electron level with such interferometers are quite challenging and have not yet been realised.
We will come back to this issue in sections \ref{Single Electron Sources} -- \ref{Quantum optics like experiments}

%%%%%%%%%%%%%%%%%%%%%%%%%%%%
%When dealing with single-electron wavepackets, it is important that there is a substantial overlap between the interfering electrons at the output of the second beam splitter. It is hence important that the two inference arms are of similar length.  
%\cite{haack_prb_2011} \\
%The case where electron wave packets are much smaller than the size of the interferometer, new types of interference affects will appear. 
%This novel physics with single-electron wave packets will be described in section \ref{Novel quantum interference experiments} 
}
%%%%%%%%%%%%%%%%%%%%%%%%%%%

%%%%%%%%%%%%%%%%%%%%%%%%%%%%
%%%%%%%%%%%%%%%%%%%%%%%%%%%% 
\subsection{Mach--Zehnder interferometry in the quantum Hall regime}
\label{MCQH}
%%%%%%%%%%%%%%%%%%%%%%%%%%%%
%%%%%%%%%%%%%%%%%%%%%%%%%%%%
%-------------------------------------------------------------
%-------------------------------------------------------------
%------------------------------------ Fig. 4b ----------------
%-------------------------------------------------------------
%-------------------------------------------------------------

%-------------------------------------------------------------
%-------------------------------------------------------------
%------------------------------------ Fig. 4b ----------------
%-------------------------------------------------------------
%-------------------------------------------------------------

In this section, we discuss the different type of electronic interferometers which have been studied in the quantum Hall regime. In GaAs/AlGaAs, several questions have been addressed to improve our understanding of electronic interferometry and electron entanglement. An edge state in the quantum Hall regime, together with the electronic beam splitter, is a building block of electronic interferometry. Historically, one of the first addressed questions was to unveil the coherence properties of the edge states in the quantum Hall regime. This has been realized using a MZI \cite{heiblum_nature_2003,neder_prl_2006,roulleau_prb_2007,roulleau_prl_2008,strunk_prb_2008}. \pr{Another big challenge was to use MZIs to make basic quantum operations.} The combination of the two MZIs lead to the two-electron Aharonov-Bohm interferometer and should allow for generation of entangled states \cite{samuelsson_prl_2004,neder_nature_2007}, assuming an appropriate tuning of the electronic beam splitters.

\subsubsection{The electronic Mach--Zehnder interferometer}

\begin{figure}[h]
\includegraphics[width=8.5cm]{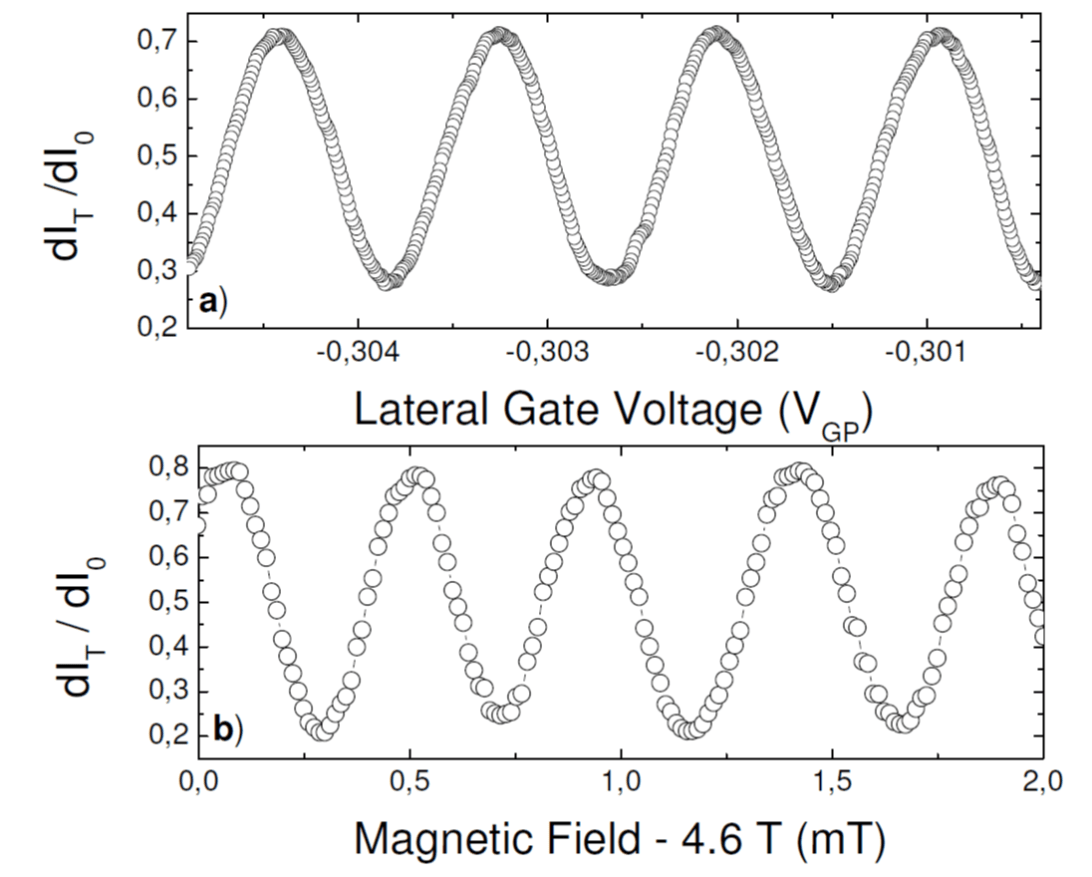}
\caption{\textbf{Interference measurements.} (a) MZI transmission as a function of the lateral gate voltage. (b) MZI transmission as a function of the magnetic field. The oscillation period is 0.46 mT which corresponds to a surface defined by the arms equal to 8.5 $\mu$m$^2$, in good agreement with the designed geometry of the MZI (7.25 $\mu$m$^2$). Interference pattern obtained at filling factor 2 and 20mK.}
\label{osc.png}
\end{figure}

The electronic MZI is the electronic counterpart of the optical one \cite{heiblum_nature_2003}, 
QPCs working as electronic beam splitters and ohmic contacts as detectors. A first QPC splits the incoming edge current to an upper path (u) and a down path (d) (see Fig. \ref{MZI.png}). The two electronic trajectories will follow the edge of the sample designed to ensure a zero length difference between the upper (u) and down (d) trajectories. 
%A non-zero length difference would be an additional source of decoherence. 
The two electronic trajectories recombine then on a second QPC. This leads to interference which is visible in the measured transmitted current I$_{\textrm{T}}$, while the reflected part of the current is collected in a grounded inner tiny ohmic contact. This is an important point in order to avoid electrons from being re-injected into the interferometer that would lead to a more complicated interference pattern.  
\pr{The transmission probability \textrm{T} through the MZI is, 
$\textrm{T}$ = $\textrm{T}_{1}\textrm{T}_{2}$ + $\textrm{R}_{1}\textrm{R}_{2}$ +$2\,\sqrt[]{\textrm{T}_{1}\textrm{T}_{2}\textrm{R}_{1}\textrm{R}_{2}}$ cos($\varphi$)
where $\varphi$ $\sim$ $\int {\bm B}\cdot {\rm d}{\bm S}$ with ${\bm S}$ the area of the interferometer, $\textrm{T}_{1}$ ($\textrm{T}_{2}$) and $\textrm{R}_{1}$ ($\textrm{R}_{2}$)  the transmission and reflection probability of the first (second) beam splitter. Consequently to observe oscillations, one  varies the Aharonov-Bohm (AB) flux through the surface defined by the two arms of the interferometer, either by varying the area defined by the paths (u) and (d) using a lateral gate or by sweeping the magnetic field. We have represented in Fig.\ref{osc.png} the two different ways to reveal oscillations. One can notice that sweeping the magnetic field to reveal quantum interferences leads to a more noisy sinusoidal curve than by sweeping the lateral gate voltage}

\pr{In the physics of quantum conductors, one of the fundamental length scales which sets an upper limit to the manifestation of quantum effects, is the quantum coherence length L$_\varphi$. It is the typical length over which an electron exchanges information with other degrees of freedom and looses its phase coherence. In the Integer Quantum Hall Regime, because of the chirality that prevent energy exchange processes, we expect a very long coherence length. To determine the coherence length, one has to measure the dependence of the visibility with different parameters (the bias, the temperature, the size of the interferometer).}

\pr{First studies focused on the effect of a DC voltage applied on the source contact.} Unexpectedly, the bias dependence of the visibility revealed an unusual lobe structure (at filling factor $\nu$=1 and 2) \cite{roulleau_prb_2007,neder_prl_2006}. This is now understood as a signature of strong Coulomb interaction between edge states and results in a separation of the spectrum of edge excitations into a slow and fast mode \cite{sukhorukov_prb_2008}. The interaction  between the two co-propagating edge states has been widely considered both theoretically and experimentally to explain coherence properties of edge states at filling factor $\nu$=2 \cite{roulleau_prl_2008b,feve_science_2013,sukhorukov_prb_2008,lee_prb_1997,cheianov_prl_2007,sukhorukov_prb_2008,berg_prl_2009,sukhorukov_prb_2012}.
\pr{Because of Coulomb interaction between the two edge states of opposite spins, new eigenmodes with different velocities arise: a fast mode that carries the charge and a slow neutral charge mode. A direct observation of this separation has been realised at filling factor $\nu$=2 where each mode can be addressed individually \cite{feve_ncom_2013,heiblum_prl_2014,fujisawa_nnano_2014,feve_ncom_2015,fujisawa_nphys_2017}.}

\pr{The systematic study based on the temperature and size dependence of the visibility came slightly later}. It enabled to extract the coherence length L$_{\varphi}$ at filling factor $\nu$=2 (B\,$\approx$\,4.6\,T), where two edge states are propagating into the interferometer \citep{roulleau_prl_2008}. 
Two conditions are necessary to measure the absolute value of L$_{\varphi}$. First one needs to prove its existence by varying the size on which interferences occur. 
Secondly, one needs to show that the interferences have a phase which does not depend on the energy of the quasiparticles (to exclude thermal smearing). This can be ensured by the geometry of the interferometer:  for equal length of interferometer arms, the phase is energy independent. 
From the temperature dependence of the visibility $\nu$, it has been shown that $\nu$ $\sim$ $e^{-\textrm{T}/\textrm{T}_{\varphi}}$ with T the electronic temperature and $\textrm{T}_{\varphi}$ $\sim$ $\nu_D$ the drift velocity of the electrons. From the size dependence of the visibility L$_{\varphi}$ $\sim$ 20$\mu$m has been extracted at T=20mK\cite{roulleau_prl_2008}. \pr{With some record visibilities equal to 90\,$\%$ \cite{neder_nature_2007}, the MZI appears as a promising brick for more complicated geometries. }

\begin{figure}[h]
\includegraphics[width=6cm]{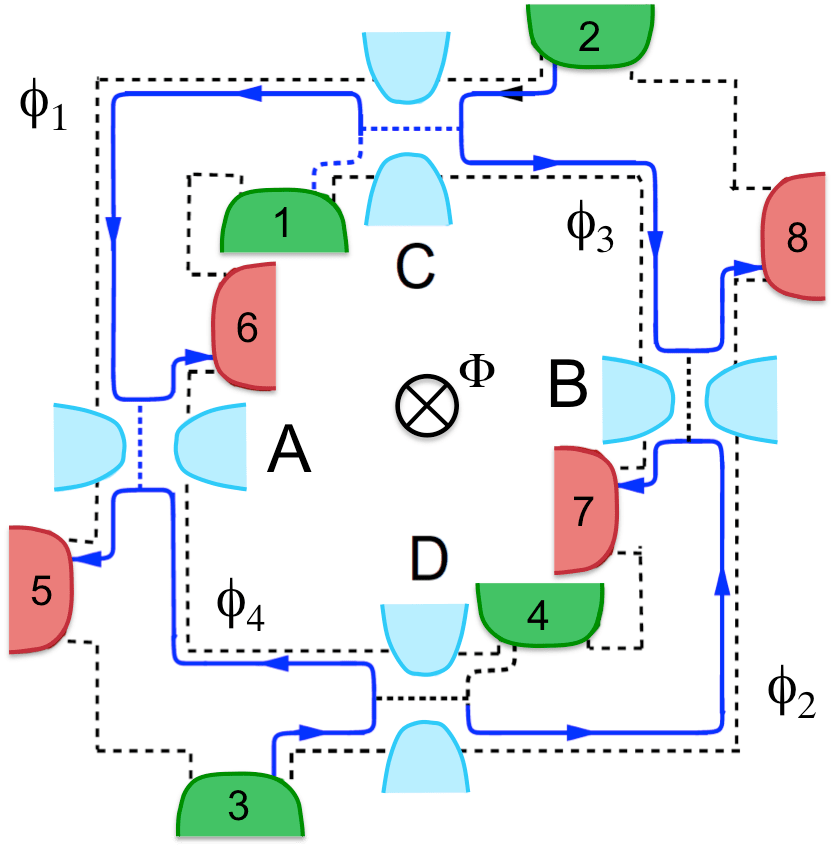}
\caption{\textbf{Schematic representation of the double MZI proposed by Samuelsson et al.\cite{samuelsson_prl_2004}}. Electrons are injected from contacts 2 and 3. Cross-correlation noise measurements are realized between contacts 5 and 8. $\Phi_{\alpha}$ is the phase accumulated along the trajectory of an edge. For example $\Phi_1$ is the accumulated phase along the outer edge between the beam splitters C and A. Cross-correlation noise measurement between contacts 5 and 8 is sensitive to the two-electron Aharonov-Bohm phase  $\Phi=\Phi_1 + \Phi_2-\Phi_3-\Phi_4$ (figure adapted from refs. \cite{samuelsson_prl_2004,samuelsson_physript_2009}).}
\label{DoubleMZI}
\end{figure}

\subsubsection{The two-electron Mach--Zehnder interferometer}

To realise quantum gates, entangled states must be generated. To create an entangled state, a two-electron interferometer where \cb{indistinguishable} electrons are injected from two independent sources is necessary \cite{samuelsson_prl_2003,buttiker_prb_2005,beenakker_prl_2005}.  As depicted in Fig. \ref{DoubleMZI}, electrons are injected from two independent sources 2 and 3. A, B, C and D are QPCs. Contacts 6 and 7 are grounded (contacts 1 and 4 are not used). The measurement is realised between contacts 5 and 8. The direction propagation is fixed by the magnetic field: electrons from source 2 (3) are partitioned by QPC C (D). Reflected electrons from 2 are sent to QPC B. Transmitted electrons from 3 are also sent to QPC B: at the output of QPC B \cb{it is} not possible to distinguish electrons coming from 2 to those from 3. This indiscernibility is the building block of the two-electron Aharonov-Bohm effect and the orbital entanglement. The quantity that will post-select the entangled part of the output state at contacts 5 and 8 is the zero-frequency current cross correlator noted $S_{58}$. When the gate transmissions are equal to 1/2, one can show that \cite{samuelsson_prl_2004,samuelsson_physript_2009}: $S_{58}=\frac{-e^2}{4h}eV(1+cos(\Phi_1+ \Phi_2-\Phi_3-\Phi_4)$  where $\Phi_1$ is the phase accumulated between QPCs C and A, $\Phi_2$ between QPCs D and B, $\Phi_3$ between QPCs C and B, $\Phi_4$   between QPCs D and A (see figure \ref{DoubleMZI}). We now assume that a magnetic flux $\Phi$ can be added through the sample. Due to the chirality of the electronic trajectories, one obtains a positive contribution of the magnetic flux for the phases $\Phi_1$ and $\Phi_2$, and a negative one for the phases $\Phi_3$ and $\Phi_4$. The global contribution related to the magnetic flux is thus equal to $\Phi_1+\Phi_2-\Phi_3-\Phi_4 = \int {\bm{B dS}}$ 
where ${\bm{B}}$
is the magnetic flux 
%%%%%%%%%%%%% flux or field --> harmonise
across the area enclosed by the four trajectories. Varying the magnetic flux through the double MZI, one should observe oscillations of  $S_{58}$. Since none of the electrons injected from 2 (or 3) can make a complete loop around $\Phi$, this effect is necessarily a two-electron Aharonov-Bohm effet. 

\begin{figure}[h]
\includegraphics[width=8.0cm]{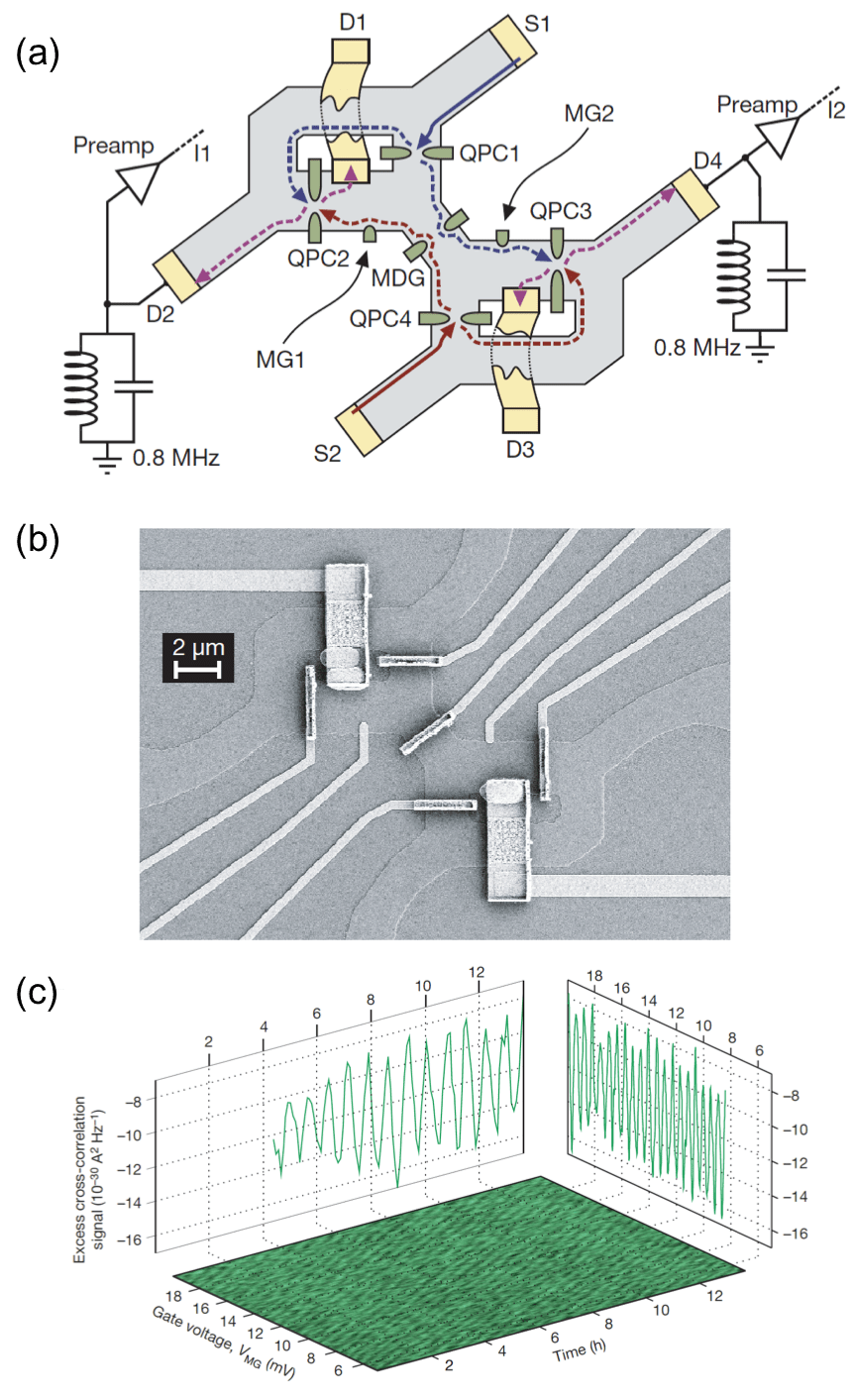}
\caption{\textbf{ \cb{Double Mach--Zehnder Interferometer.}} (a) schematic drawing of the MZI shown in (b). 
The double MZI geometry is composed of two single MZIs separated by the middle gate (MDG). The \cb{tuning} gates MG1 and MG2 are used to sweep the Aharonov-Bohm phase of the electrons in the two MZIs.
\cb{Metallic air bridges connect drains D1 and D3 to ground. (b) Scanning electron micrograph of the actual sample. Air bridges were used to contact the small ohmic contacts, the split gates of the QPCs, and the MDG.}
c) \cb{Two-electron inteference} measured as a function of the \cb{tuning} gate and time (exploiting the gradual decay of the magnetic field in persistent mode). An unexpected low visibility of 25\,$\%$ is measured. (figure adapted from ref. \cite{neder_nature_2007}).}
\label{DMZI+osc.png}
\end{figure}

\pr{The first and so far only realisation of this experiment has been done by the Weizmann team  \cite{neder_nature_2007}. As depicted in Fig.\,\ref{DMZI+osc.png}(a), the experimental double MZI was composed of two single MZIs separated by a central top gate. The central gate being closed, each single MZI was independently tuned reaching a maximum visibility of 90\,$\%$. The central gate is then fully opened to finally obtain the double MZI configuration. In Fig.\,\ref{DMZI+osc.png}(b), the cross-correlation shot noise $S_{58}$ \cb{has been measured} as a function of a lateral gate \cb{voltage} (varying the area defined by the four paths) or the magnetic field (exploiting the gradual decay of the magnetic field in persistent mode). Oscillations with a period compatible with two-electron interference have been observed, but only with a 25\,$\%$ visibility, much smaller than expected with two MZI showing 90 $\%$ visibility in single electron interference. Before going further, like performing Bell's inequalities violation\cite{bell_rmp_1966,neder_nature_2007,samuelsson_prl_2003}, one definitively needs to understand the mechanisms leading to this unexpected low visibility of the two-electron quantum interference.}

%%%%%%%%%%%%%%%%%%%%%%%%%%%%
%%%%%%%%%%%%%%%%%%%%%%%%%%%%
\subsection{Mach--Zehnder interferometry at low magnetic fields (AB ring with tunnel-coupled wires)}
\label{MC0B}
%%%%%%%%%%%%%%%%%%%%%%%%%%%%
%%%%%%%%%%%%%%%%%%%%%%%%%%%%

\begin{figure}[h]
\includegraphics[width=8cm]{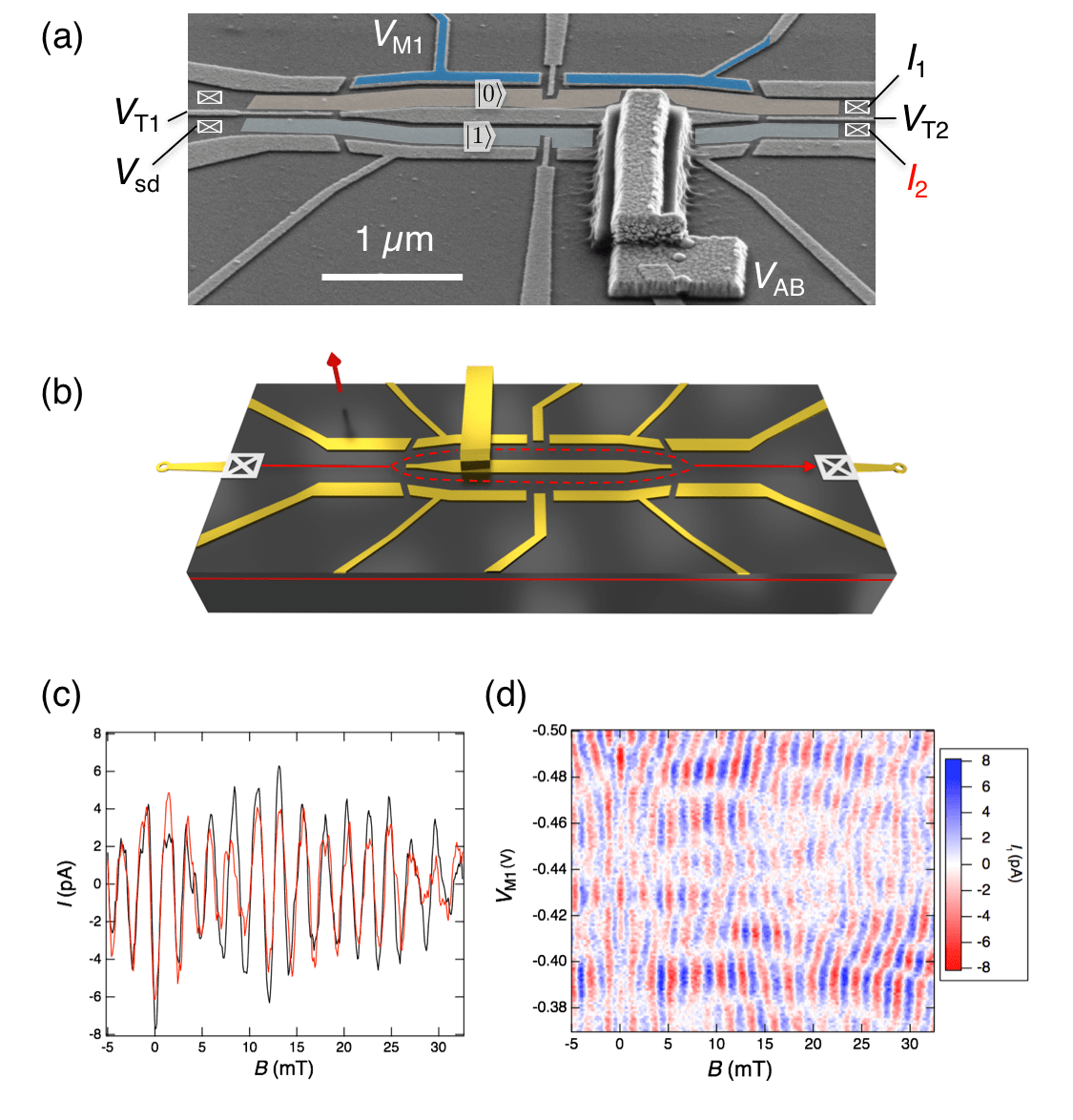}
\caption{\textbf{Flying qubit device under low magnetic field and its behaviour in the single wire regime.} (a) SEM image of the relevant device. \st{(b) Schematic of the electron trajectory in the single wire regime. (c) Oscillations of the output current $I_{\rm 1}$ (black) and $I_{\rm 2}$ (red) as a function of the perpendicular magnetic field in the single wire regime. (d) Oscillations of $I_{\rm 1}$ as a function of the perpendicular magnetic field $B$ and the side gate voltage $V_{\rm M1}$. A smoothed background is subtracted and only the oscillating components are plotted for (c) and (d). (figures adapted from refs. \cite{yamamoto_nnano_2012,edlbauer_ncom_2017}).}}
\label{fig_ABTunnelDevice}
\end{figure}

Another way to realise a MZI which works at low magnetic fields \st{ ($\lesssim 100\,$mT )} is to combine an Aharonov-Bohm interferometer with tunnel-coupled wires. In this case the chirality is not relevant for the electron transport.
As briefly mentioned before, realisation of a two-path interferometer is a direct way to realise a flying qubit.
For a two-path interferometer the two qubit states $\ket{0}$ and $\ket{1}$ are defined by the presence of an electron in either one of the two paths of the interferometer (see figure \ref{fig_ABTunnelDevice}(a)) and well defined qubit operations can be performed for the well defined electron trajectories.
For the MZI in the quantum Hall regime the chirality of the system ensures suppression of backscattering and allows for realisation of a two-path interference, however limits the system to high magnetic field (several Teslas).
Under low magnetic field it is more challenging to realise a \textit{pure} two-path interference since electrons can easily be backscattered.

\begin{figure}[h]
\includegraphics[width=8.0cm]{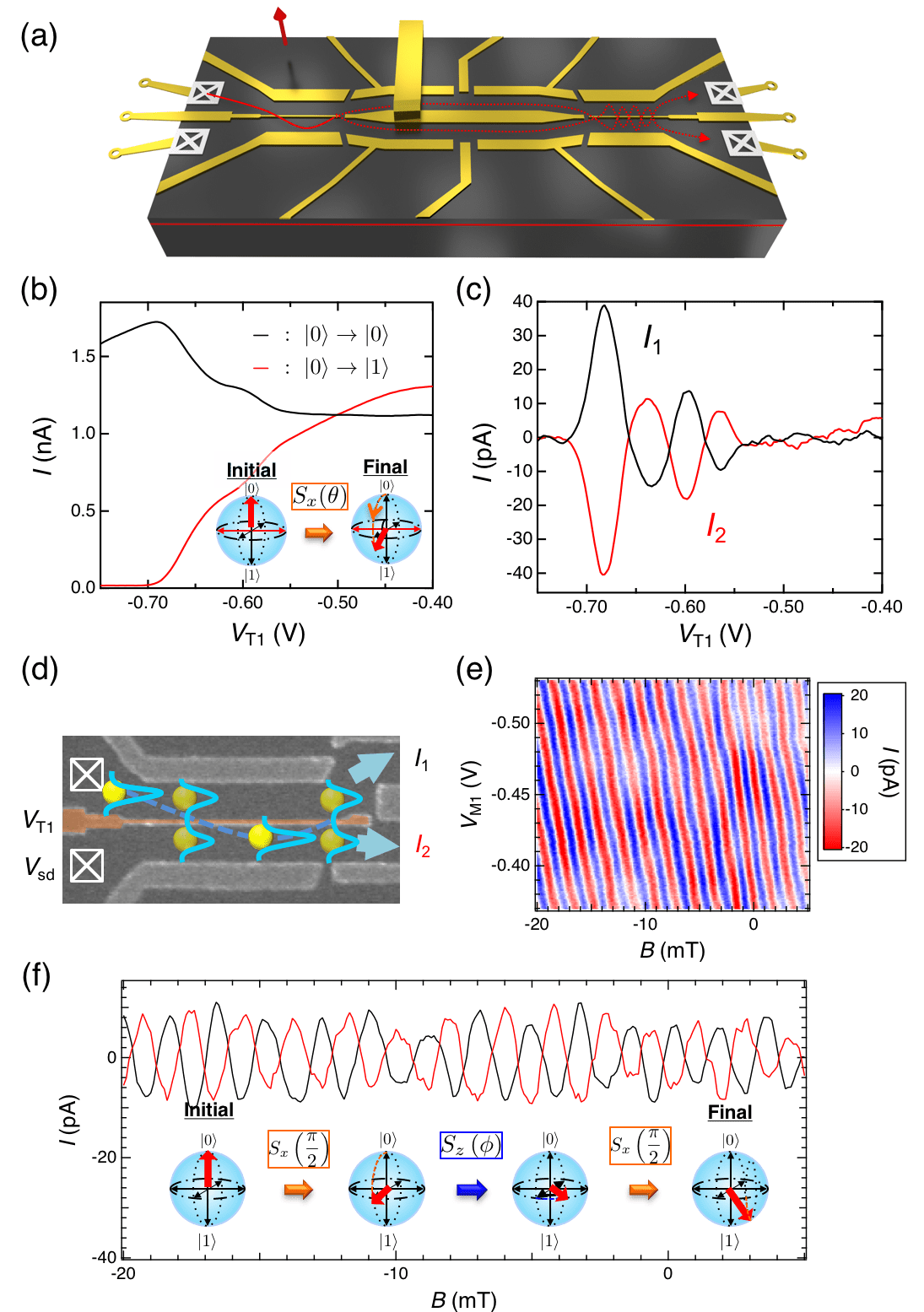}
\caption{\st{\textbf{Flying qubit operations in the tunnel-coupled wire regime.} (a) Schematic of the electron trajectory in the tunnel-coupled wire regime. (b) -- (d) Rotation about x axis $R_x (\theta)$. (b) Output currents in the tunnel-coupled wire. Black and red curves are the output currents $I_{\rm 1}$ and $I_{\rm 2}$ as a function of $V_{\rm T1}$ when the current is injected from the upper wire. (inset) Evolution of the flying qubit state when a tunnel-coupling is induced. (c) Oscillating components of the output currents. Smoothed backgrounds are subtracted from the black and the red curves of (b). (d) Schematic of an electron propagating through the tunnel-coupled wire. The tunnel-coupling between the upper and lower wire induces an oscillation of the electrons between the upper and lower wire as they propagate through this region. 
(e), (f) Rotation about z axis $R_z (\phi)$. (e) Oscillating component of the output currents $I = I_{\rm 1} - I_{\rm 2}$ as a function of the perpendicular magnetic field and side gate voltage $V_{\rm M1}$. (f) Oscillations 
of the current $I_{\rm 1}$ (black) and $I_{\rm 2}$ (red) as a function of the perpendicular magnetic field in the tunnel-coupled wire regime. The smoothed backgrounds are subtracted from the raw data. (inset) Evolution of the qubit state in Ramsey interference scheme. (figures adapted from ref. \cite{yamamoto_nnano_2012}).}}
\label{fig_FQbitOperations}
\end{figure}

\cb{This can be done by combining an AB interferometer to two tunnel-coupled wires\cite{yamamoto_nnano_2012} which act as beam splitters as shown in Figs.\,\ref{fig_ABTunnelDevice} and \ref{fig_FQbitOperations}.
The device structure is tailored into a two-dimensional electron gas made from a GaAs/AlGaAs heterostructure by electrostatic surface gates.}
 \cb{Applying a negative voltage $V_{\rm AB}$ to the bridge gate allows to deplete the central region to form the Aharonov-Bohm ring}.
\st{Electrons are injected from the lower left contact by applying an ac bias ($23.3$ Hz, $50\ \mu V$).
They are guided into the two arms of the AB ring through the first tunnel-coupled wire and accumulate a phase difference between the two arms.
Finally they are guided into the two contacts on the right through the second tunnel-coupled wire and measured as currents $I_{\rm 1}$ and $I_{\rm 2}$.}
This device shows two distinct behaviours depending on the voltage $V_{\rm T1}$ and $V_{\rm T2}$ applied on the tunnel-coupling gates:
%-------------------------------------------------------------
%-------------------------------------------------------------
%------------------------------------ Fig. 8 ----------------
%-------------------------------------------------------------
%-------------------------------------------------------------
%-------------------------------------------------------------
%-------------------------------------------------------------
%------------------------------------ Fig. 8 ----------------
%-------------------------------------------------------------
%-------------------------------------------------------------

\st{ (i) When the voltages on gate $V_{\rm T1}$ and $V_{\rm T2}$ are set to zero, both tunnel-coupled wires behave simply as single quantum wires.
In this single wire regime the two ohmic contacts on each side are equivalent and the interferometer effectively works as a two-terminal AB interferometer as shematised in Fig.\,\ref{fig_ABTunnelDevice}(b).}
\cb{The corresponding AB oscillations of the two output currents $I_{\rm 1}$ and $I_{\rm 2}$ for this situation are shown in Fig.\,\ref{fig_ABTunnelDevice}(c), where one probes the modulation of the phase difference $\Delta \phi = \int {\bm k}\cdot d{\bm l} - (e/\hbar)BS$ between the two paths of the AB ring when sweeping the perpendicular magnetic field $B$.}
The fact that $I_{\rm 1}$ and $I_{\rm 2}$ behave in the same way clearly shows that the two ohmic contacts on the right are equivalent and the tunnel-coupled wires behaves like a single quantum wires.
For such a two-terminal device Onsager's law \cite{onsager_pr_1931} as well as current conservation imposes the boundary condition $G(B)=G(-B)$ on the linear conductance \cite{buttiker-yeyati_prb_1995}.
This can be demonstrated by modifying $\Delta \phi$ via modulation of ${\bm k}$. Changing the side gate voltage $V_{\rm M1}$ locally \st{ modifies the wave vector} ${\bm k}$ of the electron in the path along gate $\rm M1$.
\st{As shown in Fig.\,\ref{fig_ABTunnelDevice}(d), the AB oscillations are clearly symmetric with respect to the magnetic field and show phase jumps as a function of $V_{\rm M1}$.}
Such an interference pattern indicates that the observed interference is \textit{not} a two-path interference but contains contributions from multiple interference paths to satisfy the boundary conditions \cite{bautze_prb_2014}.

%---------------------------------------------------------
%---------------------------------------------------------
%------------------ Fig. 9--------------------------------
%---------------------------------------------------------
%---------------------------------------------------------
%---------------------------------------------------------
%---------------------------------------------------------
%------------------ Fig. 9--------------------------------
%---------------------------------------------------------
%---------------------------------------------------------
\cb{(ii)}When $V_{\rm T1}$ and $V_{\rm T2}$ are set to large enough negative voltages to form a tunnel-coupled wire\st{ as schematised in Fig.\,\ref{fig_FQbitOperations}(a)}, the behaviour drastically changes. \cb{The phase now smoothly evolves as a function of the the side gate voltage $V_{\rm M1}$ (Fig.\,\ref{fig_FQbitOperations}(e)) while the output currents $I_{\rm 1}$ and $I_{\rm 2}$ show anti-phase oscillations (Fig. \ref{fig_FQbitOperations}(f))}.
In this tunnel-coupled wire regime any superposition state of $\ket{0}$ and $\ket{1}$ in the AB ring can transmit into the tunnel-coupled wire by being transformed into the superposition of a symmetric hybridised state $\ket{S}=(1/\sqrt{2})(\ket{0}+\ket{1})$ and an anti-symmetric hybridised state $\ket{A}=(1/\sqrt{2})(\ket{0}-\ket{1})$.
This is in clear contrast with the above case of single wire leads, where only $\ket{S}$ is transmitted into the leads.
Scattering of electrons from one path to the other at the entrance and the exit of the AB ring are therefore highly suppressed and prevent electrons from encircling the AB ring several times and contributing to the interference\cite{bautze_prb_2014}. These distinct behaviours depending on the tunnel-coupling energy have been also studied theoretically and nicely reproduced \cite{bautze_prb_2014,aharony_njp_2014}. 
Consequently the device works as a \textit{true} two-path interferometer as well as a flying qubit.  Due to this peculiarity,  this device has been exploited in recent studies to revisit a number of fundamental questions \cite{heiblum_nature_1997,heiblum_science_2000} about the phase modification of an electron when traversing a quantum dot \cite{takada_prl_2014,edlbauer_ncom_2017}. 

In the tunnel-coupled wire regime a rotation about the x-axis $S_x(\theta)$ can be performed as described in section \ref{Introduction}, where $\theta = \Delta k \cdot L = (k_{S}-k_{A}) \cdot L$.
This operation can be demonstrated by varying the voltage of gate $V_{\rm T1}$ which controls the tunnel-coupling between the upper and lower wire: A current is injected into the upper wire to prepare the initial state $\ket{0}$ and the output currents $I_{\rm 1}$ and $I_{\rm 2}$ are measured as a function of the gate voltage $V_{\rm T1}$ (Fig.\,\ref{fig_FQbitOperations}(b)).
$V_{\rm T1}$ changes $\Delta k$ and hence the rotation angle $\theta$.
Clear anti-phase oscillations of currents $I_{\rm 1}$ and $I_{\rm 2}$ are observed at $2.2$ K, as shown in Fig.\,\ref{fig_FQbitOperations}(c).
These anti-phase oscillations are a direct signature of electron tunnelling between the two wires as schematised in Fig.\,\ref{fig_FQbitOperations}(d).
An electron injected into the upper wire oscillates between the upper and lower wire depending on the tunnel-coupling set by the gate\st{ voltage} $V_{\rm T1}$.
For this device, the visibility of the oscillation, the ratio of the oscillation component to the total current, is limited to 1\%.
This is due to the existence of several transmitting channels and the high measurement temperature.
Improvement of design and lowering the temperature allowed to reach visibilities above 10\% \cite{roussely_thesis_2016}. 
%At low temperatures, disorder scattering in the quantum wires leads to complicated fluctuations of the output currents as a function of $V_{\rm T1}$, which masks the current oscillation induced by tunnel coupling.

A rotation about the z-axis $S_z(\phi)$ can be achieved in the AB ring by varying the perpendicular magnetic field or the gate voltages $V_{\rm M1}$ as already outlined in section \ref{Introduction}.
The relative phase difference between the upper path and the lower path is given by $\Delta \phi = \oint {\bm k} \cdot {\rm d}{\bm l} - (e/\hbar)BS$.
The combination of $S_x(\theta)$ and $S_z(\phi)$ enables the generation of an arbitrary vector state on the Bloch sphere. This can be achieved by controlling simultaneously the tunnel-coupling and \cb{phase difference} between the two paths.

$S_z(\phi)$ is demonstrated in a Ramsey-type interference (Schematic in Fig.\,\ref{fig_FQbitOperations}(f)).
The two sets of tunnel-coupled wires were prepared to $S_x(\pi/2)$ and the magnetic field is varied to perform $S_z(\phi)$ in the AB ring.
When the initial state is prepared to $\ket{0}$ by injecting a current from the upper wire, the final state becomes
\begin{equation}
	S_x\left(\frac{\pi}{2}\right)S_z(\phi)S_x\left(\frac{\pi}{2}\right)\ket{0} = \frac{{\rm e}^{i\phi}-1}{2}\ket{0}+\frac{i{\rm e}^{i\phi}+i}{2}\ket{1}.
\end{equation}
The two output currents are proportional to the square modulus of each coefficient and become
\begin{equation}
	I_{\rm 1 (2)} \propto \frac{\left| {\rm e}^{i\phi} \mp 1 \right|^2}{4} = \frac{1 \mp \cos \phi}{2},
\end{equation}
respectively.
The measured $I_{\rm 1}$ and $I_{\rm 2}$ plotted in Fig.\,\ref{fig_FQbitOperations}(f) indeed oscillate with exactly opposite phase for the modulation of $\phi$ by the magnetic field.
The phase $\phi$ can also be modulated by the gate voltages $V_{\rm M1}$, which changes the wave vector ${\bm k}$ of the path.
This is demonstrated in Fig.\,\ref{fig_FQbitOperations}(e), where the phase smoothly evolves over a range of $2\pi$ as a function of the side gate voltage $V_{\rm M1}$ along the vertical axis.
This is in strong contrast to Fig.\,\ref{fig_ABTunnelDevice}(d).
\st{This rotation about the z-axis by $V_{\rm M1}$ is important for qubit applications.
Combined with the rotation about x-axis by $V_{\rm T1, T2}$, the qubit can be fully operated by the gate voltages at zero magnetic field.
This allows for much faster operations than the ones controlled with a magnetic field.}

The flying qubit presented here is attractive for quantum information technology.
In addition to the ability to transfer the quantum information over a long distance, it has a much shorter operation time compared to other qubits in solid-state systems.
The operation time $L/v_{\rm F}$ ($L$, gate length; $v_{\rm F}$, Fermi velocity) is of the order of $10$ ps.
Analysing the temperature dependence of the oscillation amplitude shows that this qubit has a very long coherence length $l_{\rm \phi} = 86\ {\rm \mu m}$ at $T = 70$ mK \cite{yamamoto_nnano_2012}.
\cb { Using even higher quality heterostructures,} \st{the coherence length could be longer than $100\ {\rm \mu m}$.
Since each quantum operation is performed within a $1\ {\rm \mu m}$ scale, it would in principle be possible to perform more than 100 qubit operations.}  

On the other hand the visibility, defined as the AB oscillation amplitude divided by the total current, is limited to about $10$ \%.
Since the coherence length is found to be much longer than the interferometer length, decoherence is not the main origin of this limited visibility.
The influence of thermal smearing due to the difference in Fermi velocity between the two paths is also small at the measurement temperature.
The main limitation comes from the contribution of several transmitting channels in each part of the tunnel-coupled wires and in each arm of the AB ring \st{while} only one in each wire contributes to the main AB oscillation.
\st{Therefore the visibility could be improved by operating the interferometer with a highly coherent single transmitting channel (See Supplementary information in Ref.\,\onlinecite{yamamoto_nnano_2012} for more details).
One possible remedy towards \cb{this} direction would be to adiabatically reduce the number of transmitting channels to one at a specific point of the interferometer while keeping the number of channels (or the electron density) \cb{constant} over the other part of the interferometer.
Higher electron density is preferable to screen the potential fluctuations \cb{induced by} the gates, which is proposed to be \cb{the main} source of decoherence in ballistic AB interferometers \cite{seelig_prb_2001}, and hence to maintain the coherence.}

In addition to quantum information transfer, it should also be possible to create a non-local entanglement state following the scheme proposed in refs. \onlinecite{yu_ssc_2008} and \onlinecite{ionicioiu_pra_2001}, combined with single electron sources \cite{blumenthal_nphys_2007,feve_science_2007,hermelin_nature_2011,glattli_nature_2013} to synchronise qubits.
This flying qubit can also be used in combination with a \cb{spacially}  localised qubit \cite{schomerus_njp_2007}.
\vspace{1cm}

%%%%%%%%%%%%%%%%%%%%%%%%%%%%%%%%%%%%%%%%%%%%%
 \cb{In this section we introduced different device architectures which can be exploited to realise electronic flying qubits at the single-electron level.} For the \cb{implementation} of a flying qubit at the single-electron level, however, these architectures have to be combined with a single electron source as well as a single electron detector, which we will describe in the sections \ref{Single Electron Sources} and \ref{Single Electron Detectors}.
\cb{In addition, synchronisation of different qubits is required to realise two qubit operations.}
For that purpose MZI in the quantum Hall regime \cb{is advantageous}.
Chirality suppresses backscattering and synchronising different single electron sources can be straightforwardly achieved \cite{feve_science_2013}.
Upscaling of this system, \cb{however}, is not straightforward.
\cb{On the contrary}, the MZI interferometer for low magnetic fields is easier to scale-up by adding the basic qubit structure in parallel or in series.
\cb{On the other hand}, when the device gets longer, it will suffer from backscattering of the electrons, which prevents synchronisation between different qubits.
One possible way to avoid backscattering is using electron transport by surface acoustic waves \cite{barnes_prb_2000,hermelin_nature_2011,mcneil_nature_2011,hermelin_pss_2017,ford_pss_2017}. The biggest challenge, however, is single shot detection of such single flying electrons. \cb{For low magnetic field there are potential approaches to achieve this in the near future (see section \ref{Single Electron Detectors})  while single-shot detection under high magnetic field is a real challenge in this field of research.}

When dealing with single-electron wave packets, it is also important that there is a substantial overlap between the interfering electrons at the output of a beam splitter \cite{haack_prb_2011,janine_prb_2015}. It is hence important that the two inference arms are of similar length. For electron wave packets of a temporal width of about 100 ps, which can nowadays be routinely produced with state-of-the-art electronic equipment,  the spatial extension is still large (10 $\mu$m  for a speed of 10$^5$ m/s). This is of the same order of magnitude as the size of the present interferometers.
However, when going to smaller and smaller wave packets this issue has to be taken into account. The case where electron wave packets are much smaller than the size of the interferometer, new types of interference effects will appear \cite{gaury_ncom_2014}. This novel physics will be described in section \ref{Novel quantum interference experiments}.
%%%%%%%%%%%%%%%%%%%%%%%%%%%

%%%%%%%%%%%%%%%%%%%%%%%%%%%%%%%%%%%%%%%%%%%%%%
%%%%%%%%%%%%%%%%%%%%%%%%%%%%%%%%%%%%%%%%%%%%%%
\section{Single Electron Sources}
\label{Single Electron Sources}
%%%%%%%%%%%%%%%%%%%%%%%%%%%%%%%%%%%%%%%%%%%%%%
%%%%%%%%%%%%%%%%%%%%%%%%%%%%%%%%%%%%%%%%%%%%%%

\cb{In the preceding section we have presented proof-of-principle experiments for the realisation of a solid state flying qubit with two different types of MZIs.}
In these experiments, however, the electrons are injected as a continuous stream and the measurements are based on ensemble averages. 
The ultimate goal in this line of research is the ability to control the flying qubit at the single-electron level, which requires on-demand single electron sources (SES) as well as single electron detectors.
In this section we will focus on the single electron sources which have been developed over the last 10 years with the goal to perform quantum interference experiments at the single-electron level.

At the origin of most single electron sources is the quest for a fundamental standard of electrical current linking the ampere to the elementary charge and frequency.
Such single electron sources can be realised by high-speed, high-accuracy transport of single electrons in nanoscale devices \cite{pekola_rmp_2013, kaestner_ropp_2015, kaneko_mst_2016}.
\st{Development of an accurate single electron pump is of particular importance for metrology.
It allows for the precise determination of the value of the elementary charge, which is one of the seven reference constants in the new SI units \cb{which will be redefined in 2018} \cite{bipm_2010, mills_rsta_2011,nature-focus_2017}.
In addition it contributes to the quantum metrology triangle (QMT) experiment \cite{Keller_metrologia_2008, feltin_epj_2009} which is a consistency test of three quantum electrical standards: the single-electron current standard, the Josephson voltage standard and the quantum Hall resistance standard, which play a fundamental role in metrology.
The future redefinition of the international system of units in terms of natural constants requires a robust,
high-precision quantum standard for the electrical base unit ampere.}

The best single electron pumps reach nowadays \st{an error rate of better than 1 ppm at $0.6 \sim 1$ GHz \cite{fujiwara_apl_2016,Stein_metrologia_2017,kataoka_arxiv_2017}.}
Besides this, integrated single-electron circuits have also a great potential in quantum information processing as already motivated in the introduction.
Most implementations of today's single electron sources are based on small isolated regions of charges connected to a reservoir via tunable barriers, a prime example is semiconductor quantum dots. 
In this case one exploits the fact that in a sufficiently small isolated region, the energy \st{level is fully quantised by the charging energy originating from Coulomb interactions and the number of electrons inside the quantum dot can be controlled one by one \cite{kastner_rmp_1992, kouwenhoven_QDreview_1997, manninen_rmp_2002}. }

For completeness let us also mention that single electron transistors using superconductor or metallic islands have been developed to realise high accuracy current pumps \cite{pekola_rmp_2013}. 
In this review, however, we will focus on single electron sources made from GaAs heterostructures.

%%%%%%%%%%%%%%%%%%%%%%%%%%
\subsection{AC single electron source (Mesoscopic Capacitor)}
%%%%%%%%%%%%%%%%%%%%%%%%%%
%

\begin{figure}[h]%
\includegraphics[width=8.5cm]{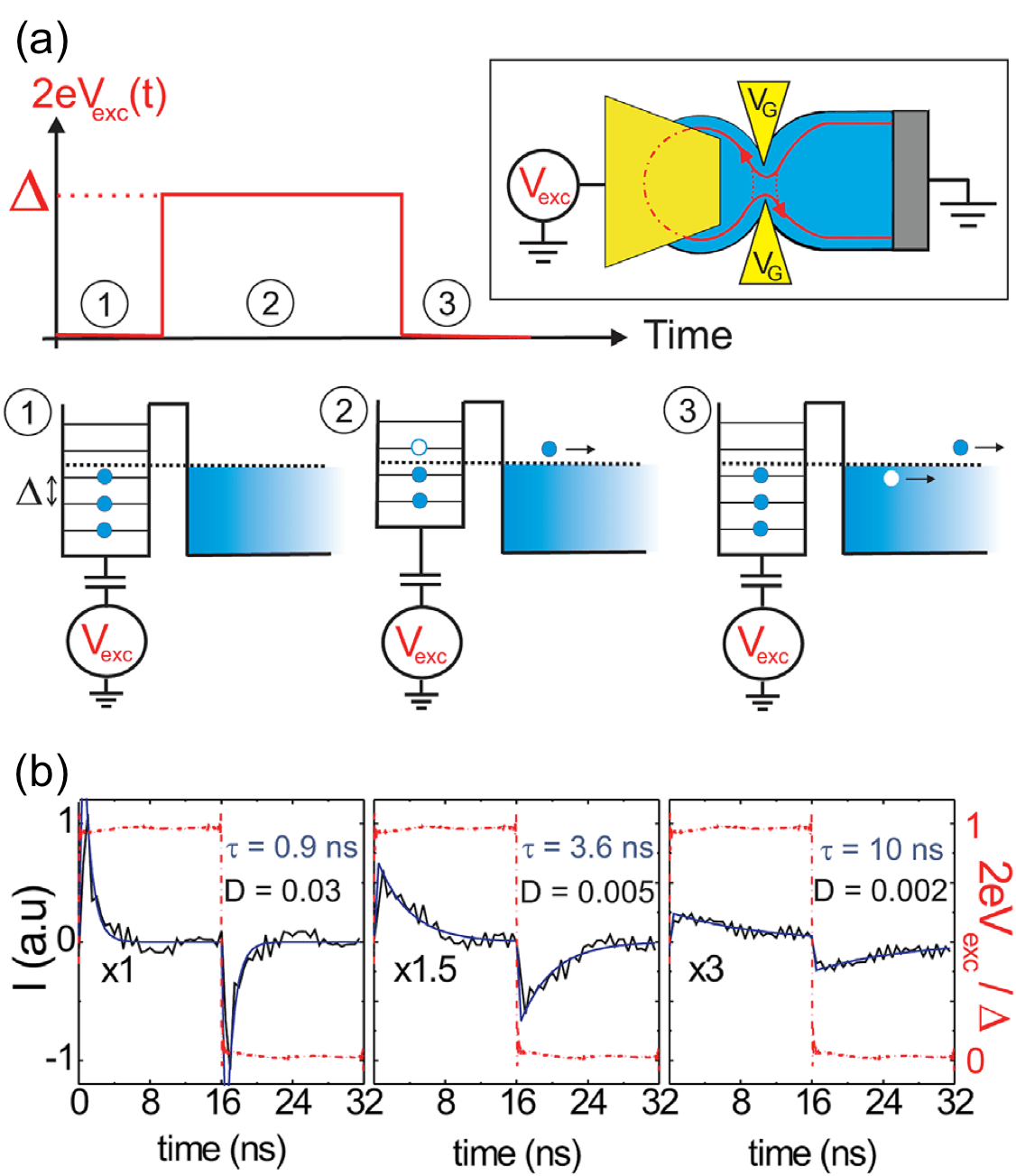}
\caption{%
{\bf{Mesoscopic Capacitor single electron source.}}
(a) Radio-frequency square-wave pulses $V_{exc}(t)$ are applied on the top gate (inset of the figure). \textcircled{1} Starting point: the Fermi level lies between two discrete energy levels of the quantum dot. \textcircled{2} 2e$V_{exc}(t)$ is equal to the level spacing $\Delta$. An electron escapes the dot at a well defined and tuneable energy. \textcircled{3} $V_{exc}(t)$ is brought back to its initial value, a hole escape at energy below the lead Fermi energy. (b) Time domain measurement of the average current (black curves) on one period of the excitation signal (red
curves) at 2e$V_{exc}(t)$ = $\Delta$ for three values of the transmission $D$. The expected exponential relaxation with time $\hbar/D\Delta$ (blue curve) fits well the data (figure adapted from ref. \cite{feve_science_2007}).}
\label{SES}
\end{figure}

A capacitor forms a simple and elegant possible realisation of a single electron source. 
The idea is to realise a RC circuit driven by an AC voltage such that the charge and discharge of the capacitor is limited to a single elementary charge $q=e$. 
The capacitor has to be weakly connected to the lead in which the elementary charges are transferred such that charge quantisation occurs between two stationary capacitor states. 
A metallic island connected to leads by a single tunnel junction could realise this device when the Coulomb charging energy $e^{2}/C$ is larger than the thermal energy $k_{B}T$. 
However, the metallic island capacitor is reputed to have a quasi-continuous density of states and the energy at which, and the state from which electrons are emitted are not well defined, while controlling the initial state is of utmost importance for quantum information applications. 
A quantum coherent electron source requires in addition using a quantum dot viewed as a mesoscopic capacitor in which electrons keep quantum coherence and their state and energy levels are well defined. \cg{The spacing of the energy levels of the dots should be much larger than the thermal energy $k_{B}T$ and the level used to emit and absorb single electrons, whose energy is closest to the Fermi energy of the leads, should be non degenerate.}
\cb{This coherent single electron source \cite{feve_science_2007} is} based on a mesoscopic capacitor \cb{which was initially} realised \cite{gabelli_science_2006} to check a prediction by M. B\"{u}ttiker \cite{buttiker_physlettA_1993} of an universal quantisation of the charge relaxation resistance (called B\"{u}ttiker's resistance $h/2e^{2})$). 
Here no DC current but only an AC current is produced. In the single electron source regime, a quantised AC current of amplitude $ef$ is made of the periodic injection of single electrons above, followed by single holes below the Fermi energy $E_{F}$. 
\cg{For ease of operation and further use in electron quantum optics, a strong perpendicular magnetic field brings the conductor in the integer Quantum Hall Effect (QHE) regime.  
In this regime, for a small dot (submicron diameter) the 1D QHE chiral edge states modes running along the dot boundary give rise to energy level quantisation with energy level spacing $\Delta =h v_{D}/p$ of typically 1 Kelvin in energy, where $v_{D}$, a few $10^{4}$ cm/s is the drift electron velocity and $p$ is the quantum dot capacitor perimeter.
The top gate, above the mesoscopic capacitor provides enough screening of the Coulomb interaction such that the charging energy is smaller than $\Delta$}, see Fig.\,\ref{SES} for a schematic description.
%, called edge channels, 
%ensure electron conduction. 
%The capacitor is made so small that energy level quantization makes it behave as a quantum dot. 
To ensure energy level and charge quantisation, the capacitor is weakly connected to the leads, the chiral edge channels, via a quantum point contact which controls the tunnel-coupling. 
The operating principle is as follows, see Fig.\,\ref{SES}. Starting from a situation where the last occupied energy level is below the Fermi energy \textcircled{1}, a sudden rise of the voltage applied on the capacitor top gate rises the occupied energy level above the Fermi energy \textcircled{2}. 
After a time of the order of the energy-level life time $\simeq\hbar/D\Delta$, which is controlled by the barrier transmission $D$, an electron is emitted at a tunable energy $\varepsilon_{e}$ above the Fermi level ($\Delta$ is the energy level spacing). 
Then restoring the top gate voltage to its initial value \textcircled{3} pulls down the energy level below the Fermi energy: an electron is captured or equivalently a hole is emitted at a definite energy $-\varepsilon_{h}$ below the lead Fermi energy. \cg{Fig.\,\ref{SES}(b) shows time domain measurements of the measured current averaged over a large number of periodic emission cycles  at a 30 MHz repetition frequency. The exponential decay of the current reflects the exponential decay rate (characteristic time $\tau_{em}=\hbar/D\Delta$) of the emission probability of electrons and holes.}  

\cg{The mesoscopic capacitor electron source is an energy-resolved electron source. It provides a convenient single electron source where electrons can be emitted at a tunable energy ($<\Delta/2$) above the Fermi energy. The emission quantum energy uncertainty $ D \Delta $ is also tunable, as reflected by the quantum emission time $\tau_{em}$.
Among major achievements obtained with the mesoscopic capacitor single source is the demonstration of single electron partitioning and the Hong--Ou--Mandel (HOM) correlations. Regarding limitations, for HOM experiments, it is technically difficult to realize identical capacitor dots due to nanolithography reliability. Another limitation is that a good energy resolution requires that the number of electrons (or holes) is limited to one per cycle because of the charging energy.}  
%---------------------------------------------------------
%---------------------------------------------------------
%------------------ Fig. 9--------------------------------
%---------------------------------------------------------
%---------------------------------------------------------
%---------------------------------------------------------
%---------------------------------------------------------
%------------------ Fig. 9--------------------------------
%---------------------------------------------------------
%---------------------------------------------------------
%
%%%%%%%%%%%%%%%%%%%%%%%%%%
\subsection{Single electron pumps based on dynamic semiconductor quantum dots}
\begin{figure}[h]
\includegraphics[width=8.5cm]{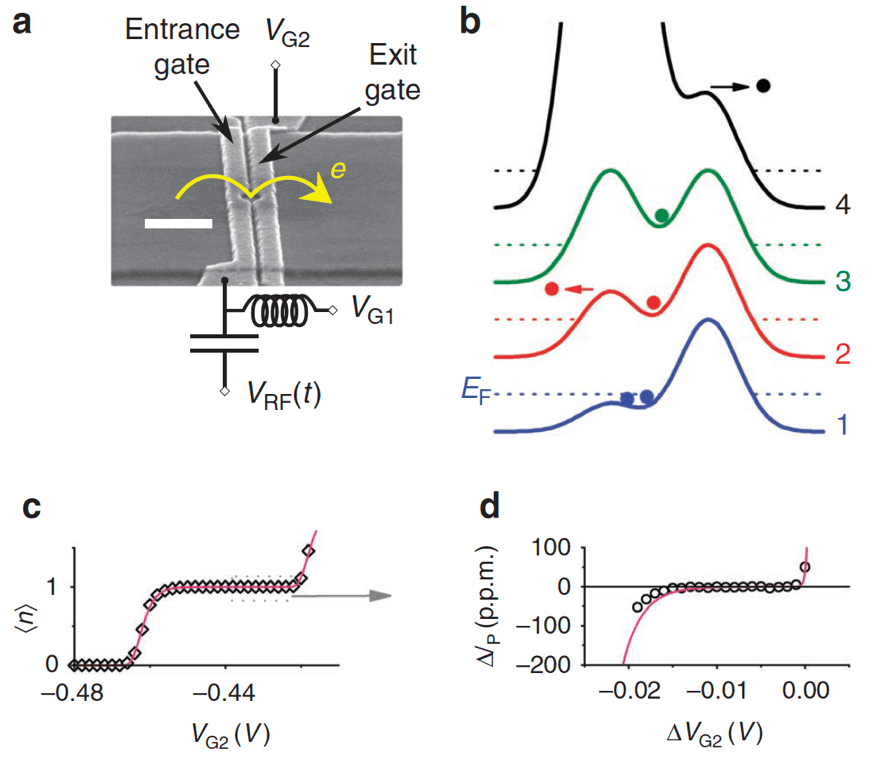}
\caption{{\bf{Single electron pump based on dynamic semiconductor quantum dots.}} \cb{(a) Scanning electron microsope (SEM) image of the device. Two electrostatic gates (light grey) allow for control of this single electron pump. Scale bar indicates 1 ${\rm \mu}$m. (b) Schematic diagrams of the potential along the channel during four phases of the pump cycle. (1) loading, (2) back-tunneling, (3) trapping and (4) emission. One cycle transports an electron from the left (source) to the right (drain) lead. (c) Average number of pumped electrons per cycle as a function of exit gate voltage $V_{\rm G2}$ at a repetition frequency of 945 MHz. (d) Comparison between the fit obtained from the data in (c) (red line) and the high-resolution data around the plateau. They are plotted on an offset gate voltage scale. (figure adapted from ref. \cite{kataoka_ncom_2012}).}}
\label{fig_SES_cambridge}
\end{figure}

%%%%%%%%%%%%%%%%%%%%%%%%%%
As mentioned in the introduction of this section, the development of single electron sources have been triggered by the quest for a fundamental standard of electrical current. 
One of such sources realised in GaAs based nanostructures will be described in the following. 
The basic building block of this device is a dynamic quantum dot (QD), where the periodically varying confining potential is varied by energy barriers.
Originally \cite{blumenthal_nphys_2007} such a source was implemented by using two barriers controlled by two independent voltage parameters.
In a more optimised version \cite{kaestner_prb_2008,fujiwara_apl_2008,kaestner_apl_2008, fletcher_prb_2012} only one of the two gate voltages is used to eject a single electron from the dynamic QD as shown in Fig.\,\ref{fig_SES_cambridge}.
%-------------------------------------------------------------
%-------------------------------------------------------------
%-------------------------------- Fig. ----------------------- --------------------------------------------------------------
%-------------------------------------------------------------
%-------------------------------------------------------------
%-------------------------------------------------------------
%-------------------------------- Fig. ----------------------- --------------------------------------------------------------
%-------------------------------------------------------------
Two parallel electrostatic gates with a small opening allow to trap a small number of electrons inside the QD.
A schematic of the one-dimensional electrostatic potential landscape is shown in the right panel.
For the loading procedure the energy of the right barrier is set well above the Fermi energy to prevent the electrons from escaping the QD.
The left barrier is then lowered to an energy close to the Fermi energy to load a small number of electrons (black curve).
By increasing the gate voltage on the left barrier (more negative voltage) the QD is progressively isolated.
During this process, some initially trapped electrons tunnel back to the reservoir before tunnelling is eventually suppressed.
The electrons, which remain trapped, are ejected from the dot, once the left barrier exceeds the potential of the right one.
By finely adjusting the voltages on the gate $V_{G1}$ and $V_{G2}$ it is then possible to eject a single electron. The single electron pumps are usually operated at a repetition frequency of 100 MHz-1GHz, limited by the tunnelling time into the QD.
Presently, current accuracy of about 1ppm have been achieved \cite{kataoka_ncom_2012}.
\cb{To obtain such a high accuracy, the experiments are usually done under a large magnetic field, which stabilises the quantised current plateaus as shown in \ref{fig_SES_cambridge}(c) and (d) most likely due to the increased sensitivity of the tunneling rate to the electrostatic potential and the suppression of non-adiabatic excitations \cite{Leicht_sst_2011, fletcher_prl_2013, kaestner_ropp_2015}. 
Working at large magnetic fields is also convenient to guide the electrons along the edge states in the quantum Hall regime}.
\cb{ A detailed review on the working principle and performance of these single electron sources can be found in ref. \cite{kaestner_ropp_2015}}. 
Let us emphasise that for this single electron source the electrons are ejected with an energy far above the Fermi sea, typically above 100 meV with an energy resolution of about 3 meV \cite{fletcher_prl_2013}. 
\cb{This is much larger than the Fermi Energy $E_F \approx$ 10 meV as well as the charging energy of the gate defined QD $E_C \approx $ 1meV.  
Naturally, this high energy will set limits to this single electron source to use it for electron optics experiments. We will come back to this issue at the end of this section.} 

It is also possible to measure the energy as well as the temporal distribution of the emitted \textit{high energy} electrons.
This can be done by adding an energy selective barrier at the arrival position \cite{fletcher_prl_2013}.
By repeating the single electron emission at the pump clock rate and by inducing a time delay between the emission and the on-off switching of the arrival barrier, one can map out the shape of the emitted wave packet with ps resolution \cite{waldie_prb_2015}.
Presently the smallest wave packet size so far detected is of the order of 5ps \cite{kataoka_arxiv_2016}.
Finally, it is also possible to load several electrons in the QD and eject them sequentially \cite{fletcher_prl_2013}.
Using in a similar manner a barrier to detect the electrons it is possible to partition them individually \cite{haug_nnano_14}. 
We will come back to this issue in section \ref{Quantum optics like experiments}.

\cb{Let us also mention that semiconductor devices made from silicon \cite{fujiwara_review_2002} become again very popular \cite{fujiwara_apl_2016} for high precision electron pumps. 
This is also the case for single electron QDs which can be operated as charge \cite{petta_science_2017} or spin qubits \cite{veldhorst_nnano_2014,defranceschi_ncom_2017}. 
Silicon has the advantage compared to GaAs that it can be \textit{relatively} easily isotopically purified. Nuclear spin free $^{28}$Si is nowadays employed in the spin qubit field and extremely long coherence times have been obtained \cite{muhonen_nnano_2014}. }

%
%%%%%%%%%%%%%%%%%%%%%%%%%%
 \subsection{ Single electron source based on voltage pulses (levitons) }
 
Here, we describe a very simple way of injecting single or multiple electrons in a quantum conductor. 
The idea is to reduce the charge emitted by the electronic reservoir to its ultimate value -- an elementary charge -- by applying an ultra-short voltage pulse.

The method presents the advantage that no lithography step is required for the electron source and the (moderate) difficulty is left to the control of a current pulse on a very short time scale. 
To describe the principle, let us first consider a perfect quantum conductor made of a single quantum channel, spin disregarded. 
According to finite frequency B\"{u}ttiker's quantum transport laws, a voltage pulse $V(t)$ applied on a contact, while other contacts are grounded, injects a current pulse $I(t)=\frac{e^{2}}{h}V(t)$ from the contact to the single channel conductor. 
To inject n electrons, one has to tune the amplitude and duration of the voltage pulse such that $\int_{-\infty}^{\infty}I(t)dt=ne$ or equivalently $\int_{-\infty}^{\infty}eV(t)dt=nh$. 
Thus realizing a single electron source ($n=1$) seems easy to perform. 
However, for quantum information application, it is important that the injected electron, the flying qubit, is the only excitation created in the quantum conductor. 
This is not the case in general, as noticed by Levitov and collaborators in a series of theoretical papers\cite{Levitov_JMathPys_1996,Ivanov_PRB_1997,Keeling_PRL_2006}.
Indeed electrons are not injected in a vacuum of quantum states, like single photons, but on a ground state full of electrons, the Fermi sea. 
The voltage pulse in general perturbs all electrons and creates extra excitations \cite{belzig_prl_2007,belzig_prb_2008,belzig_pss_2017}. 
These excitations are neutral in order to conserve the injected charge. 
Levitov's remarkable prediction was that only a special voltage pulse, a Lorentzian pulse, injecting an integer number of charge is able to provide a so-called minimal excitation state where only charge excitation is created with no extra neutral excitation \cite{Levitov_JMathPys_1996}. 
After its recent experimental realization \cite{glattli_nature_2013}, this single charge minimal excitation state has been called a leviton. 
Other pulse shapes\cite{reulet_prb_2013}, or non-integer charge injection, create non-minimal states  which are not suitable for flying qubits. 
To understand the underlying physics of the generation of levitons, one has to consider the effect of a voltage pulse on all electrons of the electrical contact subjected to the pulse. 
An electron emitted from the contact at some energy $\varepsilon$ below the Fermi energy and experiencing the potential $eV(t)$ has its phase modulated as $\phi(t)=2\pi \int_{-\infty}^{t} eV(t')dt'/h$. 
As the time dependence is breaking energy conservation, the electron will end in a superposition of quantum states of different energies. 
The probability amplitude to have its energy displaced by $\delta \varepsilon$ is $p(\delta \varepsilon)= \int_{-\infty}^{\infty}\exp(-i\phi(t))\exp(-i\delta\varepsilon t/\hbar )$. For arbitrary phase modulation (or voltage pulse shape) $p(\delta \varepsilon)$ takes finite values for both positive and negative $\delta \varepsilon$. The electrons of the Fermi sea are displaced up and down in energy and this creates electron- and hole- like excitations. To create a leviton, a pure electron excitation with no hole, one needs $p(\delta \varepsilon)=0$ for $\delta \varepsilon<0$. 
To ensure this for a single pulse, $exp(-i\phi(t))$ must have no poles in the lower half complex plane and at least one pole in the upper half, i. e. $exp(-i\phi(t))=(t+iw)/(t-iw)$. One immediately sees that the phase derivative (and so the voltage $V(t)$) must be a Lorentzian, with : 
\begin{equation}
\label{VLor}
V(t)=\frac{\hbar}{e}\frac{2w}{t^{2}+w^{2}}
\end{equation}
the parameter $w$ is the width of the Lorentzian. Adding an extra pole in the upper complex plane is equivalent to adding an extra electron. For periodic leviton injection with period $T=1/f$, the poles are regularly spaced at values $-kT+iw$, $k$ integer and $exp(-i\phi(t)=\sin(\pi(t+iw)/T)/\sin(\pi(t-iw)/T)$ and 
\begin{equation}\label{VacLor}
    eV(t)= \frac{h f}{2} \frac{\sinh(2\pi w/T)}{\sinh(\pi w/T)^{2}+ \sin(\pi t/T)^{2}}
\end{equation}

\begin{figure}[!htb]
\centering
\includegraphics[width=7.5
cm]{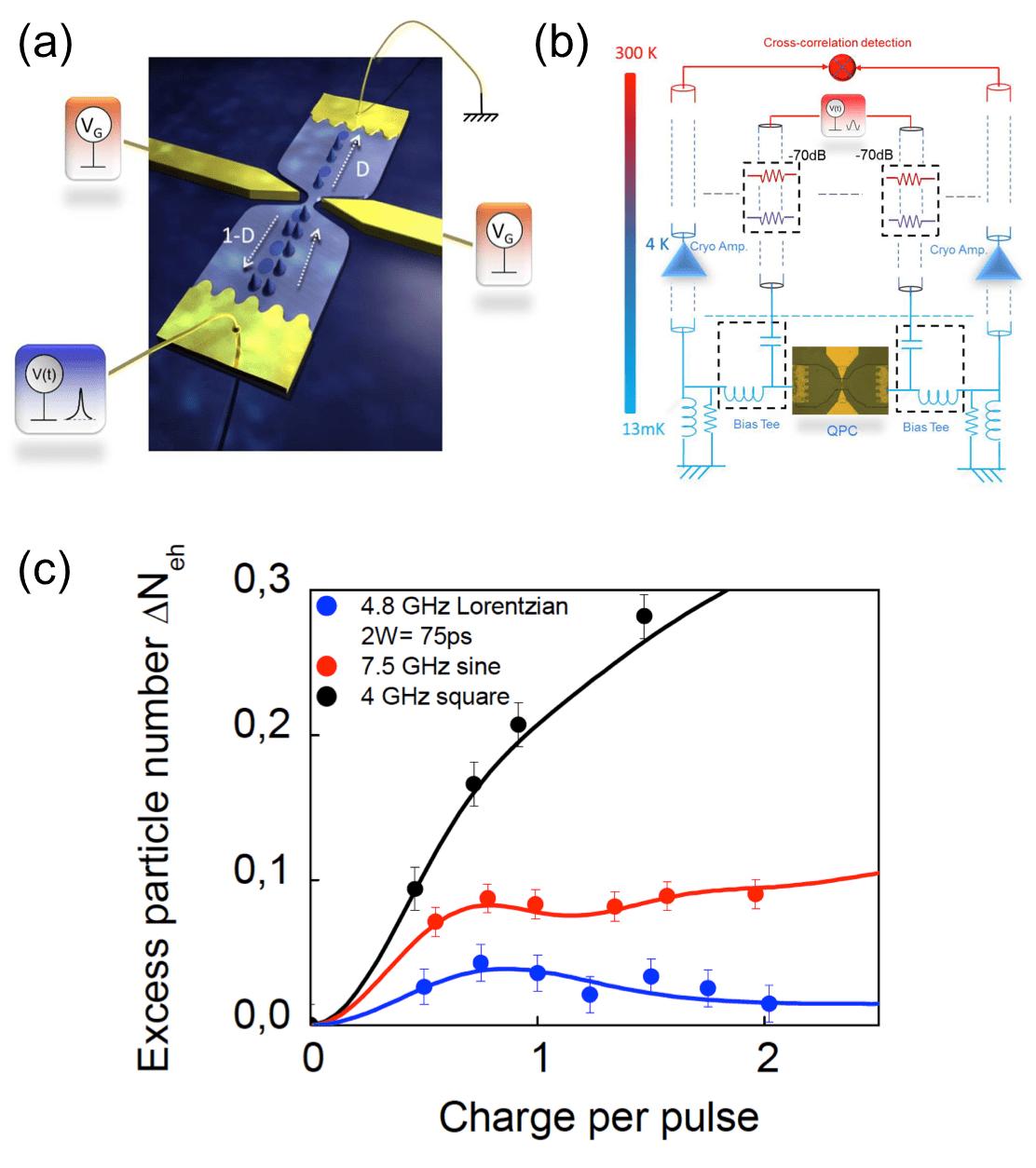}
\caption{{\bf{Leviton voltage pulse source.}} (a) schematic view of the sample, a 2D electron gas with a quantum point contact in its middle. (b) experimental cryogenic set-up designed to send short $\sim 30ps$ wide Lorentzian pulses and equipped for cross-correlation noise measurements with cryogenic HEMT amplifiers. The lower graph (c) shows the excess shot noise generated by partitioned electron pulses. Square and sine wave pulse gives finite noise due to partitioning of extra electron-hole pairs accompanying the pulse. Lorentzian pulses with integer charge give no noise except a weak thermal noise contribution. This is consistent with minimal excitation states called levitons and suitable for charge flying qubits (figures are adapted from ref. \cite{glattli_nature_2013}).}
\label{leviton}
\end{figure}

The first implementation of a single electron voltage pulse has been done recently, see ref. \cite{glattli_nature_2013}. 
The periodic injection of a single electron, using square, sine and Lorentzian pulses have been compared to evidence Levitov's prediction of minimal excitation states. 
To do this a measure of the total number of excitations created per pulses is needed. 
This is provided by sending the charge pulses towards a QPC with finite transmission $D$. 
This artificial scatterer plays the role of a beam splitter which partitions the charge into transmitted and reflected states following a binomial law. 
Assuming we have created levitons, i.e. electrons not accompanied by electron-hole pairs, the partitioning statistics of $n$ electrons arriving at frequency $\nu$ gives a low frequency current noise spectral density $S_{I}=2e^{2}f n D(1-D)$ (here $n=1$). 
If however both electron and hole excitations are incoming on the QPC, one can show that, at zero temperature, one has exactly, see ref.\cite{glattli_prb_2013} :
\begin{equation}
\label{Noise}
S_{I}=2e^{2} f (N_{e}+N_{h}) D(1-D)
\end{equation}
while the mean current is $I=e\nu (N_{e}-N_{h}) = e\nu n$. Levitons with $N_{h}=0$ and $N_{e}=n$ give minimal noise. This was experimentally demonstrated in ref.\cite{glattli_nature_2013}. 
Reduction of the shot noise has also been observed in experiments using tunnel junctions with a biharmonic drive \cite{reulet_prb_2013}.
\pr{In Fig. \ref{leviton} (a) and (b), the experimental set-up is depicted: leviton pulses are sent on the ohmic contact of the quantum point contact, while partitioned quasiparticles are detected by cross-correlation measurement techniques. In Fig. \ref{leviton} (c), excess particle number as a function of injected charge per pulse is shown: compared to the sine and square pulse, the Lorentzian pulse gives the smallest amount of electron-hole pairs. This approach ensures an excellent control of the electronic wave function that arrives at the QPC. For electron-quantum optics experiments, this source has the advantage to inject electrons at the Fermi energy and will less suffer from relaxation processes, as observed in the mesoscopic capacitor source.}  It is worth noticing that this experiment is an electron analog of a photonic Hanbury-Brown Twiss experiment \cite{henny_science_1999,yamamoto_science_1999} where single photons are sent to a beam splitter, but with electrons. We will see later that we can go further and perform Hong--Ou Mandel--interferometry by sending two periodic trains of levitons on each beam splitter input and measuring the noise correlation.
%-------------------------------------------------------------
%-------------------------------------------------------------
%-------------------------------- Fig. ----------------------- --------------------------------------------------------------
%-------------------------------------------------------------
%-------------------------------------------------------------
%-------------------------------------------------------------
%-------------------------------- Fig. ----------------------- --------------------------------------------------------------
%-------------------------------------------------------------
%%%%%%%%%%%%%%%%%%%%%%%%%%
%
%
%%%%%%%%%%%%%%%%%%%%%%%%%%
\subsection{SAW driven single electrons}
%%%%%%%%%%%%%%%%%%%%%%%%%%

\begin{figure}[h]
\includegraphics[width=7cm]{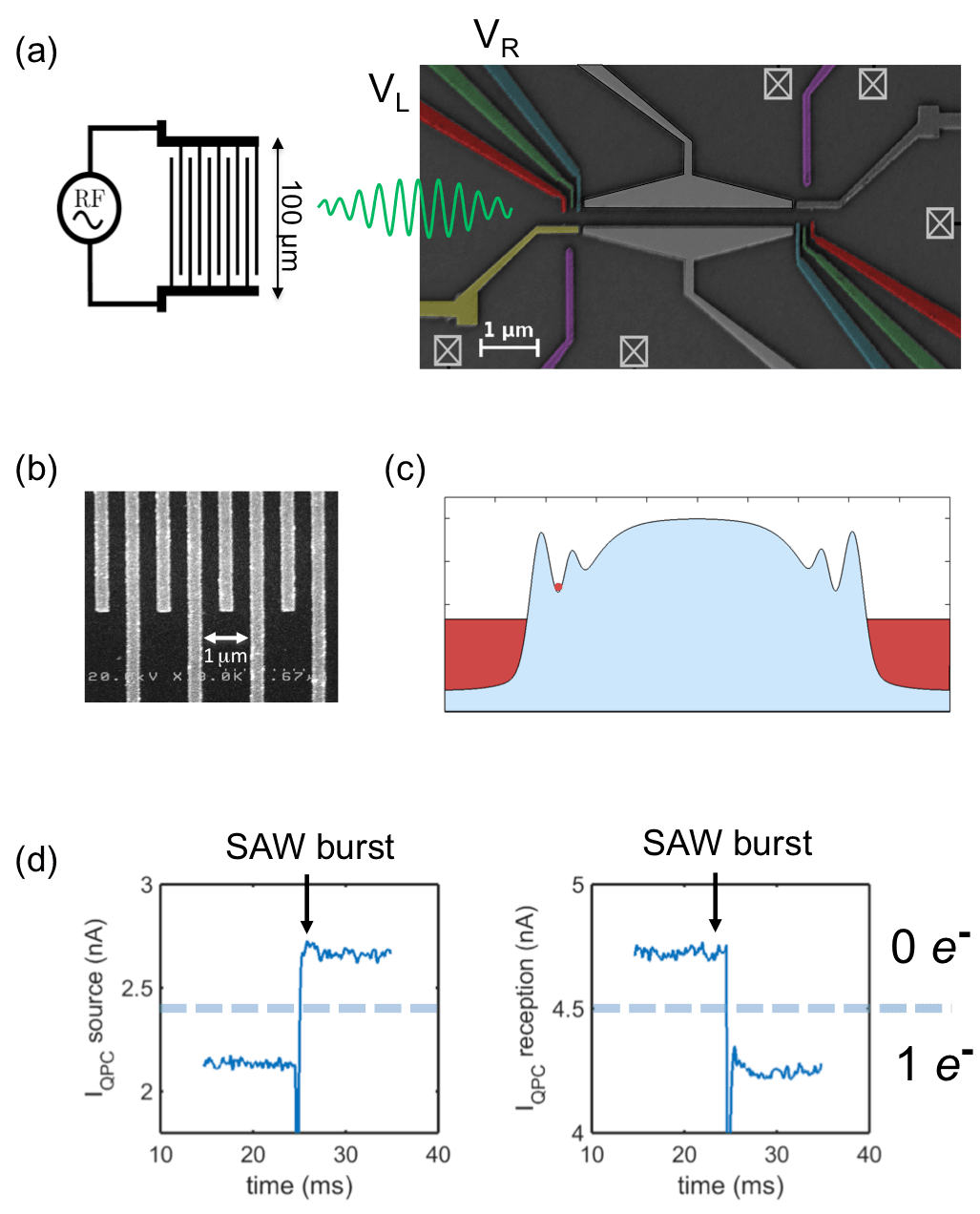}
\caption{\textbf{Single electron SAW device}. (a) SEM of the single electron source. The source and reception dot is defined by four gates highlighted in color (yellow, red, green and blue). The violet gates serves as a QPC in order to determine the electron number of each quantum dot. The two large central gates serve to guide the emitted single electron on a specific trajectory. (b) SEM image of the central part of the interdigitated transducer schematised in (a). It is composed of 70 fingers with a spacing of 1 $\mu$m. Each finger has a width of 250 nm and a length of 100 $\mu$m. The transducer is placed approximately 1 mm away from the central structure. (c) Electrostatic potential landscape for a single electron transfer experiment. (d) Coincidence measurements of a single-shot QPC trace at the source (left) and reception (right) quantum dot. An electron initially trapped in the source quantum dot is transferred to the reception dot after applying a SAW burst. Figures are adapted from ref. \cite{hermelin_nature_2011,hermelin_thesis_2012,hermelin_pss_2017}. }
\label{fig_SAW}
\end{figure}

Yet another highly efficient on-demand single electron source can be realised by transporting a single electron with a surface acoustic wave (SAW) \cite{hermelin_nature_2011,mcneil_nature_2011}.
One exploits again the fact that a single electron can be isolated in a semiconductor quantum dot (QD) and single electron transfer can be realised by transferring the single electron into a moving SAW QD.
When connecting two QDs with a quantum channel (see figure \ref{fig_SAW}(a)), it is then possible to transfer a single electron from one QD to the other with detection efficiencies much higher than 90 \% \cite{hermelin_nature_2011,mcneil_nature_2011}.
%over a distance of several tens of microns \cite{takada_coupler_SAW}. 
\\
GaAs is a piezo-electric substrate and therefore allows to generate SAWs which carry a moving electric field for certain crystal directions.
The SAW can be generated by an interdigitated transducer (IDT) which is deposited on the surface of the GaAs substrate (see figure \ref{fig_SAW}(b)). 
The IDT is usually composed of several tens of interdigitated metallic fingers with a length of about 100 $\mu$m in order to create an oscillating electric field at the surface of the GaAs crystal when applying a radio frequency signal to the two electrodes. 
Due to the piezo-electric effect, the crystal contracts periodically and generates a Rayleigh wave for specific crystal directions which travels at the surface of the GaAs crystal with a sound velocity of the order of 3000 m/s \cite{lima_jap_2003}. 
\cb{This slow speed, which is about 2 orders of magnitude slower than the Fermi velocity, is advantageous as it allows to induce gate operations of the propagating electrons on shorter length scales compared to ballistic electrons.}
%
%\cite{hayashi_prl_2003,petersson_prl_2010}.
%and moving SAW electrons in a tunnel-coupled wire \cite{kataoka_prl_2009}.  
The wavelength of the SAW can be simply engineered with the distance between the fingers of the IDT.
For such single electron transfer experiments, \cb{the IDT is operated at a frequency close to 3 GHz which translates into a wavelength of} the order of 1 $\mu$m \cite{shilton_jpcm_1996,talyanskii_prb_1997,fujisawa_prl_2006,benoit_nanotechnology_2016}.
This ensures a moving QD of a size of several hundred nanometers when propagating through an electrostatically defined one-dimensional channel of similar dimensions. 
\cb{Going to higher frequency is desirable as it increases the confinement potential and hence the level spacing of the QD. 
This however comes with technical difficulties. 
Beyond a frequency of 5-6 GHz the efficiency of the IDTs decreases drastically due to ohmic losses in the metal strips (usually made from gold) but also due to losses into the bulk of GaAs. 
The wavelength gets so small that surface roughness can excite bulk waves. 
Going to smaller wavelength puts also much higher constraints on the spatial resolution of nanofabrication.
The highest frequencies which have been achieved (f $\approx$ 20 GHz) so far have been realised with nano-imprint techniques \cite{vanderwiel_nanotech_2012}.
}
%-------------------------------------------------------------
%-------------------------------------------------------------
%-------------------------------- Fig. ----------------------- 
%-------------------------------------------------------------
%-------------------------------------------------------------
%-------------------------------------------------------------
%-------------------------------- Fig. ----------------------- 
% --------------------------------------------------------------
%-------------------------------------------------------------

\begin{figure}[h]
\includegraphics[width=7.5cm]{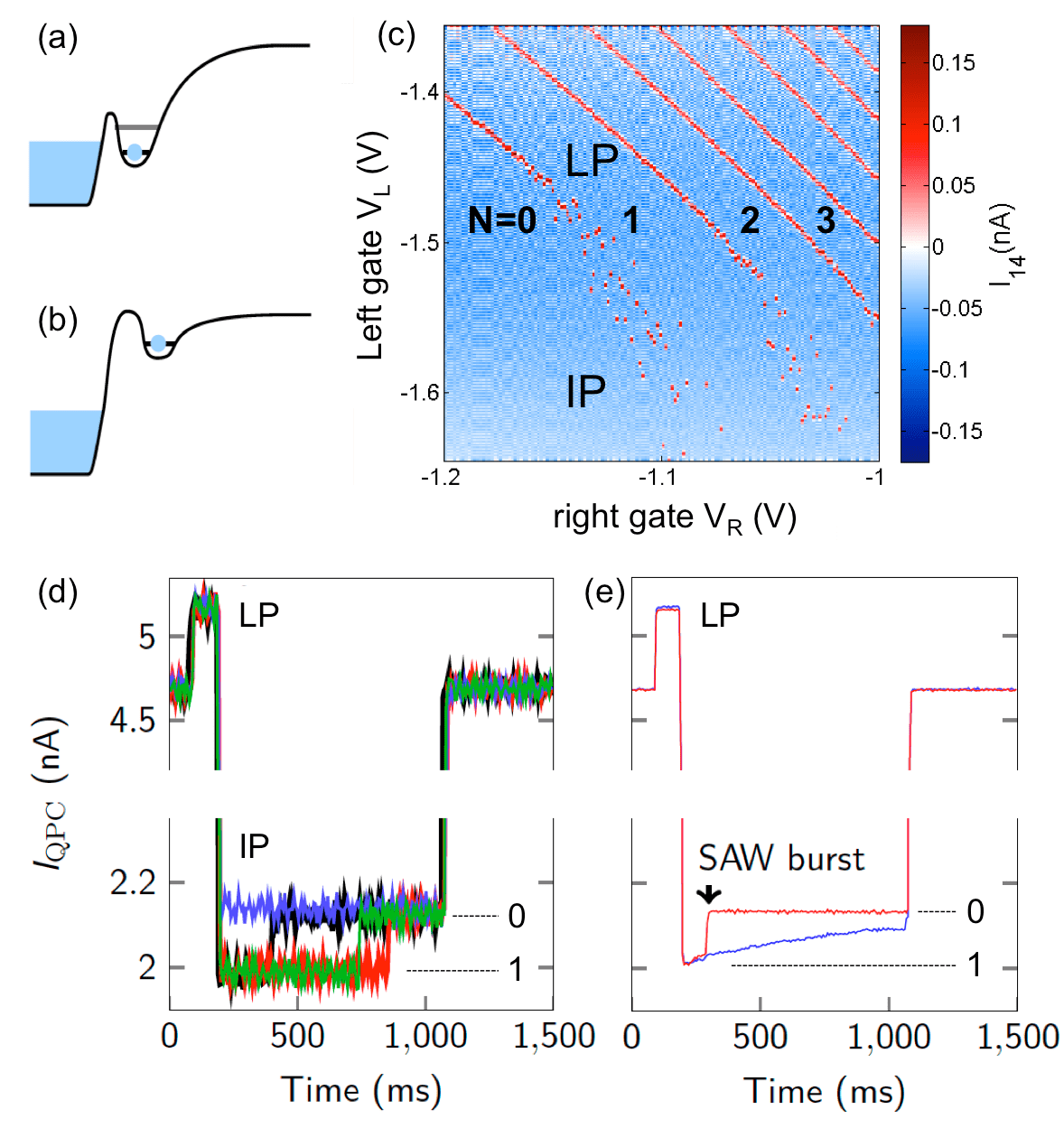}
\caption{\textbf{Single electron loading procedure and dwell time measurement of the electron}. Electrostatic potential profile for (a) the loading position LP, (b) the isolated position IP. (c) Charge stability diagram of the quantum dot obtained by varying the two barrier gate voltages $V_L$ and $V_R$. The color scale corresponds to the derivative of the QPC current. Whenever a red line is crossed, the electron number on the dot changes exactly by one. (d) Individual escape events of the electron when in the IP. (e) blue curve: average of 10000 single escape events; red curve: same measurement as for the blue curve, with the difference that  after a waiting time of 50 ms in the isolated position a 60 ns long SAW train is launched.}
\label{fig_SAW_QPC}
\end{figure}

\cb{The SAW driven single electron source is operated in the following way:} 
At first, a single electron is loaded into the QD at the loading position LP of the charge stability diagram shown in figure \ref{fig_SAW_QPC} (c).
The quantum point contact (yellow and purple gates in figure \ref{fig_SAW}(a)) allows to detect whether an electron is present inside the QD (see section \ref{Single Electron Detectors} for details).
Varying gate voltages $V_L$ and $V_R$ the electron is then moved to its isolated position IP by increasing the barriers formed by the electrostatic gates which separate the QD from the reservoir ($V_L$) and the channel ($V_R$).
At this position the electron can be trapped for a very long time \cite{bertrand_prl_2015}. 
%This is usually done in the measurement position M where the QPC is very sensitive to the electrostatic potential created by the presence of the electron (position A in figure \ref{fig_SAW_QPC} and its corresponding electrostatic potential). 
The dwell time of an electron in this isolated position can be measured by statistical average of the time\st{ when} the electron stays in the QD before it tunnels into the nearby reservoir. 
This is shown in figure \ref{fig_SAW_QPC}(d) where several individual escape events are measured. 
Averaging 10000 of such events results in an exponential decay from which one can deduce the dwell time of the electron in the isolated position, here $\approx$ 700 ms. 

If one now keeps an electron in the isolated position and launches a SAW train, one can expel the single electron from the QD with very high efficiency. 
To demonstrate this, a 60 ns long SAW train is launched from the IDT 50 ms after the single electron is brought to the isolated position. 
If the amplitude of the SAW is sufficiently strong, the moving electric field can pick-up the electron from the QD and carry it along.
This is shown by the red curve in figure \ref{fig_SAW_QPC}(e). 
Whenever the SAW arrives at the QD, the electron is ejected and the QD is depopulated.
Adjusting the SAW amplitude and the QD potential, ejection efficiencies higher than 96 \% have been achieved \cite{hermelin_nature_2011,mcneil_nature_2011}. 
\cb{By engineering one-dimensional channels with electrostatic gates, the ejected electron can be transported and guided at will to any desired position on the electronic circuit.
The single electron is then literally surfing on the SAW within the electrostatic confinement potential landscape created by the one-dimensional channel as shown in figure \ref{fig_SAW}(c).
In order to show that the electron is indeed transferred to the detector QD through the one-dimensional channel it is necessary to perform coincidence measurements on both QDs at the same time \cite{hermelin_pss_2017}.
Such coincidence measurements are shown by the QPC traces in figure \ref{fig_SAW}(d) }. 

%-------------------------------------------------------------
%-------------------------------------------------------------
%-------------------------------- Fig. ----------------------- --------------------------------------------------------------
%-------------------------------------------------------------
%-------------------------------------------------------------
%-------------------------------- Fig. ----------------------- --------------------------------------------------------------

\cb{
Now that we have presented the different single electron sources let us discuss their advantages and disadvantages.
All four SES can easily be integrated into electronic quantum circuits engineered from GaAs heterostructures. 
The leviton source has the least constraints on nano fabrication as it simply requires an ohmic contact to which a short voltage pulse is applied. 
Care has to be taken for the design of the waveguide which guides the radio-frequency (RF) signal to the contact, but this is true for any electrical connection on-chip which is operated at RF frequencies. 
The ohmic contact should be reduced to a small size, typically $ 10 \mu $m$ \times 10  \mu $m, to be able to easily guide the single-electron wave packet into a gate defined quantum rail and the contact resistance should be as small as possible, typical values are of the order of 100 $\Omega$.  
For all other sources, several additional electrostatic gates have to be implemented.
The most demanding from this point of view is certainly the SAW based single electron source as it requires a detector (QPC) as well as a QD to make it operational.
On the other hand an important advantage of this SES is that it also allows for \textit{single-shot} detection, a requirement absolutely necessary for quantum information purposes. 
For this single electron transport technique it is possible to capture the propagating single electron in another QD and measure its presence with a single-shot measurement \cite{hermelin_nature_2011,mcneil_nature_2011} with a precision higher than 99 \% \cite{takada_tbp_2017}. 
This is presently not the case for the other three electron sources. 
All electron counting experiments performed to date with these SES are based on a measurement of the average DC current or the low frequency current noise while repeating the experiment billions of times.
To reach the single-shot limit is very challenging since the propagation speed of the generated electron is very fast and the interaction time with any detector will be very short (see next section for details).

To realise a current standard the non-adiabatic charge pump is obviously the best choice as it allows quantisation of the electron charge as well as very high clock frequencies. 
SAW based single electron sources as the one presented here cannot be operated at very high repetition frequencies.
\st{Presently they are limited to below $100$ Hz due to technical issues of the experimental setup \cite{hermelin_pss_2017}.}
It is also possible to apply the SAW in a continuous manner by driving the SAW across a constriction  \cite{wixforth_prl_1986,shilton_jpcm_1996,ford_pss_2017} 
%K. Flensberg, A. A. Odintsov, F. Liefrink, and P. Teunissen, Towards single-electron metrology, Int. J. Mod. Phys. B 13 (21-22), 2651?2687 (1999).
rather than by single electron transport between QDs with very short SAW bursts. 
In this case one can confine a single electron within each minimum of the SAW. 
and allows to obtain a very high repetition frequency (several GHz).
This technique was initially developed with the motivation to realise a very precise electron pump with a current of several hundred pA (1GHz $\approx$ 160 pA).
The best precision which could be achieved with this source was, however, only of the order of 100 ppm \cite{ford_pss_2017}.
For quantum interference experiments single electron transport using a SAW wave in a continuous fashion presents also some limitations. 
The power dissipation due to the SAW itself is high, not favourable for quantum interference experiments.
%\cite{SAW-heating}
%
%add comment about duty cycle
%
Neither the leviton source nor the Mesoscopic Capacitor source can be used as a current standard. 
The leviton source does not deliver any \textit{quantised} current, that is no quantisation plateau will appear when changing for instance the amplitude of the voltage pulse, while the Mesoscopic Capacitor source delivers a zero net DC current as it periodically generates an electron followed by a hole. 
%The presence of an electron (hole) is then determined via shot noise measurements \cite{blanter $??$}.
%maybe cite blanter buttiker review here 
%Such measurements give information on the error of the 

In this review, however, we are mainly interested in discussing quantum coherent nanocircuits suitable for implementation of electron quantum optics and flying qubits with electrons. 
This requires an electron wave packet that preserves its phase while propagating throughout the entire quantum circuit. 
In addition, two electrons sent from two different sources should be well synchronised in time and should be indistinguishable.   
In this respect the four SES have quite different properties. 
Synchronization on the ps level can be achieved with all four single electron sources.
State-of-the-art arbitrary wave generators allow to induce a time difference of presently of about 1\,ps between two output signals\cite{kataoka_arxiv_2016} which is well below the actual time-spreading of the generated single-electron wave packets.
In this respect only the synchronisation of the SAW wave source seems problematic as a single electron has to be loaded into exactly the same SAW minimum for each SES.
Recent experiments have shown that this can be achieved with very high efficiency when using ps pulse triggering of the loading barrier gate to inject an electron into a specific minimum of the SAW train \cite{takada_tbp_2017}.
With this technique it is then possible to map out in a time-resolved manner the entire SAW train.

An important issue for realising quantum interference experiments at the single-electron level is phase preservation. 
For the non-adiabatic SES, the ejection energy of the electron is very high.
While propagating, the electron will relax and loose energy which results in visibility loss when performing interference experiments. 
It has been shown that electrons can be transported over several microns with small inelastic electron-electron or electron-phonon scattering \cite{fletcher_prl_2013}.
For a 170 meV emission energy the contribution due to inelastic scattering is less than 1 meV.
In order to reduce energy relaxation over longer propagation length, it is possible to completely deplete the electron gas which minimizes electron-electron interactions \cite{kataoka_prl_2016}. 
Interference experiments with such a SES, however, have not yet been reported. 
Energy relaxation is also an issue for the Mesoscopic Capacitor electron source. 
Similar to the non-adiabatic single electron source, the generated electron have a well defined energy, however with a 100 times smaller energy. 
Still, when launching the electron wave packet, it will relax in energy during propagation which will lead to decoherence \cite{degiovanni_pss_2017,degiovanni_prb_2013}.
In addition electron-electron interaction between the edge channel have to be taken into account.
Even though additional edge channels help to reduce the Coulomb interaction due to mutual screening \cite{heiblum_prb_2016}, Hong--Ou--Mandel interference measurements show that the Pauli dip does not go to zero \cite{feve_science_2013}.
We will come back to this issue in more detail in section \ref{Quantum optics like experiments}. 
%For the case of $\nu=1$ the phase coherence is usually reduces due to  
% maybe comment that in section optics
% \cite{thesis_marguerite_2017} show that the Hong?Ou?Mandel dip (see section \ref{Quantum optics like experiments}.

This issue is conceptually different for the other two sources. 
%While the Mesoscopic Capacitor source generates a well defined electron wave packet in energy it is poorly resolved in time while for the leviton source this is exactly the opposite. It has a very short time spreading. 
As mentioned above, the leviton source generates a very peculiar type of single electron excitation.
\cb{Due to its exponential energy distribution, the leviton lives very close to the Fermi sea and since all electronic states below the Fermi sea are occupied, it is well protected against energy relaxation.}
Two particle Hong--Ou--Mandel interference have shown a reduction of the Pauli dip which is in good agreement with theoretical expectations \cite{glattli_nature_2013} and nicely demonstrates that the two emitted electron wave packets are indistinguishable.
From these experiments it is however difficult to estimate a value of the phase coherence length. 
Preliminary experiments on a 40 $\mu$m long tunnel-coupled wire with electron wave packets of a temporal width shorter than 100 ps have shown that phase coherence is preserved throughout the wire\cite{roussely_thesis_2016}.
This is very promising and suggests that the leviton source is very suitable for integration into a flying qubit architecture.
Apart from being very suitable for quantum optics like experiments, the leviton source will also allow to explore novel quantum effects.
As the wave packet is propagating at the surface of the Fermi sea, novel quantum interference phenomena have been predicted when a very short charge pulse is interfering with the Fermi sea \cite{gaury_ncom_2014}.
This novel physics will be discussed in section \ref{Novel quantum interference experiments} 

The SAW electron source on the contrary eliminates completely the effect of the Fermi sea.
The electron is confined in a moving QD well above the Fermi energy. 
In this sense this technique is the closest to photon experiments.
However, the fact that the electrons are completely isolated form the Fermi see makes them also more vulnerable to external perturbations since electron screening is strongly reduced.
As mentioned above, this technique allows to transport single electrons with a very high fidelity \cite{hermelin_nature_2011, mcneil_nature_2011}.  
Recent experiments have been able to push the transfer efficiency beyond 99 \% for a transfer distance of more than 20 microns \cite{takada_tbp_2017}. 
Employing a structure similar to the one schematised in figure \ref{fig_TC+AB}(a), it has been possible to partition on-demand electrons coming from input port 1 into the two output ports, hence realising a directional coupler at the single-electron level \cite{takada_tbp_2017}.
Present research is devoted to the realisation of phase coherent transport of SAW driven electrons.
First attempts for the observation of coherent tunnelling between two tunnel-coupled wires have been done with a continuous wave approach.
In these experiments tunnelling from one quantum wire into a two-dimensional reservoir have been observed \cite{kataoka_prl_2009,ford_pss_2017}.
Coherent tunnelling between two tunnel-coupled wires with SAWs have so far not been realised.
This is subject to ongoing experimental research.
%maybe cite michi/ito work from EP2DS abstract.
% say something about possible difficulties ?
Finally, let us also mention that single electron transport assisted with SAW also allows to exploit the spin properties of the electron which is quite appealing for quantum information processing.
The spin degrees of freedom couple much less to the electromagnetic environment compared to the charge degree of freedom.
Recent measurements 
%by Bertrand et al. \cite{bertrand_nnano_2016} 
have shown \cite{bertrand_nnano_2016} that the spin polarisation is preserved during the transport with a fidelity of 70 \%.
}

%These issues will be treated in section \ref{Quantum optics like experiments}. 
%This is subject to ongoing experimental and theoretical research 
%is subject of ongoing experimental and theoretical research
%
%issues to be solved for this ambitious goal are...

%%%%%%%%%%%%%%%%%%%%%%%%%%
%%%%%%%%%%%%%%%%%%%%%%%%%%
\section{Single Electron Detectors}
\label{Single Electron Detectors}
%%%%%%%%%%%%%%%%%%%%%%%%%%
%%%%%%%%%%%%%%%%%%%%%%%%%%

The detection of a single electron \cb{can be achieved} when the electron is captured inside a static quantum dot. 
As the electron can be trapped for a sufficiently long time, it can be detected with conventional on-chip detectors \cite{field_prl_1993} \cb{and which we will detail below. }
In quantum experiments \cite{heiblum_nature_2003,roulleau_prl_2008,yamamoto_nnano_2012,feve_science_2013,glattli_nature_2013}\cb{, where one would like to detect the electrons on the fly}, this task is much more difficult: indeed the \cb{interaction time with any detector} usually does not exceed 1 ns and is fixed by the speed of the flying electrons, the size of the on-chip detector and the spatial extension of the electronic wave packet. 
\cb{This issue will be addressed in more detail in section \ref{electron detection on the fly}.}

%At present the interaction time is 2 orders of magnitude faster than the best on-chip charge detector demonstrated so far in a 2DEG \cite{barthel_prl_2009,dial_prl_2013} and hence represents an important limitation to investigate high order quantum correlations in experiments with flying electrons. 

%%%%%%%%%%%%%%%%%%%%%%%%%%
\subsection{Single electron detection in static quantum dots}
%%%%%%%%%%%%%%%%%%%%%%%%%%

\begin{figure}[h]
\includegraphics[width=8cm]{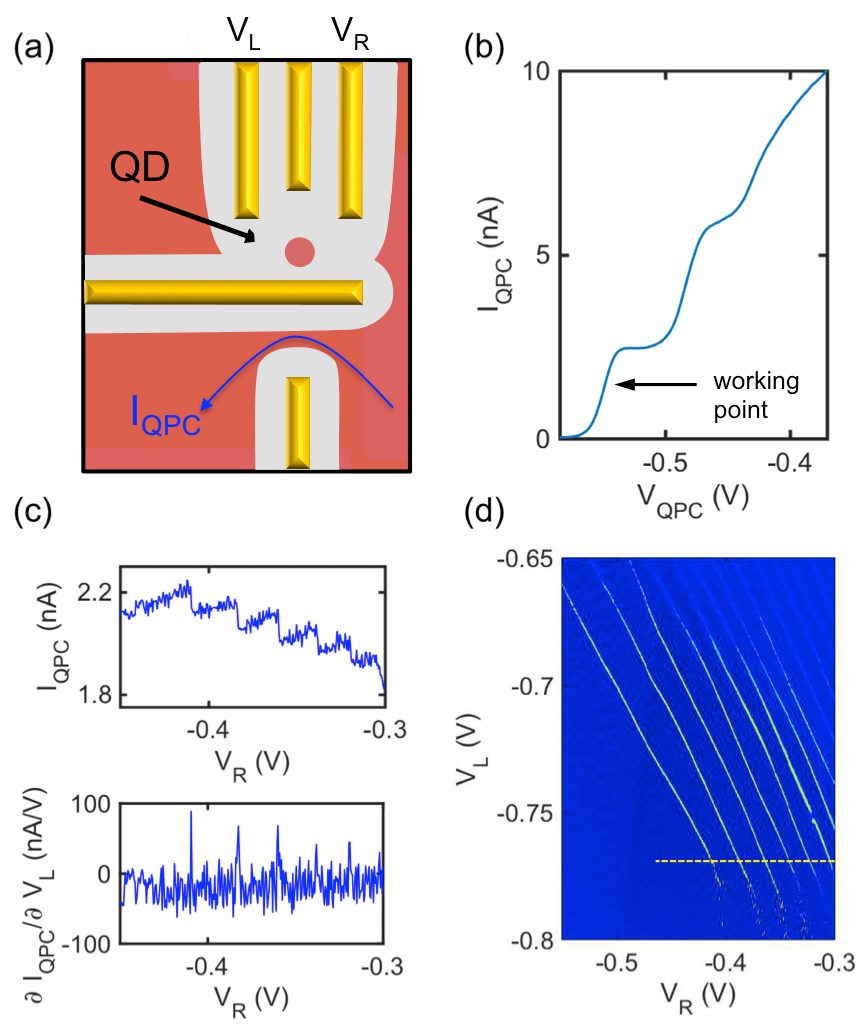}
\caption{\textbf{Principle of single charge detection with a quantum point contact}. (a) schematic of the gate structure of the quantum dot (QD) and its adjacent QPC. The (un)depleted electron gas is represented in grey (red) color, the electrostatic gates are drawn in yellow. (b) Conductance curve of the QPC. The arrow indicates the approximate working point of the QPC. (c) QPC current when the voltage of the left QD gate is changed. The abrupt jumps in the current correspond to a change of the electron number on the quantum dot by exactly one. (d) Charge stability diagram: differential conductance (white: high ; blue: low) is plotted as a function of the two side gate voltages V$_L$ and $V_R$.   
}
\label{fig_QPC}
\end{figure}

A \cb{very convenient} way to read out the electronic charge state of a static quantum dot can be realised by a quantum point contact \cite{field_prl_1993} when placed in close vicinity to the quantum dot (QD) (see figures \ref{fig_SAW} and \ref{fig_QPC}). 
\cb{The quantum point contact (QPC) is brought to a gate voltage condition where the sensitivity $\delta$G/$\delta V_G $ is the highest, usually in between the first quantised plateau and the pinch-off as shown in figure \ref{fig_QPC}(b).
The QPC is biased typically  at a few hundred microvolts to avoid back action \cite{ensslin_ssr_2009,ludwig_natphys_2012} and the current through the QPC is continuously monitored.
At this working point the current is very sensitive to the nearby electrostatic environment. 
If the electron number of the quantum dot is changed, the nearby electrostatic environment will be modified and results in an abrupt variation of the QPC current as shown in figures \ref{fig_QPC} (c).}
When sweeping the two side gates ($V_R$ and $V_L$) of the QD shown in figure \ref{fig_SAW}(a) and \ref{fig_QPC}(a), one can map out the so-called \cb{charge} stability diagram \cite{kouwenhoven_QDreview_1997} of the QD. 
Plotting the derivative of the conductance with respect to the gate voltage $V_R$ to suppress the smooth background and to enhance the conductance steps, one observes parallel lines which delimit regions where the electron number is constant.
Whenever a diagonal line is crossed, the number of electrons changes exactly by one.
Decreasing the voltage to more negative values ejects the electrons one by one until one ends up with a QD containing no electron at all (large region on the left in figure \ref{fig_QPC} (d) where the diagonal lines are absent). 

%-------------------------------------------------------------
%-------------------------------------------------------------
%-------------------------------- Fig. ----------------------- %-------------------------------------------------------------
%-------------------------------------------------------------
%-------------------------------------------------------------
%-------------------------------------------------------------
%-------------------------------- Fig. ----------------------- %-------------------------------------------------------------
%-------------------------------------------------------------
%%%%%%%%%%%%%%%%%%%%%%%%%%
%%%%%%%%%%%%%%%%%%%%%%%%%%
Nowadays the QPC is often replaced by a sensing QD where one takes advantage of a very sharp Coulomb blockade peak, which can have a higher sensitivity $\delta$G/$\delta V_G $.
In this case the sensing QD is operated at a very steep flank of a Coulomb peak.
This requires, however, an additional gate to be added to the sample design.
In addition to these improvements, fast \cb{read-out schemes} have been implemented by reflectometry in order to increase the measurement bandwidth.
To do so, the QPC or QD is integrated into a LRC tank circuit which operates at a few hundred MHz.
This is the so-called RF-QPC which allows for single-shot detection at timescales below 1\,$\mu$s. \cite{muller_icps_2007,reilly_apl_2007,cambridge_apl_2007}.

%%%%%%%%%%%%%%%%%%%%%%%%%%%%%%%%%%%%%%%%%%%%%%%%%%%
%%%%%%%%%%%%%%%%%%%%%%%%%%%%%%%%%%%%%%%%%%%%%%%%%%%
\subsection{Towards single electron detection \textit{on the fly} }
\label{electron detection on the fly}
%%%%%%%%%%%%%%%%%%%%%%%%%%%%%%%%%%%%%%%%%%%%%%%%%%%
%%%%%%%%%%%%%%%%%%%%%%%%%%%%%%%%%%%%%%%%%%%%%%%%%%%

To detect an electron \textit{on the fly} is presently the most challenging task in the field of electron quantum optics.
\cb{First, the interaction time between a submicron-size electrometer and the flying electron does not exceed 1 ns and is fixed by the speed of the flying electrons, the size of the on-chip electrometer and the width of the electronic wave packet.
This interaction time is 2 orders of magnitude faster than the typical timescale needed to detect a single electron for the best on-chip charge detector demonstrated so far in a 2DEG \cite{barthel_prl_2009,dial_prl_2013}.
Second, the experiment has to be performed at temperatures below 50 mK to avoid fluctuations of the local chemical potential of the Fermi sea larger than the influence of a single electron on the detector \cite{Levitov_JMathPys_1996}.}

Combining single electron injection and readout would open the avenue for the realisation of full quantum experiments and will bring the field of electron quantum optics to a status similar to that of quantum optics. In particular, single-shot detection of flying electrons would enable to perform electron flying qubit operations. 

Up to now most measurements are statistical: electron injection is periodically repeated and DC current or low frequency current noise measurements provide information on the average value and the fluctuations of the charge that arrives in each contact. 
Recording the detected single charge each time\st{, when a single charge is injected,} will also give unprecedented access to the Full Counting Statistics of the electron partitioned between the different output arms of an interferometer.
\cb{In addition, due to the existence of Coulomb interaction electrons provide new possibilities for linear quantum optics \cite{knill_nature_2001,beenakker_prl_2004} especially to generate entanglement \cite{ionicioiu_ijmp_2001}.}

Towards this goal several quantum systems have been identified as extremely sensitive systems to external perturbations and potentially good detectors \cite{haroche_nature_2007}. 
They have been used for example to detect a single phonon excitation of a nanomechanical system \cite{cleland_nature_2011}. 
In a 2DEG, two quantum systems have been recently proposed to detect propagating electrons: a double quantum dot charge qubit and a Mach-Zehnder interferometer \cite{neder_njp_2007}.

Another option is to exploit the extreme sensitivity of spin qubits when operated in the charge regime. 
The idea is to couple capacitively the single flying electron to a so-called Singlet-Triplet (S-T$_0$) qubit \cite{thalineau_arxiv_2014}.  
The two levels of the qubit are the two anti-parallel spin states of a double quantum dot with one electron in each dot \cite{petta_science_2005}. 
For such a spin qubit, the energy separation J is directly related to the exchange interaction. 
J is highly dependent on both, the energy detuning $\epsilon$ between the two dots and the tunnel-coupling and can be widely tuned on fast timescale in lateral double quantum dots in GaAs heterostructures \cite{hanson_rmp_2007}. In more recent experiments it has been shown that the the tunnel-coupling can be tuned over an extremely large range, from basically zero to several GHz \cite{bertrand_prl_2015,kuemmeth_prl_2016,sandia_prl_2016}.
The S-T$_0$ qubit can hence be used as an ultra-sensitive detector to probe the local electrostatic environment \cite{dial_prl_2013}. 
These properties can be exploited to imprint the passage of the flying electron on the qubit population where it can be stored for several microseconds. 

\begin{figure}[h]
\includegraphics[width=3.4in]{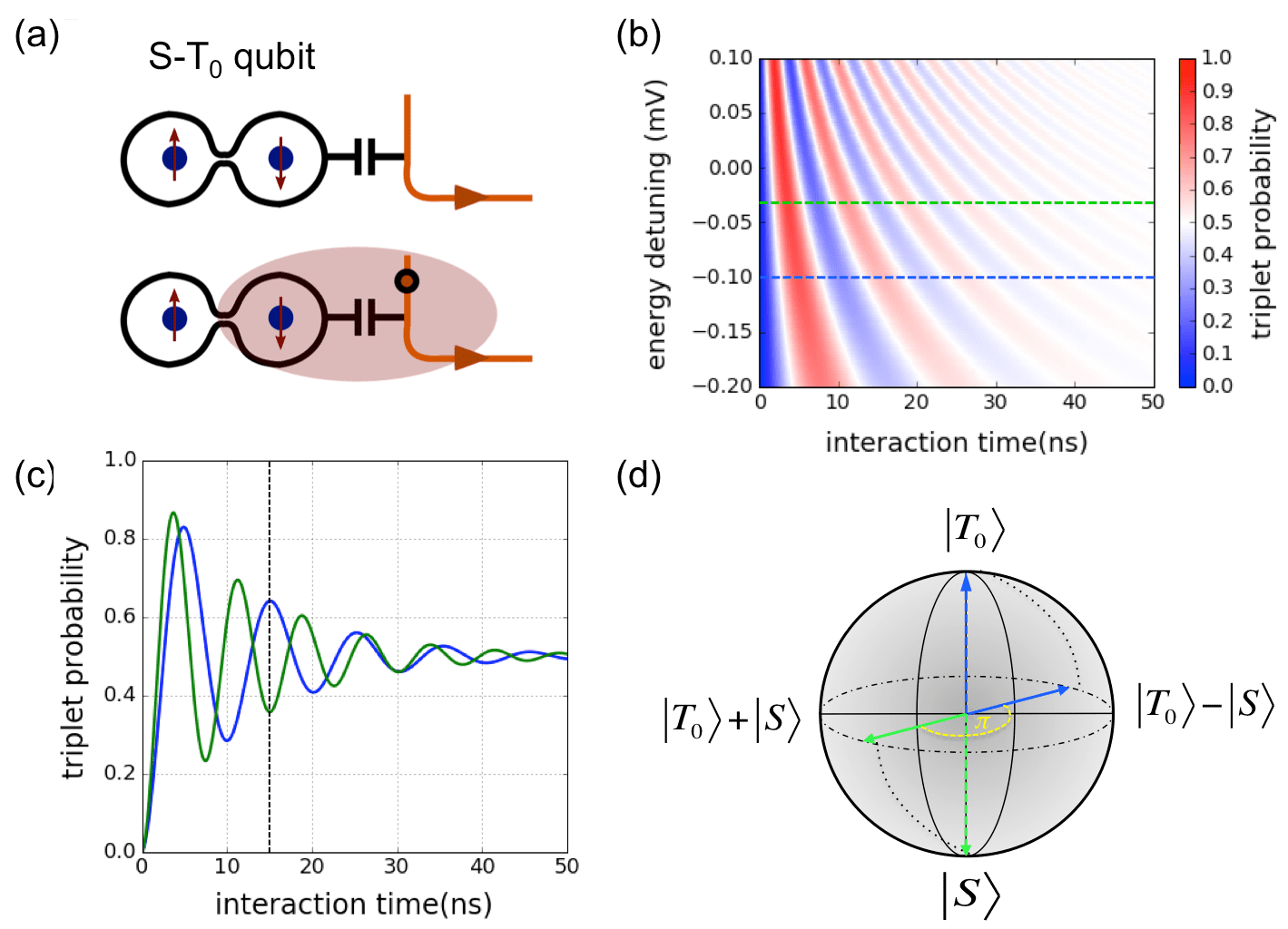}
\caption{\textbf{Principle of the S-T$_0$ qubit charge detector}. (a) An electron is passing next to a double quantum dot. Due to the capacitively coupling to the S-T$_0$ qubit the electron will locally change the electrostatic field experienced by the S-T$_0$ qubit. (b) \cb{Simulated coherent} Larmor oscillations of the S-T$_0$ qubit as a function of energy detuning $\epsilon$. The triplet probability is plotted in colour scale.  (c) \cb{Simulated coherent oscillations} for two different values (-0.1 mV and -0.035 mV) of the energy detuning $\epsilon$ indicated by the green and blue dotted lines where the induced phase shift is $\pi$ after an interaction time of 15 nanoseconds. (d) Bloch sphere for the S-T$_0$ qubit. Projection of the qubits states (solid arrows) into the read-out basis $T_0$ and $S$ (dotted arrows) for an induced phase shift of $\pi$.}
\label{fig_S-T}
\end{figure}

The principle of this flying electron detector is depicted in figure \ref{fig_S-T}.
An electron, which is guided via electrostatic gates or the edge channels in the QH regime is passing by the S-T$_0$ charge detector schematically shown as a double dot. 
Due to the capacitive coupling between the electron-wave packet and the detector, the electrostatic environment of the quantum dot will be slightly modified. This leads to a change of the qubit energy splitting during the interaction time. 
As the Rabi oscillations (see figure  \ref{fig_S-T}(b)) accelerates for more positive energy detuning, the passage of an electron close by to the detector will induce a phase shift in the rotating frame of the S-T$_0$. The accumulated phase shift corresponds to a population change of the S-T$_0$ qubit. If the accumulated phase is $\pi$ this would correspond to a complete flip of the S-T$_0$ population.  
Such a change in population can be stored for several tens of microseconds in the spin degrees of freedom of the qubit and then be measured single-shot by spin-to-charge conversion. 
Ideally, if the coupling is large enough to ensure a phase shift, a one-to-one correspondence between the qubit state and the presence of the flying electron is expected and therefore a single-shot detection of the flying electron can be performed. 
The challenge here is to reach a large enough phase shift to have the sensitivity to detect a single flying electron in a sub-nanosecond timescale.
First attempts have been done in this direction \cite{thalineau_arxiv_2014} \cb{and report a $\pi$ phase shift when a packet of 80 flying electrons is passing by the detector.} Improved sample design should allow to reach in the near future the detection of a single electron on the fly.
%-------------------------------------------------------------
%-------------------------------------------------------------
%-------------------------------- Fig. ----------------------- 
%-------------------------------------------------------------
%-------------------------------------------------------------

%-------------------------------------------------------------
%-------------------------------------------------------------
%-------------------------------- Fig. ----------------------- %-------------------------------------------------------------
%-------------------------------------------------------------
%%%%%%%%%%%%%%%%%%%%%%%%%%
%%%%%%%%%%%%%%%%%%%%%%%%%%
\section{Quantum optics like experiments with single electrons}
\label{Quantum optics like experiments}
%%%%%%%%%%%%%%%%%%%%%%%%%%
%%%%%%%%%%%%%%%%%%%%%%%%%%

Photons are very interesting flying particles to study quantum effects such as entanglement, non locality or quantum teleportation \cite{zeilinger_nature_1997,zeilinger_swap_prl_1998,zeilinger_nonloc_nature_2000}.
One can produce single photons on-demand, \cb{single-photon} detection can be realised after propagation and most importantly its coherence can be preserved over hundred kilometres \cite{zeilinger_nature_2012,zbinden_nphot_2015}.
Important tools with which to infer complex photon correlations inaccessible from ensemble measurements are single photon sources and single photon detectors. 
They are also the elementary building blocks for the manipulation of information coded into a quantum state, a qubit. 
When combined with beam splitters, polarisers etc., photonic qubits can be manipulated to process quantum information. 
A well-known example is quantum cryptography \cite{bennet_j-crypt_1992,gisin_rmp_2002}, a secure way to transmit information. 
On the other hand, \cb{it is very difficult to make two photons interact} and this represents a crucial technological hurdle for producing more complex quantum operations with photons.
Several strategies towards this goal have been proposed based on optical cavity QED concepts, however it turned out to be very challenging experimentally \cite{forchel_nature_2004,deppe_nature_2004}. 
An interesting concept has been brought forward to realise efficient quantum computing with linear optics. 
In this case the combination of beam splitters, phase shifters, single photon sources and photo-detectors with feed \st{forward} control allows to implement efficient quantum computation \cite{knill_nature_2001}\st{, where two qubit operations can be performed probabilistically by measurement of photons \cite{obrien_nature_03,obrien_science_2008,obrien_science_2015}}.
% but relies on post selection 
In analogy with photons, similar experiments should be possible with single flying electrons in a solid-state device. 
The advantage of performing quantum optics experiments with flying electrons is the existing Coulomb coupling between the electrons. 
Photons are basically non-interacting quantum particles and they therefore have a longer coherence time than electrons. 
However, due to the absence of interactions it is more difficult to construct a two-qubit gate, which operates at the single photon level. 
This represents some fundamental limitation to the development of quantum computation with photons. 
\st{In contrast Coulomb interaction allows to envision deterministic two qubit operations for electrons at the single-electron level \cite{ionicioiu_ijmp_2001,beenakker_prl_2004}.}
\cb{Naturally, strong interactions come with a short coherence time and may set limitations to this system for quantum information processing. We have seen in section \ref{Flying qubit circuits} that quantum operations can be done on a very fast time scale for ballistic electrons, allowing in principle to perform about one hundred quantum operations on the fly. Yet, this research field is only at its beginning and much technological development is still needed in order to bring it to the level of its photonic counter part. 
We have seen, however, that single electron sources have an extremely high efficiency, well above what it is possible at present with single photon sources. 
The best on-demand single photon sources are based on quantum dots integrated into optically resonant micro-cavities \cite{claudon_nphot_2010,senellart_nphot_2016,ding_prl_2016}. They offer a high degree of indistinguishability and a collector efficiency of approximately 65\,$\%$.
%,but the real(pratical) efficiency is only of the order of 15 $\%$. 
Integration and synchronization of multiple single photon sources into multiple photonic waveguides\cite{lohdahl_prl_2014} by keeping at the same time a high emission efficiency, a high degree of indistinguishability as well as a high brightness is a real challenge. The same is true for the integration of single photon detectors into photonic wave guide structures\cite{photon_detector_2015}.
On the other hand, single electron detection is much more challenging compared to photons when dealing with single electrons propagating at the Fermi sea. In addition, an important question is how good a quantum state one can prepare and how well it can be preserved upon propagation. 
This is subject to ongoing experimental as well as theoretical research. Nevertheless the possibility to exploit single electron circuitry for quantum information processing merits to be explored to see how far this technology can be pushed. Besides that, we will also see that a lot of other interesting physics arises due to the interaction of the single electrons and the Fermi sea \cite{blanter_physrep_2000}.}
In the following we will present the present state of the art concerning such \textit{electron quantum optics} experiments. 
%The advent of on-demand single electron sources and the ability to transport single electrons in a controlled way has opened the route towards the realisation of quantum optics experiments with flying electrons in solid state on-chip devices. 
%Naturally the quantum state of an electron is much more fragile than I It looks therefore possible to implement several tens. But of course this has to be demonstrated first.

%%%%%%%%%%%%%%%%%%%%%%%%%%
\subsection{HOM interference in quantum Hall edge channels}
%%%%%%%%%%%%%%%%%%%%%%%%%%
Since the seminal work of Hong, Ou and Mandel (HOM) \cite{hom_prl_1987}, HOM interferometry is commonly used in photonic quantum optics.
HOM interferometry is a two-particle interference (or second-order coherence) which tests the indistinguishably and the quantum statistics of single particles. It consists in sending repeatedly two particles at the two separate inputs of a $1/2$ beam splitter and looking at the arrival coincidence in the two separate outputs. A time delay $\tau$ is introduced between the incoming particles. For $\tau=0$, the particles are fully indistinguishable and fully mix in the beam splitter. 
%-------------------------------------------------------------
%-------------------------------------------------------------
%-------------------------------- Fig. ----------------------- 
%-------------------------------------------------------------
%-------------------------------------------------------------
\begin{figure}[h]
\includegraphics[width=8cm]{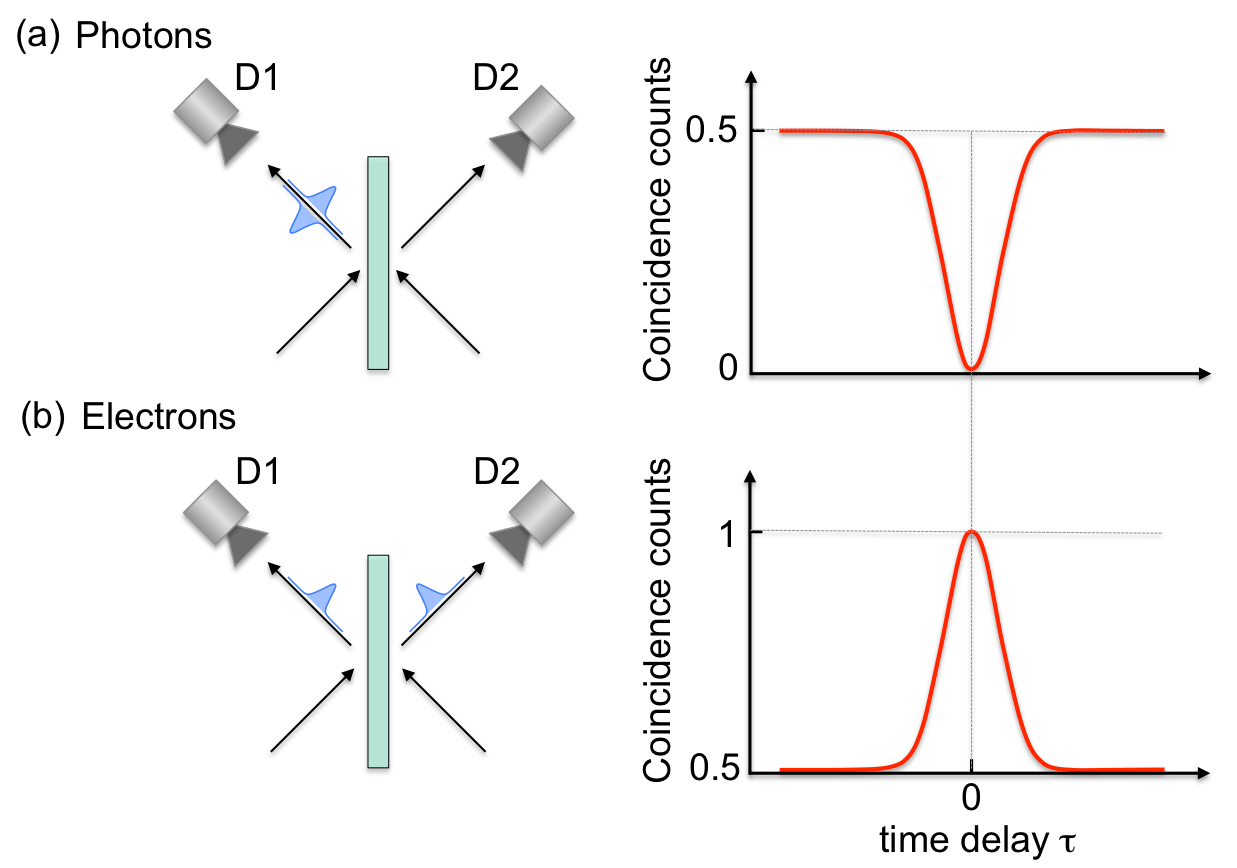}
\caption{\textbf{Schematic of Hong--Ou--Mandel interference}. Left: two photons (a) or two electrons (b) collide on a beam splitter and are collected at the detectors D1 and D2. Right: Normalized coincidence counts of detectors D1 and D2. 
The normalized coincidence counts is 1 when for each collision event exactly one particle is detected in D1 and one particle in D2. The coincidence count rate is zero when for each event the two particles are detected together in either D1 or D2. 
(a) At zero time delay, indistinguishable bosons (photons) always exit in the same output (\textit{bunching}) and results in a suppression of the coincidence counts at zero time delay. 
(b) The opposite behavior is expected for indistinguishable fermions (electrons). At zero delay the two electrons exit into the opposite outputs (\textit{anti-bunching}) and the coincidence count rate doubles. }
\label{fig_HOM}
\end{figure}
%-------------------------------------------------------------
%-------------------------------------------------------------
%-------------------------------- Fig. ----------------------- 
%-------------------------------------------------------------
%-------------------------------------------------------------
According to the fact that photons are bosons, the positive quantum statistical interference makes them to appear simultaneously in one of the two detectors. This is the so-called \textit{photon bunching}. As a consequence the coincidence counts between the two output ports are fully suppressed. By varying the time delay between the emitted photons as shown in figure \ref{fig_HOM}, the overlap of the wave function between the two arriving single particle states can be measured. 
The overlap is maximal when the particles arrive simultaneously at the beam splitter while it tends to zero when the time delay becomes larger than the wave packet. 
Looking at the particle number fluctuations, the noise in each output corresponds to binomial partitioning of two particles and is doubled with respect to the case where only a single particle were sent to the beam splitter. 
This is the limit which is recovered if the time delay $\tau$ is much longer than the extension of the photon wave packet $\psi$ . In the intermediate range of $\tau$ the noise is $\propto (1+|\langle\psi(0)|\psi(\tau)\rangle|^2)$. For particles obeying fermionic statistics, a similar effect occurs, which however leads to the opposite behaviour \cite{blanter_physrep_2000}.
Destructive quantum statistical interference makes them never appearing in the same output for $\tau=0$. This is the Pauli exclusion as opposed to photon bunching. 
Thus, if we send periodically single electrons in an electron beam splitter one expects a low frequency current noise $\propto (1-|\langle\psi(0)|\psi(\tau)\rangle|^2)$. 

\begin{figure}[h]
\includegraphics[width=8 cm]{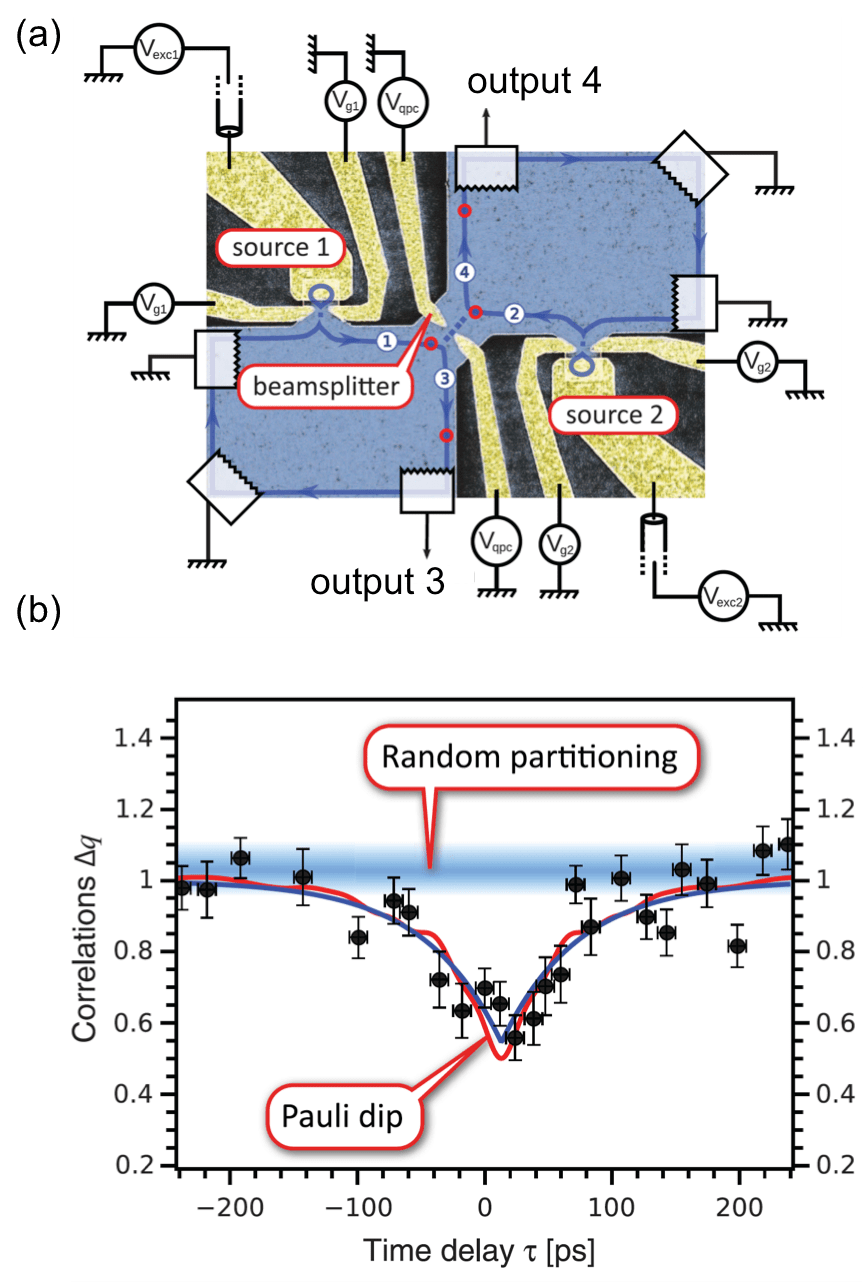}
\caption{\textbf{Hong--Ou--Mandel interference in the quantum Hall regime}. (a) False colour SEM image of the sample.  
Two Mesoscopic Capacitor sources emit a single electron which are guided towards a beam splitter (QPC).  
The transparency of the beam splitter partitioning the inner edge channel (blue line) is tuned to a transmission T = 1/2. The average ac current generated by sources 1 and 2 is measured on output 3, and the low-frequency output noise $S_{44}$ is measured on output 4. (b) Measurement of the excess noise $\Delta q = S_{44}/e^2f$ as a function of the delay $\tau$ and normalised by the value of the plateau observed for long delays. The blue line corresponds to the partition noise of both sources. The experiments are done at filling factor $\nu = 3$ (adapted from ref. \cite{feve_science_2013}). }
\label{fig_feve}
\end{figure}

Electronic HOM like experiments have been done using DC voltage sources\cite{henny_science_1999,yamamoto_science_1999,tarucha_nature_1998}
but they were incomplete as there was no way to provide time control. 
In contrast, the advent of on-demand single electron sources allow performing full HOM interferences. 
The first HOM experiment at the single-electron level has been done using the mesoscopic capacitor single electron source \cite{feve_science_2013} operated in the quantum Hall regime.
%Another way of realising two-particle interference at the single-electron level is to combine two ac single particle sources when operated in the quantum Hall regime. 
Two electrons are injected into the quantum Hall edge channel from each single electron source as shown in figure \ref{fig_feve}. 
It is possible to guide the electrons along the edges towards a beam splitter where they interact and then detect them at the two output ports. 
This set-up is the electronic equivalent of the photonic Hong--Ou--Mandel (HOM) \cite{hom_prl_1987} experiment, where two photons are colliding on a beam splitter. 

%-------------------------------------------------------------
%-------------------------------------------------------------
%-------------------------------- Fig. ----------------------- 
%-------------------------------------------------------------
%-------------------------------------------------------------
%-------------------------------------------------------------
%-------------------------------------------------------------
%-------------------------------- Fig. ----------------------- 
%-------------------------------------------------------------
%-------------------------------------------------------------

Electrons arriving at the same time at the beam splitter will exit in different output ports (anti-bunching). However, single-shot measurements of the arriving electrons, as for the case of photons, is at present not achievable. 
For this reason, the anti-bunching is probed by the measurement of low-frequency fluctuations of the electrical current in the two output leads rather than by coincidence counts. 
Two identical quantum dots are placed at a distance of approximately 3 $\mu$m away from a quantum point contact tuned to a transmission $T=1/2$ and which acts as a beam splitter. 
For the present experiment, the average emission time for both single electron sources were fixed to 58 $\pm$ 7 ps and the low frequency partition noise is measured by repeating the measurement at a frequency f $\approx$ 2 GHz. 
As mentioned in section \ref{Single Electron Sources}, for this single electron source, at each cycle one electron is emitted followed by a hole.  The low frequency current noise \cite{blanter_physrep_2000} at one of the outputs is then given by \cite{martin_prb_2012} $ S_{33} = S_{44} = e^2 f \times [1-|\langle\psi_1|\psi_2\rangle|^2] $.
By changing the time delay $\tau$ of the electron emission of the two sources, one can then probe the indistinguishability of the electrons coming from the two single electron sources. 
This is shown in figure \ref{fig_feve}b, where the excess partition noise $\Delta q$ is plotted as a function of the time delay $\tau$. 
At large delays, one essentially measures the random classical partition noise as indicated by the blue shaded line.
For short time delay one observes a dip in $\Delta q$, which corresponds to the fermionic equivalent of the HOM dip, and, which clearly demonstrates two-particle interference effects.
The dip, however, does not go to zero for zero delay as for the case of indistinguishable photons but stays at a finite value. The particles arriving at zero time delay at the beam splitter have hence lost a certain degree of indistinguishability.
While the two-particle interference effect was clearly observed, the HOM dip at $\tau=0$ was not fully developed as decoherence occurred and prevented full indistinguishably. 
% question: why don't they observe a dip for time delays when electrons and holes  from one cycle overlap (probably around $\delta \tau \approx 250 ps$. 
The origin of the finite excess noise at zero delay originates from Coulomb interaction effects\cite{martin_prl_2014,martin_jstat_2016}  between the different edge channels as the experiment has been realised in the quantum Hall regime at filling factor $\nu = 3$. 
%maybe a word why this experiment is not performed at $\nu = 1$.
Even though a single electron is injected into the edge channel \cb{closest to the edge} of the sample, during propagation inter-channel coupling leads to a loss of coherence towards the other channels and results in a reduction of the HOM dip \cite{martin_prl_2014,martin_jstat_2016}. 
This has been investigated in more detail recently \cite{feve_ncom_2015} for the case of filling factor $\nu = 2$ .
When injecting an electron into the outer edge channel, due to Coulomb interaction the electron wave packet decomposes into new propagation modes that couple both channels: a slow neutral mode where the charge is anti-symmetrically distributed over the two channels and a fast charge mode which has a symmetric charge distribution \cite{sukhorukov_prb_2008}. 
This leads to additional dips in the HOM curve at time delays which correspond to the difference of the propagation velocities of the two different modes \cite{feve_annalen_2014}. 
%Intra-channel Coulomb interaction will also lead to such a fractionalisation process. 
This has also been addressed recently by time resolved measurements \cite{fujisawa_nnano_2014}. We will come back to this point in the outlook section.

%%%%%%%%%%%%%%%%%%%%%%%%%%
\subsection{HOM interferometry with levitons}
%%%%%%%%%%%%%%%%%%%%%%%%%%

Another HOM experiment, which we describe in this section has been performed with levitons \cite{glattli_nature_2013} propagating in the single channel of a QPC at zero magnetic field, a situation less sensitive to interaction induced decoherence. In addition, levitons live close to the Fermi energy, which increases their coherence time. 
%-------------------------------------------------------------
%-------------------------------------------------------------
%-------------------------------- Fig. ----------------------- 
%-------------------------------------------------------------
%-------------------------------------------------------------
\begin{figure}[!htb]
\centering
\includegraphics[width=8.0cm]{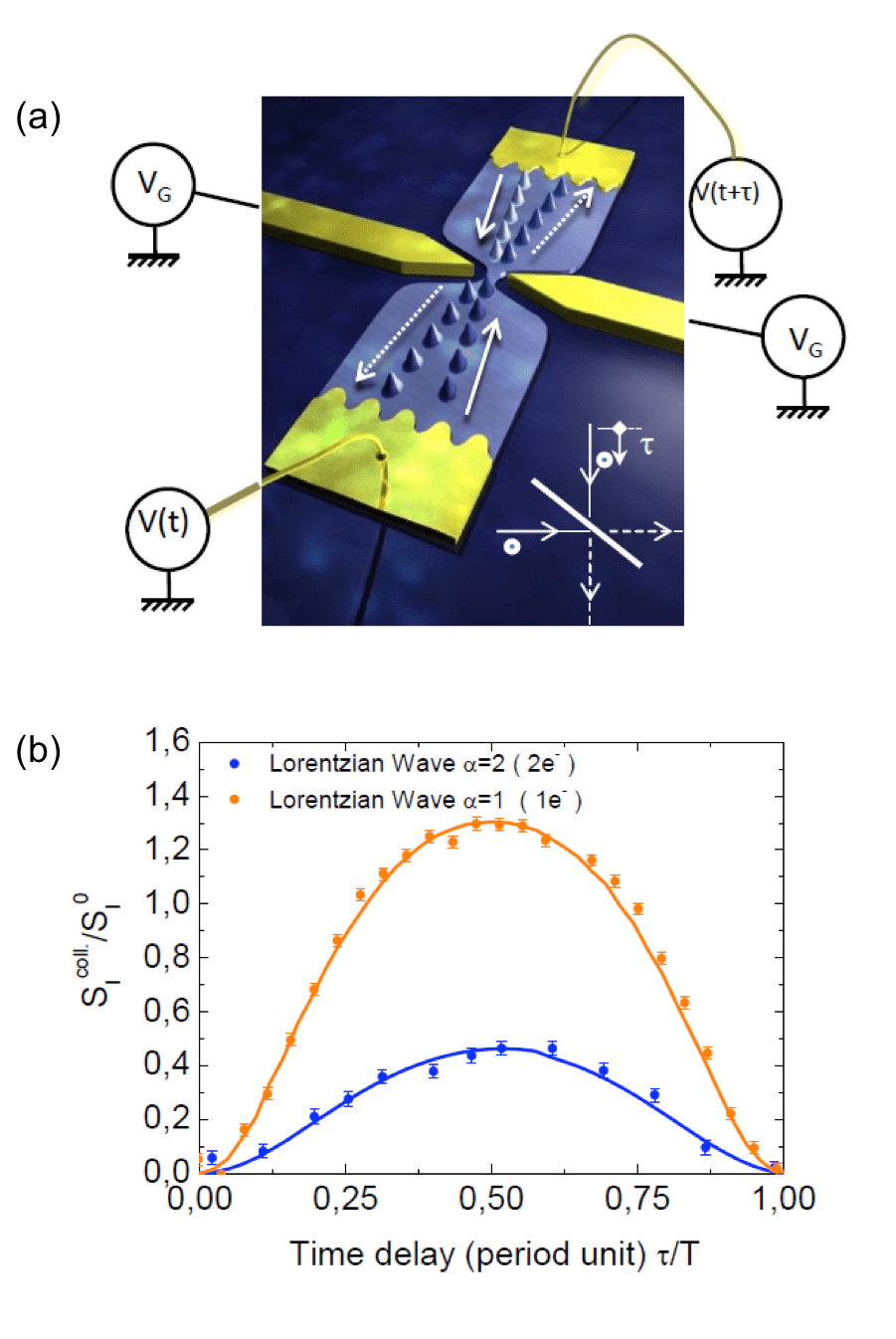}
\caption{{\bf{ Hong--Ou--Mandel interference using single electron voltage pulses.}} (a) Schematic principle of HOM measurements. (b) Two-particle partition shot noise versus time delay $\tau$ resulting from the interference of identical periodic trains of levitons in the QPC beam splitter. For $\tau=0$ perfectly indistinguishable levitons and absence of decoherence give zero noise. (figure adapted from Ref.\cite{glattli_nature_2013}).}
\label{HOM_png}
\end{figure}
%-------------------------------------------------------------
%-------------------------------------------------------------
%-------------------------------- Fig. ----------------------- 
%-------------------------------------------------------------
%-------------------------------------------------------------

Figure \ref{HOM_png} shows the schematic principle where two periodic Lorentzian pulses  are injected with a controlled time-delay $\tau$ on opposite contacts of a QPC beam splitter. The two trains of levitons interfere and mix in the electron beam splitter and the cross-correlated noise is measured. Here single and doubly charged levitons are considered. At zero temperature one expects that the noise gives direct information on the overlap of the leviton wave functions. It is given by:
\begin{equation}
\label{ShotHOM}
S_{I}=2e^{2}\nu D(1-D)2(1-|\langle\psi(x)|\psi(x-v_{F}\tau)\rangle|^2)
\end{equation}
where $\psi$ is the leviton wave function. If doubly charged levitons are sent, one expects:
\begin{equation}
\begin{split}
\label{ShotHOM2}
S_{I}=2e^{2}\nu D(1-D)2(2-|\langle\psi_{1}(x)|\psi_{1}(x-v_{F}\tau)\rangle|^2-\\
|\langle\psi_{2}(x)|\psi_{2}(x-v_{F}\tau)\rangle|^2)
\end{split}
\end{equation}
where $\psi_{1,2}$ are the first two orthogonal levitonic wave functions \cite{glattli_physicaE_2016}
of a Slater determinant describing the two incoming electrons. Finite temperature only slightly reduces the amplitude of noise variations with $\tau$, not the shape. Remarkably, the voltage pulse electron injection technique allows the generation of an arbitrary number of electrons and to perform a N-electron HOM correlation. In figure \ref{HOM_png} experimental HOM measurements are shown (points) and compared with theory (solid curves) with no adjustable parameters for single and doubly charged interfering levitons.
%-------------------------------------------------------------
%-------------------------------------------------------------
%-------------------------------- Fig. ----------------------- 
%-------------------------------------------------------------
%-------------------------------------------------------------
\begin{figure}[!htb]
\centering
\includegraphics[width=8cm]{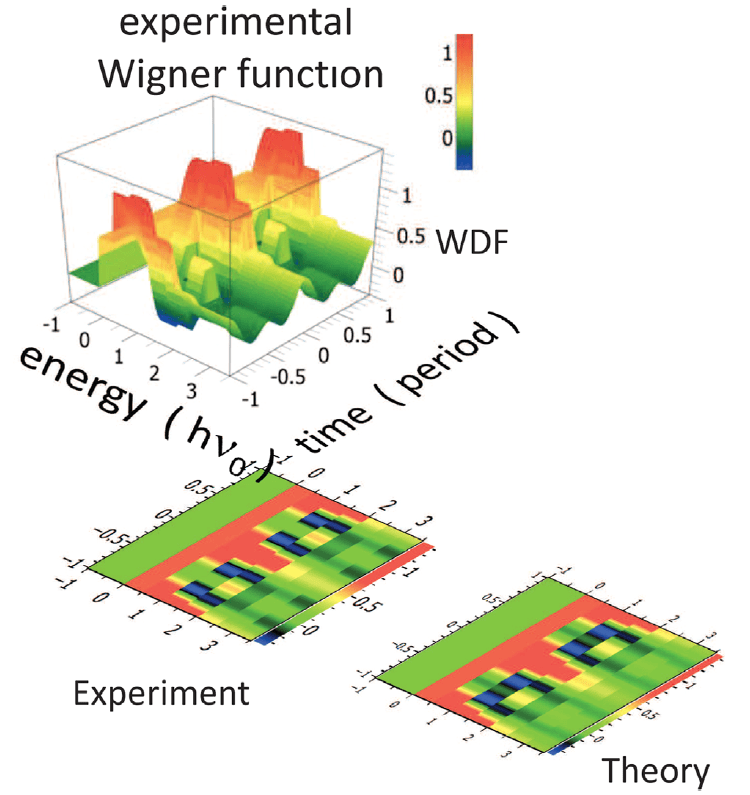}
\caption{{\bf{Experimental Wigner function of a leviton and comparison to the theory}}. The time periodicity of the Wigner function arises from the periodic injection of levitons. As expected for levitons no value is found at negative energies. Both experimental and theoretical Wigner functions have been truncated to the first two harmonics. (figure adapted from ref. \cite{glattli_nature_2014}).}
\label{Wigner_png}
\end{figure}
%-------------------------------------------------------------
%-------------------------------------------------------------
%-------------------------------- Fig. ----------------------- 
%-------------------------------------------------------------
%-------------------------------------------------------------

HOM experiments have also been performed with a tunnel junction driven by a harmonic time-dependent voltage\cite{reulet_prb_2016}.  
Beyond HOM experiments, electron quantum optics with single electron sources allows to perform all quantum experimental standards of quantum optics. Closely related to HOM interference, is the Quantum State Tomography (QST). The goal of QST is to give a complete view of the electron wave function. It consists in measuring the energy density matrix $\varrho(\varepsilon',\varepsilon)=<\psi^{\dag}(\varepsilon') \psi(\varepsilon)>$ from which the Fermi sea contribution has been subtracted. From this, one reconstructs the Wigner function which, in the representation of the conjugate energy-time variables $(\varepsilon,t)$, is given by 
\begin{equation}
\label{WignerEq}
W(\overline{t},\varepsilon)=\int_{-\infty}^{\infty}d\delta\varrho(\varepsilon+\delta/2,\varepsilon-\delta/2)e^{(-i\delta \overline{t}/\hbar)}.
\end{equation}
For periodic injection, only the non-diagonal elements of $\varrho$ differing by a multiple $k$ of the fundamental frequency $\varepsilon'-\varepsilon=k h f$ are non zero. This implies that the Wigner function is a periodic function of time and this conveniently restricts the number of measurements to be done. A quantum state tomography procedure to measure $\varrho(\varepsilon',\varepsilon)$ has been theoretically proposed by Grenier and collaborators \cite{degiovanni_njp_2011,degiovanni_prb_2013}.
This has been experimentally realised in Ref. \cite{glattli_nature_2014}. The measurements were only able to sample the first two harmonics of the Wigner function, although the third might have been accessible as well. The theoretical Wigner function restricted to two harmonics is compared with the experimental one in figure \ref{Wigner_png}. 
This example shows the degree of maturity that has reached electron quantum optics today.

%%%%%%%%%%%%%%%%%%%%%%%%%%
\subsection{Electron partitioning experiments (non-adiabatic quantised charge pump)}
%%%%%%%%%%%%%%%%%%%%%%%%%%

\begin{figure}[h]
\includegraphics[width=8 cm]{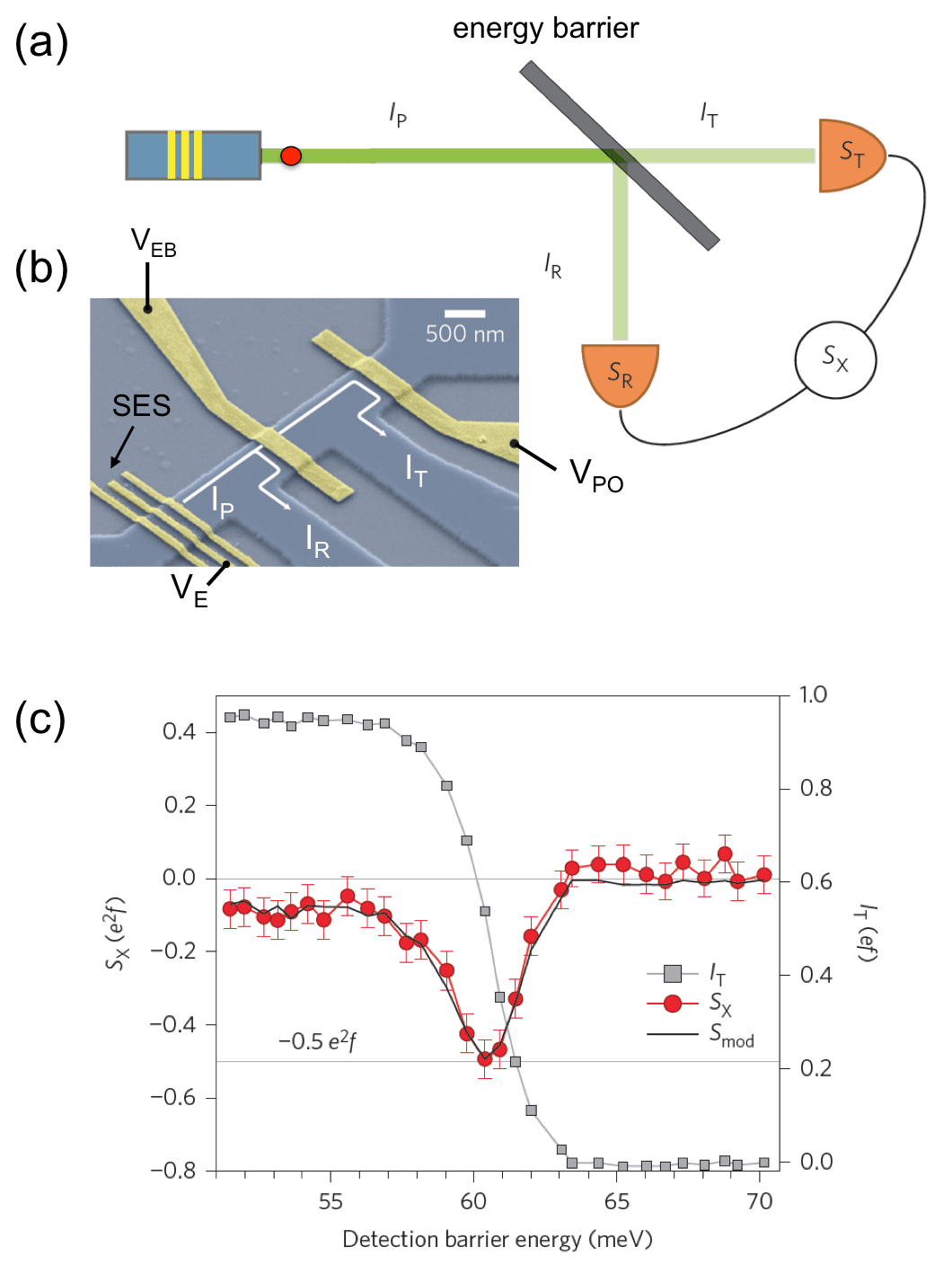}
\caption{\textbf{Partitioning experiment using \textit{hot} electrons}. 
a) Schematics: an electron is emitted from the non-adiabatic single electrons source (SES) with a repetition frequency of 280 MHz and the emitted current ($I_{\rm P}=2ef$) is partitioned at the energy barrier. 
The reflected current $I_{\rm R}$ is directed towards output $S_R$ while the transmitted current is collected at output $S_{\rm T}$.  (b) Micrograph of the sample. $V_{\rm E}$ sets the emission energy of the electron source while $V_{\rm EB}$ sets the energy of the the barrier. 
The experiment is performed in the quantum Hall regime in order to guide the electrons along the edge channels. The surface gate $V_{\rm PO}$ is fully pinched off, such that all transmitted electrons are guided towards output $I_{\rm T}$. (c) Transmitted current $I_{\rm T}$ (grey symbols) and cross-correlation noise power $S_{\rm X}$  (red symbols) of the two outputs $S_{\rm T}$ and $S_{\rm R}$ as a function of  barrier height energy. (figure adapted from ref. \cite{haug_nnano_14}).}
\label{fig_haug_1}
\end{figure}

Finally, let us also mention another way of partitioning electrons that can be realised by using an energy selective barrier. 
In this case the single-parameter non-adiabatic quantised charge pump presented in section \ref{Single Electron Sources} is connected to a two-dimensional electron gas operated in the quantum Hall regime. The ejected non-equilibrium electrons will then travel along the edge of the sample. A surface gate which is operated as an energy selective barrier is placed on the trajectory of the electrons by applying an appropriate gate voltage. 
% with minimal inelastic scattering. 
When arriving at the energy barrier, electrons having a lower energy will be reflected to output $S_{\rm R}$ while electrons with higher energy will arrive at output $S_{\rm T}$ as shown in figure \ref{fig_haug_1}. 
To partition the electron one fixes the emission energy by setting the gate voltage $V_{\rm E}$ to the desired energy and scans the energy of the barrier hight. 
For single electron emission one nicely observes that the transmitted current is given by $I_{\rm T}(E) = T(E)ef$, where $T(E)$ is the transmission probability of the barrier as shown in figure \ref{fig_haug_1}(c).
Measuring the cross-correlation between  the current in the transmitted and reflected output signal one obtains the expected partitioning noise $S=-2T(1-T)e^2f$. 
As the two detectors are completely uncorrelated, the repetition frequency (280 MHz) being sufficiently low, no correlation in the electron stream is expected.  
%check on what depends the inelastic scattering -> phonons
%-------------------------------------------------------------
%-------------------------------------------------------------
%-------------------------------- Fig. ----------------------- 
%-------------------------------------------------------------
%-------------------------------------------------------------
%-------------------------------------------------------------
%-------------------------------------------------------------
%-------------------------------- Fig. ----------------------- 
%-------------------------------------------------------------
%-------------------------------------------------------------

An interesting feature arises when sending electron pairs rather than a single electron. It is possible to load two electrons into the electron pump and emit them within one pumping cycle. 
When the pump is operated with a sinusoidal drive, the electron are emitted \cite{haug_nnano_14} sequentially at the \textit{same} energy since the electron source is operated in the adiabatic limit. 
The electrons have initially an energy separation due to the charging energy, but due to the slow operation of the pump, the electron has time to compensate the charging energy before emission. 
Independent scattering will arise at the potential barrier and each of the two emitted electrons will be partitioned with the same transmission probability $T$ resulting in the same binomial partitioning distribution, $P_2=T^2$, and $P_0=(1-T)^2$. Here $P_i$ corresponds to the probability that $i$ (2-$i$) electrons are detected in the transmitted (reflected) current. 
This is very similar to the single electron emission case. One observes that the current evolves smoothly from an almost perfect transmitted current a low barrier hight towards a fully reflected current at high barrier height, with the appearance of a single dip in the partitioning noise for a transmission probability of $T=1/2$. 
This corresponds to a 50 \% probability that the electron pair is partitioned in such a way that they arrive in the opposite outputs. The other 50 \% correspond to detecting the two electrons either at output $S_{\rm R}$ or at output $S_{\rm T}$, each event having a probability of 25 \%. 

\begin{figure}
\includegraphics[width=8.5 cm]{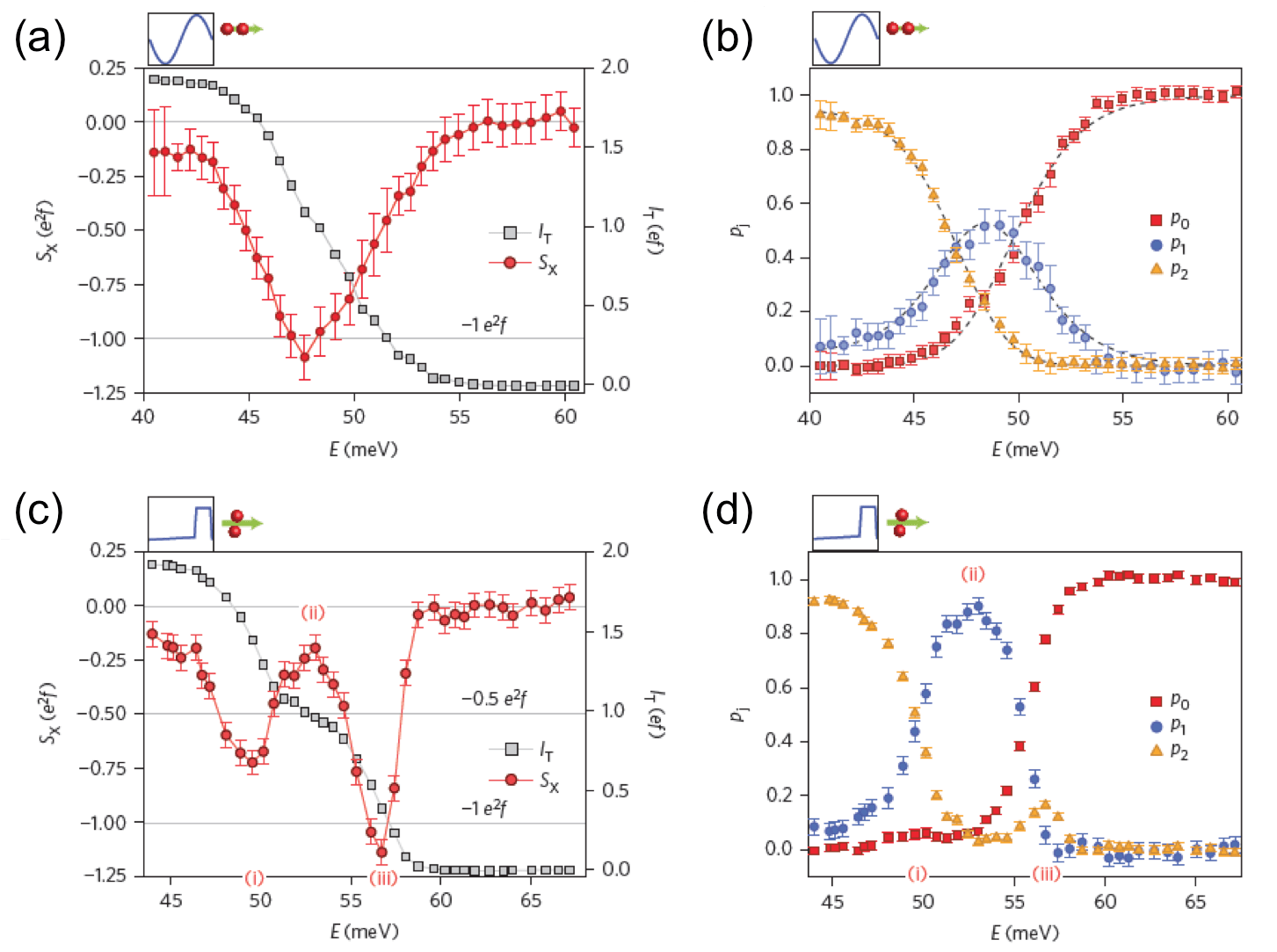}
\caption{\textbf{Partitioning experiment with electron pairs}. 
(a),(b) Two electrons are emitted from the single electrons source with a sinusoidal drive. Due to the slow variation of the sinusoidal drive, the two electrons are emitted sequentially having the same energy. (a) transmitted current $I_{\rm T}$ (grey symbols) and partitioning noise $S_{\rm X}$ (red symbols) as a function of energy barrier. (b) $P_i$ probability of detecting  $i$ electrons in the transmitted current . 
c,d) same as a,b, but for a pumping cycle with a sharp rising edge. In this case the two electrons are emitted with different energy as the charging energy can no longer be neglected. (figure adapted from ref. \cite{haug_nnano_14}).}
\label{fig_haug_2}
\end{figure}

The situation is quite different when operating the current source with a short emission pulse rather than a sine wave in order to reduce the time delay between the emission of the two electrons. 
In this case the charging energy is important and the two electrons cannot be treated as independent any more. 
They are emitted almost simultaneously, but with \textit{different} energies.
This is seen in the noise correlation measurements which show two distinct dips as a function of barrier hight indicating that the electrons are emitted at different energies.
In addition, this is accompanied by a plateau observed for $I_{\rm T} = 1ef $ at energies E $\approx$ 52-55 meV. 
As a consequence, the probability distribution $P_1$ in this energy interval is enhanced, (see figure \ref{fig_haug_2} (d)) reaching a value of 90 \%. The barrier therefore acts as an efficient energy filter. Indeed, electrons arriving with different energies are partitioned to different ports and the probability distribution $P_1$ should reach 1 if the energy selection of the barrier were perfect. 
One also observes a small feature of positive cross-correlation, a small peak in $P_2$ at an energy of $\approx$ 57meV, which means that the electrons show bunching. This feature is presently not understood. It would also be interesting to see whether coherent manipulations of such high energy electrons is possible due to their high energy.
% ask Haug why they mention in their paper (SOM) that one can get this enhacement of P_1 by statistical effects ?
%-------------------------------------------------------------
%-------------------------------------------------------------
%-------------------------------- Fig. ----------------------- 
%-------------------------------------------------------------
%-------------------------------------------------------------
%-------------------------------------------------------------
%-------------------------------------------------------------
%-------------------------------- Fig. ----------------------- 
%-------------------------------------------------------------
%-------------------------------------------------------------

%%%%%%%%%%%%%%%%%%%%%%%%%%
%\subsection{Directional Coupler using SAW driven electrons (Chris, Tristan + Shintaro)}
%%%%%%%%%%%%%%%%%%%%%%%%%%

%%%%%%%%%%%%%%%%%%%%%%%%%%
%%%%%%%%%%%%%%%%%%%%%%%%%%
\section{ Novel quantum interference experiments with short voltage pulses}
\label{Novel quantum interference experiments} 
%%%%%%%%%%%%%%%%%%%%%%%%%%
%%%%%%%%%%%%%%%%%%%%%%%%%%

With the ability to generate very short voltage pulses comes the possibility to probe the quantum dynamics
at scales faster than the propagation time through the sample. 
This is a new direction for quantum nanoelectronics
that already has interesting promises, that we illustrate below: the non-adiabatic probing of the quantum dynamics. 
These non-adiabatic regimes require very fast electronics. Indeed, typical Fermi velocities in a 2DEG are of the order of $v_F \approx 10^5 \,m/s$ while the typical system size that can remain coherent is of the order of 10$\,\mu m$, so that the typical dwell time inside such a ballistic system is 100 $ps$. 
When one uses longer pulses, the propagation inside the sample cannot be probed. 
In fact, in order to observe the novel interference effect described later in this section, the duration of the pulse must be significantly shorter than 100 $ps$ as the Coulomb interaction induces a renormalisation of the Fermi velocity into the larger plasmonic velocity \cite{chaplik_1985}.

\begin{figure}[h]
\includegraphics[width=7 cm]{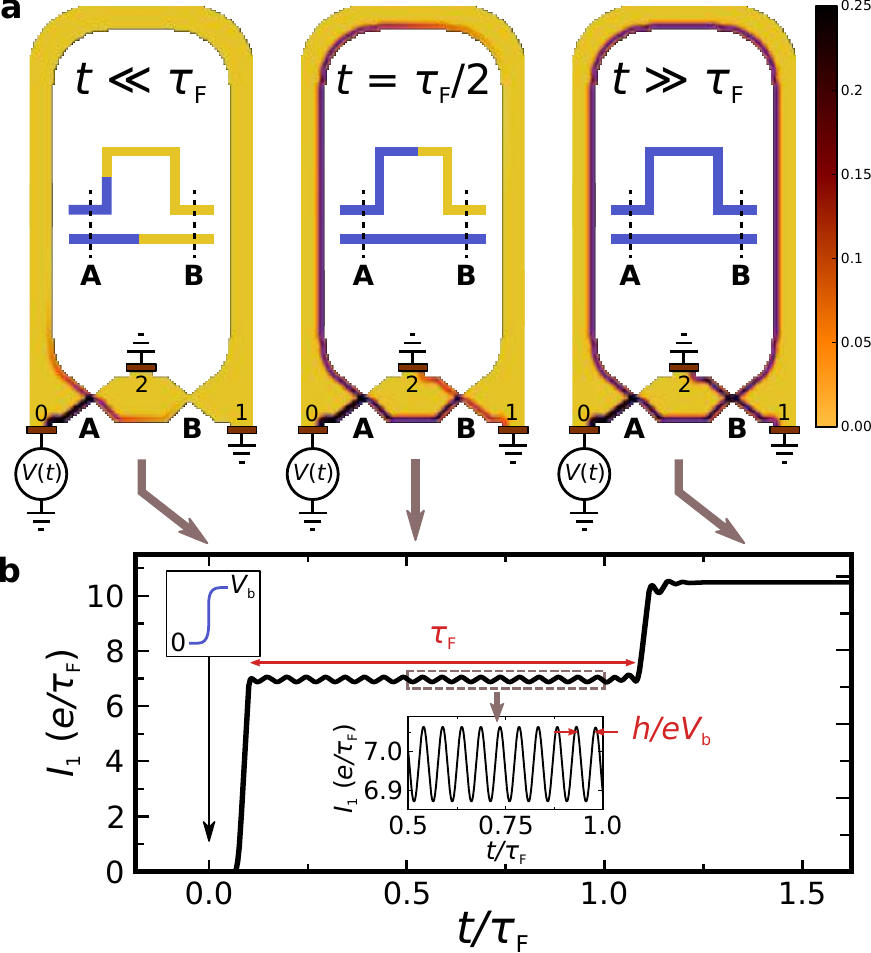}
\caption{\textbf{}
(a) Snapshots of the simulation of the propagation of an abrupt raise of voltage in an electronic Mach Zehnder interferometer in the quantum Hall regime. From left to right the snapshot has been taken: just after the voltage raise, in the transient regime where the current front has arrived through the lower arm but not the upper one, in the stationary regime. (b) corresponding current as a function of time measured in contact 1. (figure adapted from ref. \cite{gaury_ncom_2015}). 
}
\label{MZsim_snap}
\end{figure}

To illustrate the effects that can occur in the non-adiabatic regime, let us focus on numerical simulations of the Mach--Zehnder
interferometer of figure \ref{MZI.png} in the time domain. A typical snapshot of a simulation is shown in figure \ref{MZsim_snap}a: at $t=0$, one abruptly raises the voltage at contact 0 ($V(t)\approx V_b \theta (t)$, where $\theta (t)$ is the Heavyside function) and studies the propagation of the current front along the two arms of the interferometer. The three snapshots correspond to different times: in the first, the front has reached the first QPC, in the second the front has arrived to contact 1 through the short lower arm, but is still traveling along the upper arm (this is the transient regime on which we focus below), in the third, the front has arrived through both arms and one has reached a stationary state. The corresponding current observed at contact 1 is shown in figure \ref{MZsim_snap}b. In the transient regime, one observes a very striking effect: The current oscillates in time with a frequency set by the voltage bias, $I(t) \propto \cos (eV_b/\hbar)$. This is the mesoscopic analog to the AC Josephson effect in superconducting junctions \cite{gaury_ncom_2015}. An effect of similar nature can be observed if one replaces the abrupt raise of voltage by short voltage pulses. In a typical experiment, the protocol would be repeated at rather low frequency $f$ and one would measure the DC current $I_{\rm dc}$ flowing through contact 1. One finds $I_{\rm dc} = e n_1 f$ where $n_1$ is the average number of particle transmitted to contact 1 per pulse. In the "standard" adiabatic regime (pulses long with respect to the internal time scales of the device), one expects $n_1 \propto \bar n$ where $\bar n =\int dt eV(t)/h$ is the average number of particles sent in the pulse. However, for short pulses, one finds a rather different behavior \cite{gaury_ncom_2014} with $n_1 \propto \sin\bar n$, i.e. the number of electrons collected at contact 1 oscillates with the number of electrons sent. 
%-------------------------------------------------------------
%-------------------------------------------------------------
%-------------------------------- Fig. ----------------------- 
%-------------------------------------------------------------
%-------------------------------------------------------------
%-------------------------------------------------------------
%-------------------------------------------------------------
%-------------------------------- Fig. ----------------------- 
%-------------------------------------------------------------
%-------------------------------------------------------------

Let us now build a simple theory that allows one to understand these two effects in a rather simple way. Let us start by considering the propagation of a voltage pulse in a simple one-dimensional wire. The first thing to realise, is that such a voltage pulse cannot be simply associated with the classical propagation of a ballistic excitation as would be the case in vacuum. Indeed, we are considering an electronic system with a Fermi sea, so that before one sends the pulse, the system is not empty; it already sustains stationary plane waves of the form
\begin{equation}
\Psi_E(x,t) = e^{ikx-iEt}
\end{equation}
We now assume that the voltage pulse $V(t)$ leads to an abrupt drop of electric potential at $x=0$ (i.e. $V(x,t) = V(t) \theta (-x)$). We also suppose, for simplicity, that the voltage is small with respect to the Fermi energy so that one can linearise the dispersion relation $E = v_F k$. Under this condition, the effect of the increased voltage reduces to a faster oscillation of the wave function phase for $x<0$ which leads to a "phase domain wall" which propagates ballistically inside the system. The corresponding wave function reads,
\begin{equation}
\label{pulse_wf}
\Psi_E(x,t) = e^{-iE(t-x/v_F)}e^{-i\Phi(t-x/v_F)}
\end{equation}
where the phase $\Phi(t)$ is the total phase accumulated due to the time dependent voltage,
\begin{equation}
\Phi(t) = \int_0^t du \frac{eV(u)}{\hbar}
\end{equation}
Equation (\ref{pulse_wf}) has an interesting structure which was first analysed by Levitov and co workers, in particular in the context of Lorentzian pulses (levitons) and the associated (lack of) quantum noise\cite{Keeling_PRL_2006}. What is particularly appealing is the fact that the wave function phase is entirely engineered by the physicist through the shape of the voltage pulse, i.e. a quantum object (a wave function phase) is controlled by a classical, experimentally tunable quantity (a voltage). Phases cannot be observed by themselves, so that the next step is to feed this wave function to an electronic interferometer in order to probe this special feature. We focus on the two path interferometer of figure \ref{MZsim_snap}a, with an "upper" $U$ and "lower" $L$ arm, described by a transmission amplitude
\begin{equation}
d(\epsilon) = d_L + d_U e^{i \epsilon \tau}
\end{equation}
where the energy dependency is controlled by the delay of propagation $\tau$ of the upper arm with respect to the lower one.
For the Mach--Zehnder interferometer of figures \ref{MZI.png} and  \ref{MZsim_snap}a, $d_U=\sqrt{T_AT_B}$ and $d_L=\sqrt{(1-T_A)(1-T_B)}$ ($T_A$, $T_B$ transmission probability of the corresponding QPC) while for the
flying qubit of section figure \ref{fig_ABTunnelDevice}, $d_U=\cos \left(\frac{k_A-k_S}{2} L\right)$ and 
$d_L=\sin \left(\frac{k_A-k_S}{2} L\right)$. The wave function after the interferometer takes the form (up to a global plane wave phase).
\begin{equation}
\label{pulse_wf2}
\Psi_E(x,t) =  d_L e^{-i\Phi(t-x/v_F)} + d_U e^{-I E \tau/\hbar} e^{-i\Phi(t-x/v_F - \tau)}
\end{equation}
The interesting aspect of Eq.(\ref{pulse_wf2}) is the fact that the two phases $\Phi(t)$ are delayed by the time $\tau$ which
enables the possibility to control the interference pattern hence to access the wave function experimentally. Let us discuss a
  few specific example. In the "adiabatic" limit where $V(t)$ varies very slowly with respect to $\tau$, $\Phi(t)$ and 
$\Phi(t-\tau)$ are essentially the same and one is back to the usual DC theory. In the opposite limit, one sends a pulse
which is very short with respect to $\tau$ so that $\Phi(t)$ abruptly goes from $0$ to $2\pi\bar n$ where $\bar n$ is the
average number of electrons sent by the pulse, according to the Landauer formula $e\bar n \equiv \int dt I(t)=(e^2/h)\int dt V(t)$. Let us focus on the wave function somewhere after the interferometer. At both small and large time, one recovers the usual interference pattern,
\begin{equation}
\Psi_E(t)\propto  d_L + d_U e^{-i E \tau/\hbar}
\end{equation}
but in the transient regime where the pulse has arrived through the lower arm but not yet through the upper one, 
one gets a \textit{dynamically modified} interference pattern:
\begin{equation}
\Psi_E(t)\propto  d_L + d_U e^{-i E \tau/\hbar}e^{-i2\pi\bar n}
\end{equation}
which \textit{oscillates} with the number of particles sent (i.e. with the amplitude of the pulse). This effect is even more drastic if one considers an abrupt raise of potential from $0$ to $V_b$ instead of a pulse: $V(t) = V_b \theta(t)$. One gets,
\begin{equation}
\Psi_E(t)\propto  d_L + d_U e^{-i E \tau/\hbar}e^{-ieV_bt/\hbar}
\end{equation}
The interference pattern in the transient regime now oscillates in time with frequency $eV_b/h$. This is the normal effect analogous to the ac Josephson effect in superconductors advertised above. To conclude this section, the dynamical control of coherent conductors opens the possibility for a direct manipulation of quantum mechanical objects - the phase of the electronic wavefunction  -  which gives rise to physical phenomena that have no DC equivalents. 

%%%%%%%%%%%%%%%%%%%%%%%%%%
%%%%%%%%%%%%%%%%%%%%%%%%%%
\section{Outlook and future developments}
\label{Outlook} 
%%%%%%%%%%%%%%%%%%%%%%%%%%
%%%%%%%%%%%%%%%%%%%%%%%%%%

The field of \textit{single electron electronics} is now at a point where quantum interference experiments at the single-electron level are possible \cite{feve_science_2013,glattli_nature_2013}. As we have seen above, on-chip experiments with individual electrons can now be performed, similar to those realised with single photons on optical tables. 
When combined with quantum interferometers, these building blocks will form the basis for the next generation of quantum electronic circuits and a multitude of novel experiments not accessible with standard DC voltage sources can be envisioned.
For instance when combining single electron voltage pulses with MZ type interferometers, strikingly different behaviours are predicted compared to the static case. 
Depending on the number of electron charges contained in a voltage pulse the visibility of the current oscillations can be higher than for the static case \cite{flindt_prb_2014,bertoni_jpcm_2015}. 

An interesting feature of the physics of voltage pulses is that it allows probing quantities which are hardly accessible by other means. One example is the full counting statistics (FCS) of a mesoscopic conductor. 
When a single charge can be trapped for a relatively long time, such as in quantum dots, single charge detection allowed for measurements of cumulants up to 15th order. For single \textit{flying} electrons this is presently out of reach. 
Up to now, only cumulants up to third order have been accessible for mesoscopic conductors \cite{reulet_prl_2003, reznikov_prl_2005,pekola_prl_2007}. Using a dynamical scheme, however, it has been proposed that by simply measuring the average current, the FCS should be accessible \cite{flindt_arxiv_2016}. 
This can be realised by injecting periodic voltage pulses into a  quantum conductor and a MZ interferometer which are capacitively coupled through one arm of the interferometer. 
The propagating electron wave packet in the quantum conductor will couple to the electron wave packet propagating in one arm of the MZ interferometer and induce an additional phase shift which can be accessed via the average current.
By changing the relative time delay between the two electron wave packets, the effective coupling can be controlled and which in turn controls the counting field of the FCS. 

The controlled emission of single electron excitations constitutes also an important step forward towards the on-demand generation and detection of entangled single- and few-electron states in mesoscopic structures \cite{hofer_pss_2017}.  
The coherent partitioning at a QPC or tunnel-coupled wire leads to an entanglement between the electron wave functions leaving the two output paths \cite{flindt_njp_2016}. Entanglement could then be demonstrated via violation of Bell's inequality \cite{belzig_arxiv_2017} by measurements of the mean current and the zero frequency noise. 
%%%%%%%%%%%%%%%%%%%%%%%%%%
%%%%%%%%%   linear quantum computing:
% In addition, obeying fermionic statistics, electrons provide new opportunities for linear quantum optics entanglement hardly possible with photons obeying bosonic statistics. 

%%%%%%%%%%%%%%%%%%%%%%%%%%
%%%%%%%%%%%%    physics of non-integer  charge pulses
%%%%%%%%%%%%%%%%%%%%%%%%%%
The ability to generate non-integer charge pulses open another interesting perspective. 
As mentioned above, for the case of integer charge pulses of Lorentzian shape one can generate a pure electronic excitation \cite{Levitov_JMathPys_1996,glattli_nature_2013}.
However, by simply varying the voltage pulse amplitude one can generate any value of the injected charge. In this case the generated excitations \cite{belzig_prl_2007,belzig_prb_2008,belzig_pss_2017} cannot be considered as a \textit{pure} leviton and interesting new features appear.
For instance, for the case of \textit{half-levitons}, a Lorentzian charge pulse containing half an electron charge, remarkable zero-energy single particle states can be generated \cite{moskalets_prl_2016}. 
The energy distribution function of such a state is symmetric and sharply peaked at the Fermi energy. 
As a consequence, one can annihilate effectively a half leviton and its anti-particle  - an anti-half leviton - which can be created by reversing the sign of the voltage pulse. 
Such an annihilation effect, not possible with ordinary quasiparticles can then be realised by colliding such anti-particles on a beam splitter.
Similarly, when non-integer charge pulses are injected into a MZ interferometer unusual features compared to the static case arise \cite{flindt_prb_2014}.
%
%
%%%%%%%%%%%%%%%%  interactions +  FQHE

\begin{figure}[h]
\includegraphics[width=7 cm]{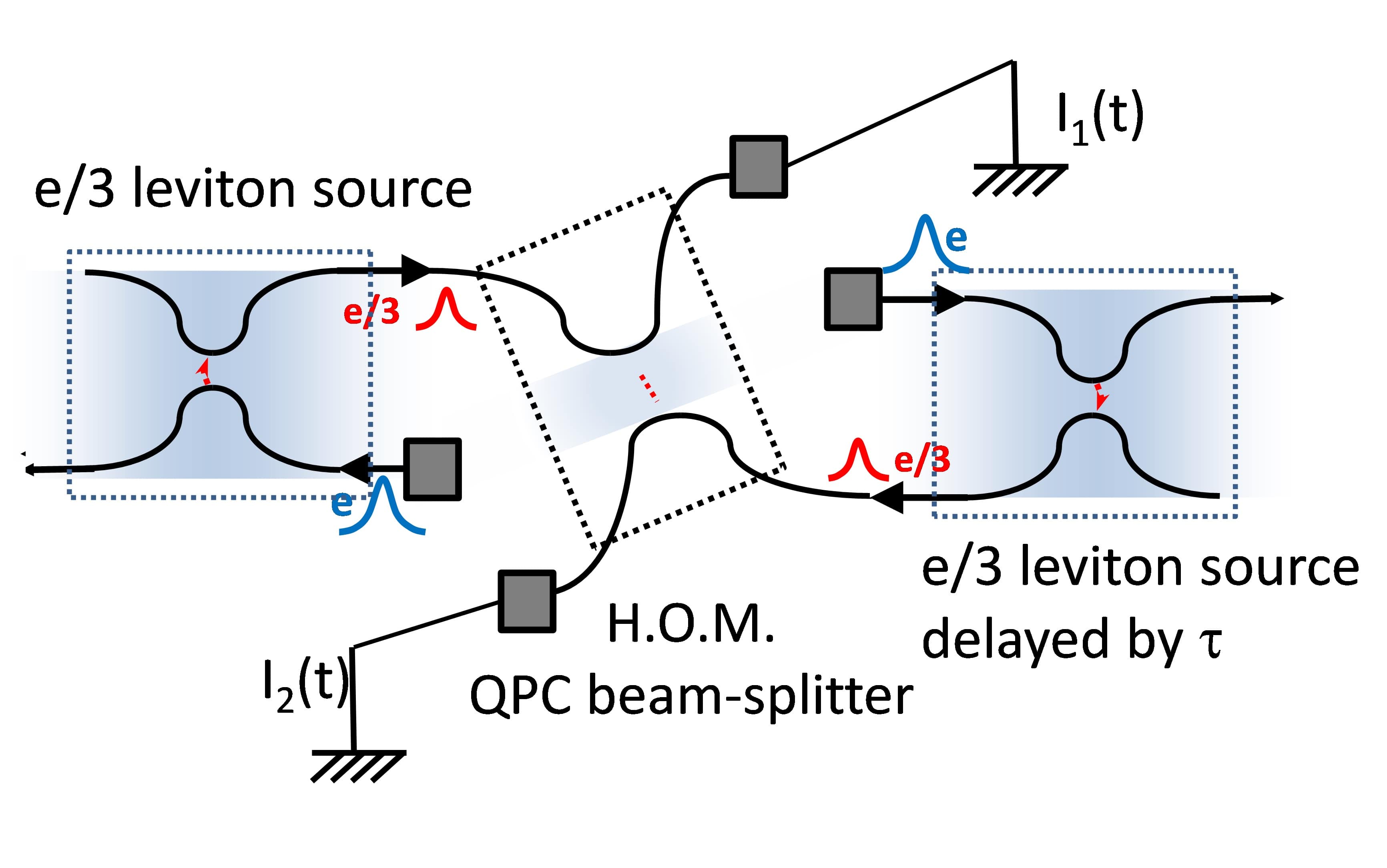}
\caption{\textbf{}
Two Poissonian sources of $e/3$-anyons are realized by weakly backscattering integer charge levitons (minimal excitation states). The anyons are sent to a QPC-beam splitter to provide HOM interference analysed by the cross-correlation noise spectrum of the current $I_{1}$ and $I_{2}$. Introducing a time-delay $\tau$ between the two sources would provide unambiguous evidence of anyonic bunching properties 
}
\label{HOManyon}
\end{figure}

Another research line is the study of interactions in one-dimensional electron systems \cite{giamarchi_book,yacoby-glazman_nature_2010}. 
% usually referred to Tomonaga-Luttinger liquids 
A typical example are the chiral edge states of the fractional quantum Hall effect (FQHE). 
The possibility to inject time-resolved wave packets, in particular minimal excitation states by means of levitons, provides a new tool to study the anyonic statistics. 
This regime has been considered theoretically recently \cite{safi_epl_2010,martin_arxiv_2016,martin_pss_2017,sassetti_arxiv_2017} and the question at present is whether these minimal excitation states can indeed probe the fractional charge excitations. 
As levitons must carry an integer charge, hence involving several Laughlin quasiparticles, they are believed to leave a signature in the partition noise. 
A source of $e/3$ fractional charges made of integer charge levitons weakly backscattered by a QPC at FQHE filling factor $\nu=1/3$ has been proposed recently \cite{Glattli_PSS_2017}, which may be used to probe anyonic statistics. 
This on-demand source of anyons would generate a Poissonian flux of $e/3$ charges keeping the time resolved properties of the integer charge levitons. 
As shown in Fig. \ref{HOManyon}, synchronizing two anyon sources and sending them to a QPC-beam-splitter should evidence their quantum statistics via HOM correlations or other two particle interferometry \cite{martin_prb_2015}.
%-------------------------------------------------------------
%-------------------------------------------------------------
%-------------------------------- Fig. ----------------------- 
%-------------------------------------------------------------
%-------------------------------------------------------------
%-------------------------------------------------------------
%-------------------------------------------------------------
%-------------------------------- Fig. ----------------------- 
%-------------------------------------------------------------
%-------------------------------------------------------------

Interactions in one-dimensional electrons systems can also be probed by time of flight measurements. 
Injecting an electron into a one-dimensional electron system, Coulomb interactions lead to spin-charge separation \cite{yacoby_science_2005, jompol_science_2009} and charge fractionalisation \cite{safi_prb_1995,safi_proceeding_1995,yacoby_nphys_2008}.
This charge fractionalisation as well as the propagation speed of the created charge excitations depend on the strength of the Coulomb interaction \cite{yacoby-glazman_nature_2010}. 
In pioneering experiments the speed of such excitations has been measured indirectly from the dispersion relation exploiting a momentum and energy conserving tunnelling process between two quantum wires\cite{yacoby_science_2005}.
Such samples are quite challenging to realise as two two-dimensional electron gas systems have to be brought into close contact to allow tunnelling in between them.
More recently this charge fractionalisation has also been investigated in the quantum Hall edge channels \cite{feve_ncom_2013,heiblum_prl_2014,fujisawa_nnano_2014,feve_ncom_2015,fujisawa_nphys_2017}. 
A conceptually simpler way to determine the propagation speed of such excitations is to measure directly the time of flight. This however requires an extremely precise control of emission and detection of the injected wave packet. With the development of faster and faster RF electronics this is now possible \cite{kataoka_prl_2016,fujisawa_nphys_2017}.
It would hence be interesting to see what happens to a leviton when injected into a quantum wire defined by electrostatic gates or into a quantum Hall edge state. Due to Coulomb interaction this initial wave packet should fractionalise \cite{degiovanni_prb_2013} and the resulting quasiparticle excitations will propagate with different speeds. 
For instance for a quantum wire defined by electrostatic gates, it should be possible to access the propagation speed of the individual modes of the quantum wire. One expects that the propagation velocity is strongly renormalised due to Coulomb interactions \cite{glazman_prl_1993,glazman_physicaB_1993}. This has been experimentally demonstrated very recently \cite{roussely_arxiv_2017}.
%This fast charge mode has never been observed in the pioneering experiments on spin-charge separation.

%%%%%%%%%%%%%%%
%%%%%%% ultrafast nanoelectronics %%%%%%%%%%%%%%%%%%%
%%%%%%%%%%%%%%%
\cb{Finally let us mention that with} the development of faster and faster radio-frequency (RF) techniques it should be possible to reach in the near future frequencies that are comparable to the internal characteristic time scales that set the quantum dynamics of the \cb{quantum nanoelectronic} devices. 
This field of \textit{ultra-fast quantum nanoelectronics} is only at its very beginning as voltage pulses of 10 ps or shorter have to be generated. 
This is highly non-trivial as the generation of such short pulses using standard RF techniques is already challenging from a technological point of view. 
At present RF technology is limited to the 100 GHz bandwidth. In addition, limits are set due to the dispersive character of the electrical lines, which connect the room temperature microwave electronics with the quantum electronic devices situated at very low temperatures. 
An alternative approach is to adopt well known techniques from THz optics and make them compatible with nanoelectronic circuits. Sub-picosecond electrical pulses can be generated on chip using photoconductive switches \cite{auston_apl_1975}: The photo switch is illuminated by a laser pulse of sub-picosecond duration. The \cb{THz} field is then confined near a lithographically defined metallic transmission line and converted on-chip into an electrical signal. The width of the generated voltage pulse is limited by the electron hole recombination time which can be as short as 1 ps or even below, depending on the material properties of the photo switches.  
Several low temperatures transport experiments using this scheme have been carried out in the past \cite{mceuen_nnano_2008,cunningham_scirep_2015}, however to reach the quantum regime of a nanoelectronic
conductor is still a big step away. 
The challenge here is to develop efficient photo-switches in order to reduce the heat generated by the femto second laser. 
First attempts have been done in this direction \cite{cunningham_rsi_2013,cunningham_nletters_2014} and in future interesting developments can be expected. 
%
%conclusion
%
%From the development of 

\cb{From coherent single electron devices we can certainly expect in the future many diverse and fascinating developments. 
A strong motivation for the field remains to exploit the fermionic nature of the single particle excitations as well as the study of more exotic single particle excitations, such as anyons.
With the possibility to combine single electron injection and readout will open the avenue for the realisation of \textit{full} quantum experiments. It will allow to access the Full Counting Statistics (FCS) of electrons partitioned by a well-controlled interferometer and will bring the field of electron quantum optics to a status similar to that of quantum optics with photons. 
We expect the emergence of new concepts at the crossroads between quantum optics and solid state nanoelectronics.
From the quantum information point of view, new opportunities for linear quantum optics entanglement will emerge that are hardly possible with photons. Concepts such as universal quantum computing with loop based architectures as recently proposed for photonic systems \cite{furusawa_prl_2017} might also be more easily accessible in nanoelectronic devices using ultrashort voltage pulses.
Promising applications of the single-electron physics are also the possibility to verify very precisely signal shapes on-chip \cite{kataoka_arxiv_2016}, which may find applications in the on-chip control of quantum systems.}

%%%%%%%%%%%%%%%%%%%%%%%%%%%%%%%%%%%%%%%%%%%%%%%%%%%%%%%%%%%%%%%%%%%
\bibliography{ropp_biblio.bib}
%%%%%%%%%%%%%%%%%%%%%%%%%%%%%%%%%%%%%%%%%%%%%%%%%%%%%%%%%%%%%%%%%%%
%
%
%
%
%
%
\end{document}